\documentclass[prd,superscriptaddress,nofootinbib,floatfix,notitlepage]{revtex4-1}
\pdfoutput=1

\usepackage{amsmath,amssymb,epsfig}
\usepackage{hyperref}
\usepackage{natbib,ifthen}

\usepackage{mathrsfs}
\usepackage{simplewick}
\usepackage[lofdepth,lotdepth,caption=false]{subfig}
\usepackage{color}
\usepackage{bbold}

\begin{document}

\newcommand{\jcap}{JCAP}
\newcommand{\apjl}{APJL~}

\def\Btheo{{B_\delta^{\textrm{theo}}}}
\newcommand{\todo}[1]{{\color{red}{TODO: #1}}} 
\newcommand{\lbar}[1]{\underline{l}_{#1}}
\newcommand{\drm}{\mathrm{d}}
\renewcommand{\d}{\mathrm{d}}
\newcommand{\gaensli}[1]{\lq #1\rq$ $}
\newcommand{\bartilde}[1]{\bar{\tilde #1}}
\newcommand{\barti}[1]{\bartilde{#1}}
\newcommand{\ti}{\tilde}
\newcommand{\oforder}[1]{\mathcal{O}(#1)}
\newcommand{\D}{\mathrm{D}}
\renewcommand{\(}{\left(}
\renewcommand{\)}{\right)}
\renewcommand{\[}{\left[}
\renewcommand{\]}{\right]}
\def\<{\left\langle}
\def\>{\right\rangle}
\newcommand{\mycaption}[1]{\caption{\footnotesize{#1}}}
\newcommand{\hattilde}[1]{\hat{\tilde #1}}
\newcommand{\mycite}[1]{[#1]}
\newcommand{\mnras}{Mon.\ Not.\ R.\ Astron.\ Soc.}
\newcommand{\apjs}{Astrophys.\ J.\ Supp.}

\def\uk{{\bf \hat{k}}}
\def\un{{\bf \hat{n}}}
\def\ur{{\bf \hat{r}}}
\def\ux{{\bf \hat{x}}}
\def\bk{{\bf k}}
\def\bn{{\bf n}}
\def\br{{\bf r}}
\def\bx{{\bf x}}
\def\bK{{\bf K}}
\def\by{{\bf y}}
\def\bl{{\bf l}}
\def\bkp{{\bf k^\pr}}
\def\brp{{\bf r^\pr}}

\newcommand{\fixme}[1]{{\textbf{Fixme: #1}}}
\newcommand{\detD}{{\det\!\cld}}
\newcommand{\clh}{\mathcal{H}}
\newcommand{\ud}{{\rm d}}
\renewcommand{\eprint}[1]{\href{http://arxiv.org/abs/#1}{#1}}
\newcommand{\adsurl}[1]{\href{#1}{ADS}}
\newcommand{\ISBN}[1]{\href{http://cosmologist.info/ISBN/#1}{ISBN: #1}}
\newcommand{\vort}{\varpi}
\newcommand\ba{\begin{eqnarray}}
\newcommand\ea{\end{eqnarray}}
\newcommand\be{\begin{equation}}
\newcommand\ee{\end{equation}}
\newcommand\lagrange{{\cal L}}
\newcommand\cll{{\cal L}}
\newcommand\cln{{\cal N}}
\newcommand\clx{{\cal X}}
\newcommand\clz{{\cal Z}}
\newcommand\clv{{\cal V}}
\newcommand\cld{{\cal D}}
\newcommand\clt{{\cal T}}

\newcommand\clo{{\cal O}}
\newcommand{\cla}{{\cal A}}
\newcommand{\clp}{{\cal P}}
\newcommand{\clr}{{\cal R}}
\newcommand{\uD}{{\mathrm{D}}}
\newcommand{\calE}{{\cal E}}
\newcommand{\calB}{{\cal B}}
\newcommand{\curl}{\,\mbox{curl}\,}
\newcommand\del{\nabla}
\newcommand\Tr{{\rm Tr}}
\newcommand\half{{\frac{1}{2}}}
\newcommand\fourth{{1\over 8}}
\newcommand\bibi{\bibitem}
\newcommand{\kf}{\beta}
\newcommand{\kfprod}{\alpha}
\newcommand\calS{{\cal S}}
\renewcommand\H{{\cal H}}
\newcommand\K{{\rm K}}
\newcommand\mK{{\rm mK}}
\newcommand\synch{\text{syn}}
\newcommand\opacity{\tau_c^{-1}}

\newcommand{\Psil}{\Psi_l}
\newcommand{\bsigma}{{\bar{\sigma}}}
\newcommand{\bI}{\bar{I}}
\newcommand{\bq}{\bar{q}}
\newcommand{\bv}{\bar{v}}
\renewcommand\P{{\cal P}}
\newcommand{\numfrac}[2]{{\textstyle \frac{#1}{#2}}}

\newcommand{\la}{\langle}
\newcommand{\ra}{\rangle}
\newcommand{\lla}{\left\langle}
\newcommand{\rra}{\right\rangle}

\newcommand{\Omtot}{\Omega_{\mathrm{tot}}}
\newcommand\xx{\mbox{\boldmath $x$}}
\newcommand{\phpr} {\phi'}
\newcommand{\gam}{\gamma_{ij}}
\newcommand{\sqgam}{\sqrt{\gamma}}
\newcommand{\delk}{\Delta+3{\K}}
\newcommand{\dph}{\delta\phi}
\newcommand{\om} {\Omega}
\newcommand{\dom}{\delta^{(3)}\left(\Omega\right)}
\newcommand{\rar}{\rightarrow}
\newcommand{\Rar}{\Rightarrow}
\newcommand\gsim{ \lower .75ex \hbox{$\sim$} \llap{\raise .27ex \hbox{$>$}} }
\newcommand\lsim{ \lower .75ex \hbox{$\sim$} \llap{\raise .27ex \hbox{$<$}} }
\newcommand\bigdot[1] {\stackrel{\mbox{{\huge .}}}{#1}}
\newcommand\bigddot[1] {\stackrel{\mbox{{\huge ..}}}{#1}}
\newcommand{\Mpc}{\text{Mpc}}
\newcommand{\Al}{{A_l}}
\newcommand{\Bl}{{B_l}}
\newcommand{\eAl}{e^\Al}
\newcommand{\ix}{{(i)}}
\newcommand{\ixp}{{(i+1)}}
\renewcommand{\k}{\beta}
\newcommand{\HD}{\mathrm{D}}

\newcommand{\nonflat}[1]{#1}
\newcommand{\Cgl}{C_{\text{gl}}}
\newcommand{\Cgltwo}{C_{\text{gl},2}}
\newcommand{\He}{{\rm{He}}}
\newcommand{\Mhz}{{\rm MHz}}
\newcommand{\vx}{{\mathbf{x}}}
\newcommand{\ve}{{\mathbf{e}}}
\newcommand{\vv}{{\mathbf{v}}}
\newcommand{\vk}{{\mathbf{k}}}
\newcommand{\vn}{{\mathbf{n}}}
\newcommand{\vPsi}{{\mathbf{\Psi}}}

\newcommand{\vnhat}{{\hat{\mathbf{n}}}}
\newcommand{\vkhat}{{\hat{\mathbf{k}}}}
\newcommand{\taueps}{{\tau_\epsilon}}

\newcommand{\vgrad}{{\mathbf{\nabla}}}
\newcommand{\fbarln}{\bar{f}_{,\ln\epsilon}(\epsilon)}

\newcommand{\secref}[1]{Section \ref{#1}}
\newcommand{\expt}{\mathrm{expt}}
\newcommand{\eq}[1]{(\ref{eq:#1})} 
\newcommand{\eqq}[1]{Eq.~(\ref{eq:#1})} 
\newcommand{\fig}[1]{Fig.~\ref{fig:#1}} 
\renewcommand{\to}{\rightarrow}
\renewcommand{\(}{\left(}
\renewcommand{\)}{\right)}
\renewcommand{\[}{\left[}
\renewcommand{\]}{\right]}
\renewcommand{\vec}[1]{\mathbf{#1}}
\newcommand{\vy}{\vec{y}}
\newcommand{\vz}{\vec{z}}
\newcommand{\vq}{\vec{q}}
\newcommand{\VPsi}{\vec{\Psi}}
\newcommand{\vecv}{\vec{v}}
\newcommand{\vnabla}{\vec{\nabla}}
\newcommand{\vl}{\vec{l}}
\newcommand{\VL}{\vec{L}}
\newcommand{\dl}{\d^2\vl}
\newcommand{\valpha}{\vec{\alpha}}
\renewcommand{\L}{\mathscr{L}}

\newcommand{\abs}[1]{\lvert #1\rvert}

\newcommand{\ul}{\underline{l}}


\thispagestyle{empty}

\title{Near optimal bispectrum estimators for large-scale structure}

\author{Marcel Schmittfull}
\affiliation{Berkeley Center for Cosmological Physics, Department of
  Physics and Lawrence Berkeley
  National Laboratory, University of California, Berkeley, CA 94720, USA}

\author{Tobias Baldauf}
\affiliation{School of Natural Sciences, Institute for Advanced Study, Einstein Drive, Princeton, NJ 08540, USA}

\author{Uro\v{s} Seljak}
\affiliation{Department of Physics, Department of Astronomy and Lawrence Berkeley National Laboratory,
University of California, Berkeley, CA 94720, USA}

\date{\today}

\begin{abstract}
Clustering of large-scale structure provides significant cosmological information through the power spectrum of density perturbations. Additional information can be gained from higher-order statistics like the bispectrum, especially to break the degeneracy between the linear halo bias $b_1$ and the amplitude of fluctuations $\sigma_8$. We propose new simple, computationally inexpensive bispectrum statistics that are near optimal for the specific applications like bias determination. Corresponding to the Legendre decomposition of nonlinear halo bias and gravitational coupling at second order, these statistics are given by the cross-spectra of the density with three quadratic fields: the squared density, a tidal term, and a shift term. For halos and galaxies the first two have associated nonlinear bias terms $b_2$ and $b_{s^2}$, respectively, while the shift term has none in the absence of velocity bias (valid in the $k \rightarrow 0$ limit). Thus the linear bias $b_1$ is best determined by the shift cross-spectrum, while the squared density and tidal cross-spectra mostly tighten constraints on $b_2$ and $b_{s^2}$ once $b_1$ is known. Since the form of the cross-spectra is derived from optimal maximum-likelihood estimation, they contain the full bispectrum information on bias parameters. Perturbative analytical predictions for their expectation values and covariances agree with simulations on large scales, $k\lesssim 0.09h/\mathrm{Mpc}$ at $z=0.55$ with Gaussian $R=20h^{-1}\mathrm{Mpc}$ smoothing, for matter-matter-matter, and matter-matter-halo combinations. For halo-halo-halo cross-spectra the model also needs to include corrections to the Poisson stochasticity.
\end{abstract}

\maketitle

\section{Introduction}

Large-scale structure (LSS) surveys constrain cosmology with ever higher
precision (e.g.~\cite{Anderson1312} and references therein). However,
the unkown bias between invisible dark matter and observable tracers
introduces degeneracies which weaken constraints on cosmological
parameters. In particular, the normalization of the matter power
spectrum $\sigma_8$ is degenerate with the unknown linear bias $b_1$
if only the $2$-point correlation function or the power spectrum in
Fourier space are measured, because they only depend on the
combination $b_1^2\sigma_8^2$ (on large scales). This degeneracy is
reduced if anisotropic redshift-space distortions are
included. Further improvements can be obtained by probing statistics
beyond the $2$-point level such as the $3$-point correlation function
or the bispectrum in Fourier space, exploiting the fact that different
combinations of cosmological and bias parameters are associated with
different functional dependencies of the bispectrum on triangle shape.
This has been demonstrated on galaxy survey data in
\cite{2001ApJ...546..652S,2001PhRvL..86.1434F,2002MNRAS.335..432V,2004ApJ...607..140J,2004MNRAS.353..287W,2013MNRAS.432.2654M,2011ApJ...739...85M,2011ApJ...726...13M}. Recently,
Gil-Marin et al.~\cite{Hector1407Data1,Hector1408Data2} measured the
bispectrum of SDSS DR11 BOSS galaxies with unprecedented accuracy
obtaining constraints on the growth rate $f$ and the clustering
amplitude $\sigma_8$ from galaxy clustering data alone. An additional
important motivation to measure large-scale structure bispectra is to
constrain primordial non-Gaussianity that can help to distinguish
inflation models.

All analyses of real data to date have estimated the bispectrum or
$3$-point correlation function in a brute-force approach, going
manually over triangle configurations and orientations.  The number of
these possible triangles is very large because it is cubic in the
number of grid points per dimension (e.g.~choosing ten $k$-bins in
every direction gives $\mathcal{O}(10^3)$ triangles).  This leads to
several problems. First, the estimation of the bispectrum itself is
computationally expensive, which is typically overcome by considering
only a subset of possible triangles which comes however at the expense
of losing a potentially significant fraction of the signal in the
data. Second, the covariance between thousands or more triangles is
hard to estimate from simulations because it would require a large
number of independent realizations.\footnote{Analytical covariances can only
partially overcome this problem because it is hard to include
real-world effects from e.g.~the survey window function.  A possible
work-around is to neglect non-trivial bispectrum covariances in the
estimation procedure and only estimate the covariance between final
cosmological and bias parameters from simulations
\cite{hector1407theory,Hector1407Data1,Hector1408Data2}. However, this
leads to a sub-optimal estimator that is only unbiased in the limit of
infinitely many observations.}  Third, the task of testing theoretical
models of the bispectrum against simulations is rather complex because
it is not obvious which triangle configurations are relevant for final
parameter constraints,  making it nontrivial to find a suitable
metric for comparing models against simulations.  Lastly, modeling
anisotropic effects due to redshift-space distortions or survey
geometry at the level of triangles can be cumbersome.

To avoid some of these issues, the goal of this paper is to propose
alternative 
simpler statistics that do not require direct
bispectrum estimation for individual triangles but still include the full information of the
bispectrum on bias and cosmological parameters in an optimal way.  The
basic idea to achieve this is to square the density and cross-correlate it
with the density itself. This cross-spectrum then depends on the
bispectrum. 

In fact, the separability of dark matter (DM) and halo bispectra can be exploited to
show that optimal bispectrum estimation is equivalent to
cross-correlating three fields that are quadratic in the density with
the density itself. Due to the particular form of DM and halo
bispectra, these three quadratic fields are the
squared density in configuration space $\delta^2(\vx)$, a shift term that contracts the density
gradient with the displacement field,
$\Psi^i(\vx)\partial_i\delta(\vx)$, and a tidal term, $s^2(\vx)$. 
The cross-spectra of these quadratic fields
with the density encode the full bispectrum information but require
only the computational cost of typical power spectrum analyses, which
is much faster than brute-force bispectrum estimation.  Additionally,
since the cross-spectra only depend on a single scale $k$, covariances
are much simpler to estimate from simulations than covariances between
all triangle configurations.

The quadratic fields themselves can be computed efficiently as the
product of fields in configuration space at the same location $\vx$. The
factors are given by the density itself or related quantities like
e.g.~the density gradient $\partial_i\delta$ or the displacement field
$\Psi_i=-(\partial_i\partial^{-2})\delta$.  
The displacement field is already conventionally
computed from real data when employing reconstruction to sharpen the
BAO peak, so it should be straightforward to compute the other fields
we propose in a similar fashion.

One of the three proposed statistics, the cross-spectrum between the
squared density and the density, has been studied previously in the
literature \cite{Bernardeau:1996,Bel:2012,Hoffmann1403}, but mostly in
configuration space rather than Fourier space and in the large separation
limit. Some of these studies speculated that this cross-spectrum contains
information not present in the full three point function. The same
statistic was also analysed in Fourier space in \cite{Pollack1309}.
We will point out that the inverse variance weighted integral over
this statistic in Fourier space is in fact an optimal estimator for the
amplitude of the angle-independent part of the bispectrum, which is
sensitive to a combination of $b_1$ and $b_2$, and we calculate this
statistic consistently using one loop perturbation theory.
Furthermore, we are extending these studies in considering a more
realistic bias model and will show in detail that considering the
other two cross-spectra involving other transformations than squaring
are instrumental in constraining the bias parameters.  

Our proposed procedure is also
related to (but different from) a method recently proposed by Chiang et
al.~\cite{chiang1403}, which similarly aims at compressing bispectrum
information to simpler observable quantities. Their work considers the
correlation between locally measured power spectra with the local mean
density, which is determined by the squeezed limit of the bispectrum.
However, the squeezed limit of bispectra generated by nonlinear
gravity is suppressed in absence of primordial non-Gaussianity.  In
contrast, we systematically derive the types of cross-spectra that
explicitly probe bispectrum shapes that are optimal to estimate bias
parameters, arriving at different statistics than the one proposed in
\cite{chiang1403}. 

To some extent 
our method can  be regarded as a simplification of the
separable expansion method to estimate general bispectra
\cite{shellard1008,shellard1108,marcel1207} by tailoring it to estimate
bias parameters from large-scale structure bispectra, keeping the
scale-dependence more obvious by not integrating over
scales $k$.  The method proposed here could be combined with separable
expansion estimators to probe bispectrum contributions beyond those
included in our modeling, but we leave this for future work.

In the
context of the CMB, the fact that separable
bispectra and trispectra can be estimated more efficiently than
non-separable bispectra has been exploited for a long time. For
example, the KSW estimator \cite{ksw} is often used
to estimate separable primordial bispectra directly
(e.g.~$f_\mathrm{NL}^\mathrm{local}$), primordial bispectra are often
approximated by separable templates to facilitate their estimation,
the ISW-lensing bispectrum can be measured by cross-correlating the
lensing reconstruction field (which is quadratic in the CMB
temperature) with the CMB temperature \cite{lewis1101,Planck1303ISW},
and primordial and non-primordial general bispectra can be estimated
by expanding in separable basis functions
\cite{shellard0912,shellard1006,Planck1303NG}. Similarly, the
separability of the lensing-induced trispectrum of the CMB temperature
is exploited when computing the auto-power spectrum of the CMB
lensing reconstruction field, which is a quadratic function in the CMB
temperature with derivative operations dictated by the form of the CMB
trispectrum
\cite{ZaldarriagaSeljak98LensingRec,okamotoHu0301,duncan1008}.
However, the explicit separability of bispectra or trispectra induced
by nonlinear gravity and bias relations has so far not been exploited
in the field of large-scale structure.  The aim of this paper is to
provide a theoretical framework for this and test it with simulations.

We restrict ourselves to real space in this paper, but note
  that redshift space distortions (RSDs) should be included before
  applying the proposed technique to real data. Since RSDs
  change the expected bispectrum signal, in particular rendering it
  non-isotropic, the specific estimators we derive in real space
  should be modified and extended in order to be optimal in redshift
  space.  While this will likely increase the number of cross-spectra,
  making the analysis somewhat more complicated, 
  we do not anticipate any new conceptual challenges because the
  leading-order RSDs and certain Fingers-of-God models are still
  product-separable.  Since venturing into redshift
  space is beyond the scope of this paper we leave it for future
  work.  

The paper is organized as follows.  We
start with general definitions of quadratic fields and
bispectra in Section \ref{se:QuadraticFields}. Section \ref{se:MaxLikeliBispEsti} describes the
relationship between optimal bispectrum estimation and cross-spectra
of quadratic fields. Theoretical predictions for the cross-spectra are
computed in Section \ref{se:TheoryCrossSpectra}. Section
\ref{se:Simulations} describes simulation results in comparison with
theoretical predictions. An extension to primordial
non-Gaussianity is briefly discussed in Section
\ref{se:Extensions}.  Finally we conclude in Section
\ref{se:Conclusions}.  Two appendices provide technical details
of large-scale limits and analytical covariances.

\section{Quadratic fields and bispectrum decomposition}
\label{se:QuadraticFields}

\subsection{Quadratic fields}

As will be shown in Section \ref{se:MaxLikeliBispEsti}, maximum-likelihood bispectrum
estimators for bias parameters can be cast in form of cross-spectra of
the density field with three fields that are quadratic in the
configuration-space density, with different dependencies on the cosine $\mu$
 between the Fourier space wavevectors $\vq$ and $\vk-\vq$,
\begin{equation}
  \label{eq:mu_def}
  \mu \equiv \frac{\vq\cdot(\vk-\vq)}{q|\vk-\vq|}.
\end{equation}
Explicitly, these three quadratic fields are:
\begin{itemize}
\item
The \emph{squared
density} $\delta^2(\vx)$, which can be written as a convolution in
Fourier space,
\begin{equation}
  \label{eq:delta2_x}
  \delta^2(\vx) = \int\frac{\d^3 k}{(2\pi)^3}e^{i\vk\vx}
  \int\frac{\d^3 q}{(2\pi)^3} \mathsf{P}_0(\mu)\delta(\vq)\delta(\vk-\vq),
\end{equation}
where $\mathsf{P}_0(\mu)=1$ is the Legendre polynomial for
$l=0$.\footnote{$\mathsf{P}_l(\mu)$ with $l\in\{0,1,2\}$ always denotes Legendre
  polynomials in this paper and should not be confused with power spectra $P(k)$.} 
\item
The \emph{shift-term}
\begin{equation}
  \label{eq:shift_x}
-\Psi^i(\vx)\partial_i\delta(\vx)=
 -\vPsi(\vx)\cdot\nabla\delta(\vx)
= -\int\frac{\d^3 k}{(2\pi)^3}e^{i\vk\vx}\int\frac{\d^3 q}{(2\pi)^3}
F_2^1(q,|\vk-\vq|)\mathsf{P}_1(\mu)  \delta(\vq)\delta(\vk-\vq),
\end{equation}
which depends on the $l=1$ Legendre polynomial $\mathsf{P}_1(\mu)=\mu$
and is obtained by contracting the density gradient $\nabla\delta$
with the displacement field
\begin{equation}
  \label{eq:vPsi_def}
  \vPsi(\vk) = -\frac{i\vk}{k^2}\delta(\vk).
\end{equation}
The symmetric kernel $F_2^1$ in \eqq{shift_x} is defined as
\begin{equation}
  \label{eq:40}
  F_2^1(k_1,k_2) = \frac{1}{2}\left( \frac{k_1}{k_2} + \frac{k_2}{k_1} \right).
\end{equation}
\item
The \emph{tidal term}\footnote{An overall factor of $3/2$ is absorbed
  compared to e.g.~\cite{tobias1201},
  i.e.~$s^2_\mathrm{here}(\vx)=\frac{3}{2}s^2_\mathrm{there}(\vx)$ and
  $\mathsf{P}_{2}^\mathrm{here}(\mu)=\frac{3}{2}S_2^\mathrm{there}(\vq,\vk-\vq)$.}
\begin{equation}
  \label{eq:s2_x}
  s^2(\vx)\equiv \frac{3}{2} s_{ij}(\vx)s_{ij}(\vx) =
\int\frac{\d^3 k}{(2\pi)^3}e^{i\vk\vx}
\int \frac{\d^3 q}{(2\pi)^3}\mathsf{P}_2(\mu)
\delta(\vq)\delta(\vk-\vq),
\end{equation}
which is defined by contracting 
the tidal tensor
\begin{equation}
  \label{eq:sij_vk}
  s_{ij}(\vk) = \left(
\frac{k_ik_j}{k^2} - \frac{1}{3}\delta^\mathrm{(K)}_{ij}
\right)\delta(\vk)
\end{equation}
with itself. $\delta^\mathrm{(K)}_{ij}$ denotes the Kronecker delta.
The corresponding convolution kernel in \eqq{s2_x} is given by the $l=2$ Legendre polynomial
\begin{equation}
  \label{eq:46}
  \mathsf{P}_2(\mu) = \frac{3}{2}\left(\mu^2-\frac{1}{3}\right).
\end{equation}
\end{itemize}

While these three quadratic fields will be derived more rigorously
below, some intuition for their appearance can be gained as
follows. In standard perturbation theory (see \cite{BernardeauReview}
for a review) the dark matter density is expanded in powers of the
linear perturbation $\delta_0$.
In configuration space, truncating at second order, this can be written as
(e.g.~\cite{Bouchet1992ApJ...394L...5B,tobias1201,SherwinZaldarriaga1202})
\begin{equation}
  \label{eq:delta_2ndorder_x}
  \delta_\mathrm{m}(\vx) = \delta_0(\vx) + \frac{17}{21} \delta^2_0(\vx) + \vPsi_0(\vx)\cdot
  \nabla \delta_0(\vx) + \frac{4}{21}s_0^2(\vx).
\end{equation}
The biased halo density can be modeled by
\cite{McDonaldRoy09,tobias1201,KwanScoccimarro1201}
\begin{equation}
  \label{eq:bias_relation}
  \delta_\mathrm{h}(\vx) = b_1\delta_\mathrm{m}(\vx) + b_2\left[
\delta_\mathrm{m}^2(\vx)-\langle \delta_\mathrm{m}^2(\vx)\rangle\right]
+\frac{2}{3}b_{s^2}\left[s_\mathrm{m}^2(\vx)-\langle s_\mathrm{m}^2(\vx)\rangle\right],
\end{equation}
where the normalization of the bias parameters $b_1$, $b_2$ and
$b_{s^2}$ is the same as in \cite{tobias1201}.  Since the linear
perturbation $\delta_0$ is assumed to be Gaussian, the leading order
$3$-point function is due to expectation values of the form
$\langle\delta^{(2)}\delta_0\delta_0\rangle$, where $\delta^{(2)}$ can
be any of the terms in Eqs.~\eq{delta_2ndorder_x} and
\eq{bias_relation} that are quadratic in $\delta_0$. There are only
three types of such terms; the squared density \eq{delta2_x}, the
shift term \eq{shift_x} and the tidal term \eq{s2_x}. The
corresponding $k$-dependencies that these fields imprint on the
bispectra will later be used to reduce optimal bispectrum estimation
to cross-spectra of these quadratic fields with the density.

To simplify calculations in the rest of the paper we define a general
quadratic field $D[\delta](\vk)$ obtained from a density realization
$\delta(\vk)$ and some kernel $D(\vq, \vk-\vq)$, always assumed to be
symmetric in its arguments,
by\footnote{$D[\delta]$ with square brackets denotes the functional
  that turns some density realization $\delta(\vk)$ into the r.h.s.~of
  \eqq{D_functional}. $D(\vq,\vk-\vq)$ denotes the corresponding
  kernel.}
\begin{equation}
  \label{eq:D_functional}
  D[\delta](\vk) \equiv \int\frac{\d^3 q}{(2\pi)^3}D(\vq,\vk-\vq)\delta(\vq)\delta({\vk-\vq}).
\end{equation}
The squared density \eq{delta2_x} corresponds to the identity kernel
$D(\vq,\vk-\vq)=\mathsf{P}_0(\mu)=1$, the shift term \eq{shift_x} to
$D(\vq,\vk-\vq)=-F_2^1(q,|\vk-\vq|)\mathsf{P}_1(\mu)$ and the tidal term \eq{s2_x} to
$D(\vq,\vk-\vq)=\mathsf{P}_2(\mu)$.  We will usually suppress the arguments and
just write $D\in\{\mathsf{P}_0, -F_2^1\mathsf{P}_1, \mathsf{P}_2\}$.

\subsection{Cross-spectra}

The cross-spectrum of a quadratic field $D[\delta]$ with the density is
\begin{equation}
  \label{eq:32}
  \langle
  D[\delta](\vk)\delta(\vk')
\rangle \equiv (2\pi)^3 \delta_D(\vk+\vk')P_{D[\delta],\delta}(k),
\end{equation}
where $\delta_D$ always denotes the Dirac delta function. Given a
realization of the density, an unbiased estimator for the
cross-spectrum is given by 
\begin{equation}
  \label{eq:hat_cross_power}
  \hat P_{D[\delta],\delta}(k) = \frac{1}{4\pi L^3}\int
  \d\Omega_{\hat\vk}
D[\delta](\vk)\delta(-\vk),
\end{equation}
where $L$ is the box size.
The normalization is fixed by noting that
$(2\pi)^3\delta_D(\mathbf{0})=L^3$  and requiring 
\begin{equation}
  \label{eq:39}
\langle  \hat P_{D[\delta],\delta}(k)\rangle = P_{D[\delta],\delta}(k).
\end{equation}
On a discrete grid the angular integral is replaced by a sum over
discrete $\vk$ vectors with wavenumber $|\vk|$ belonging to the bin
centered at $k$,
\begin{equation}
  \label{eq:discrete_sum_over_modes}
\frac{1}{4\pi}  \int \d\Omega_{\hat\vk} \quad
\rightarrow
\quad
\frac{1}{N_\mathrm{modes}(k)}\sum_{\vk, [k-\Delta k/2\le |\vk|\le k+\Delta
k/2]},
\end{equation}
where $N_\mathrm{modes}(k)= 4\pi (k/\Delta k)^2$ at high $k$ if the binning
width is $\Delta k$. In practice, we count the number of
modes manually in the code because this is more accurate at low $k$.

\subsection{Legendre decomposition of gravitational bispectra}
At leading order in perturbation theory the DM bispectrum is given
by\footnote{The $\vk_i$ must be such that they form a closed triangle,
  $\vk_1+\vk_2+\vk_3=0$.  Up to the overall orientation, the triangle
  can be specified e.g.~by three sidelengths $k_1$, $k_2$ and $k_3$,
  which fixes the angle between any two sides,
  e.g.~$\vk_1\cdot\vk_2=(k_3^2-k_1^2-k_2^2)/2$.  We will parametrize
  triangles by sidelengths and/or angles, using whichever is most
  convenient in the context.}
\begin{equation}
\label{eq:BDM}
  B_\mathrm{mmm}(k_1, k_2, k_3) = 2P_\mathrm{mm}^\mathrm{lin}(k_1)P_\mathrm{mm}^\mathrm{lin}(k_2)F_2(\vk_1,\vk_2)+\mbox{ 2 perms in }\, k_1,k_2,k_3,
\end{equation}
where $P^\mathrm{lin}_\mathrm{mm}$ is the
linear DM power spectrum and  $F_2$ denotes the symmetrized kernel for
the second order density perturbation,
\begin{equation}
  \label{eq:F2}
  F_2(\vk_1,\vk_2) = \frac{17}{21} + \frac{1}{2}\left(
\frac{k_1}{k_2}+\frac{k_2}{k_1}
\right) \hat\vk_1\cdot\hat\vk_2
+
\frac{4}{21}\,\frac{3}{2}\left((\hat\vk_1\cdot\hat\vk_2)^2 -\frac{1}{3}\right),
\end{equation}
where $\hat \vk_i=\vk_i/k_i$.
The decomposition of this kernel in Legendre polynomials in the cosine
$\hat\vk_1\cdot\hat\vk_2$ is 
\begin{equation}
  \label{eq:F2LegendreDecomp}
F_2(\vk_1,\vk_2) = \sum_{l=0}^2 F_2^{l}(k_1,k_2)\mathsf{P}_l(\hat\vk_1\cdot\hat\vk_2)
\end{equation}
with  coefficients $F_2^l$ given by
\begin{eqnarray}
  F_{2}^{0}(k_1,k_2) &=& \frac{17}{21},\\
  F_{2}^{1}(k_1,k_2) &=&
\frac{1}{2}\left(\frac{k_1}{k_2}+\frac{k_2}{k_1}\right),
\\
  F_{2}^{2}(k_1,k_2) &=& 
\frac{4}{21}.
\end{eqnarray}
The bispectrum \eq{BDM} can thus be split
 in parts that depend on different Legendre
polynomials in the angle $\hat\vk_1\cdot\hat\vk_2$ (and permutations),
\begin{eqnarray}
  \label{eq:B_split_in_Bmu_parts}
  B_\mathrm{mmm}(k_1,k_2,k_3)
&=&\sum_{l=0}^2 B^{(l)}_\mathrm{mmm}(k_1,k_2,k_3),
\end{eqnarray}
where
\begin{equation}
  \label{eq:BDM_l}
  B_\mathrm{mmm}^{(l)}(k_1,k_2,k_3) =
  2P_\mathrm{mm}^\mathrm{lin}(k_1)P_\mathrm{mm}^\mathrm{lin}(k_2)F_2^{l}(k_1,k_2)\mathsf{P}_l(\hat\vk_1\cdot\hat\vk_2) + \mbox{2 perms}.
\end{equation}

The leading-order unsymmetrized matter-matter-halo bispectrum can be
obtained from the bias relation \eq{bias_relation},
\begin{align}
\label{eq:Bmmh_unsymm}
  B_\mathrm{mmh}^\mathrm{unsym}(k_1,k_2,k_3) =
&\;b_1B_\mathrm{mmm}(k_1,k_2,k_3)
+ 2b_2P_\mathrm{mm}^\mathrm{lin}(k_1)P_\mathrm{mm}^\mathrm{lin}(k_2) 
+
\frac{4}{3}b_{s^2}P_\mathrm{mm}^\mathrm{lin}(k_1)P_\mathrm{mm}^\mathrm{lin}(k_2)\mathsf{P}_2(\hat\vk_1\cdot\hat\vk_2),
\end{align}
where $k_3$
is for the halo density. 
This can similarly be decomposed into $l=0,1,2$
Legendre polynomials:
\begin{eqnarray}
  \label{eq:Bmmh_l0}
  B_\mathrm{mmh}^{\mathrm{unsym},(0)}(k_1,k_2,k_3) &=& b_1 B_\mathrm{mmm}^{(0)}(k_1,k_2,k_3) + 
2b_2P_\mathrm{mm}^\mathrm{lin}(k_1)P_\mathrm{mm}^\mathrm{lin}(k_2),\\
  \label{eq:Bmmh_l1}
  B_\mathrm{mmh}^{\mathrm{unsym},(1)}(k_1,k_2,k_3) &=& b_1
  B_\mathrm{mmm}^{(1)}(k_1,k_2,k_3),\\
  \label{eq:Bmmh_l2}
  B_\mathrm{mmh}^{\mathrm{unsym},(2)}(k_1,k_2,k_3) &=& b_1 B_\mathrm{mmm}^{(2)}(k_1,k_2,k_3) + 
\frac{4}{3}b_{s^2}P_\mathrm{mm}^\mathrm{lin}(k_1)P_\mathrm{mm}^\mathrm{lin}(k_2)\mathsf{P}_2(\hat\vk_1\cdot\hat\vk_2).
\end{eqnarray}

Similarly, the halo-halo-halo bispectrum
\begin{equation}
  \label{eq:Bhhh}
  B_\mathrm{hhh}(k_1,k_2,k_3) 
=
2P_\mathrm{mm}^\mathrm{lin}(k_1)P_\mathrm{mm}^\mathrm{lin}(k_2)\left[
b_1^3F_2(\vk_1,\vk_2)+b_1^2b_2+\frac{2}{3}b_1^2b_{s^2}\mathsf{P}_2(\hat\vk_1\cdot\hat\vk_2)
\right]+\mbox{2 perms}
\end{equation}
 can be decomposed into
\begin{eqnarray}
  \label{eq:Bhhh_l0}
  B_\mathrm{hhh}^{(0)}(k_1,k_2,k_3) &=& b_1^3 B_\mathrm{mmm}^{(0)}(k_1,k_2,k_3) + 
2b_1^2b_2\left[P_\mathrm{mm}^\mathrm{lin}(k_1)P_\mathrm{mm}^\mathrm{lin}(k_2)+\mbox{2
    perms}\right],\\
  \label{eq:Bhhh_l1}
  B_\mathrm{hhh}^{(1)}(k_1,k_2,k_3) &=& b_1^3
  B_\mathrm{mmm}^{(1)}(k_1,k_2,k_3),\\
  \label{eq:Bhhh_l2}
  B_\mathrm{hhh}^{(2)}(k_1,k_2,k_3) &=& b_1^3 B_\mathrm{mmm}^{(2)}(k_1,k_2,k_3) + 
\frac{4}{3}b_1^2b_{s^2}\left[P_\mathrm{mm}^\mathrm{lin}(k_1)P_\mathrm{mm}^\mathrm{lin}(k_2)\mathsf{P}_2(\hat\vk_1\cdot\hat\vk_2)
+\mbox{2 perms}\right].
\end{eqnarray}
Since the $l=1$ part depends only on $b_1$ and is not
  contaminated by
the nonlinear bias parameters  $b_2$ and $b_{s^2}$ we expect it to be
most powerful for determining $b_1$. The $l=0$ contribution depends on a
mixture of $b_1$ and $b_2$, so one could constrain $b_2$ if $b_1$ was
known. Similarly, the $l=2$ part depends on $b_1$ and $b_{s^2}$, so
$b_{s^2}$ can be constrained once $b_1$ is known.

Note that all bispectrum contributions are product-separable and only
depend on $l=0,1,2$ Legendre polynomials.  In the following section we
will discuss how this can be used to simplify estimators for the
amplitude of bispectrum contributions, which can in turn be used to
estimate bias parameters.

\section{Maximum likelihood bispectrum estimation} 
\label{se:MaxLikeliBispEsti}
\subsection{General bispectra}
Assuming a fiducial theoretical power spectrum $P_\delta(k)$ and
bispectrum $f_\mathrm{NL}^{B_\mathrm{theo}}\Btheo$ for the density
perturbation $\delta$, the maximum likelihood estimator (in the limit
of weak non-Gaussianity) for the amplitude\footnote{In our 
  notation $f_\mathrm{NL}^B$ denotes the nonlinearity amplitude
  of an arbitrary bispectrum $B$, no matter if it is generated by
  primordial non-Gaussianity or nonlinear gravity.}
$f_\mathrm{NL}^{B_\mathrm{theo}}$ of the bispectrum is given by
\cite{shellard1008,shellard1108}
\begin{align}
  \label{eq:fnl_esti}
  \hat{f}_\mathrm{NL}^{B^\mathrm{theo}} =&
  \frac{(2\pi)^3}{N_\mathrm{theo}} \int
\frac{\d^3 k}{(2\pi)^3}\int\frac{\d^3q}{(2\pi)^3}
\frac{
  \Btheo(\vq,\vk-\vq,-\vk)[\delta(\vq)\delta(\vk-\vq)\delta(-\vk)-3\langle
\delta(\vq)\delta(\vk-\vq)\rangle \delta(-\vk)]}{P_\delta(q)P_\delta(|\vk-\vq|)P_\delta(k)}
  \end{align}
  where $P_\delta$ factors in the denominator represent
  inverse-variance weighting of the observed density perturbation
  $\delta$, and $N_\mathrm{theo}$ is a normalization factor depending
  on the theoretical bispectrum whose amplitude we aim to measure (see
  \cite{marcel1207} for the explicit definition). The term
 linear in the observed density will be
  omitted in the following because it is only relevant if the field is
  statistically inhomogeneous (e.g.~due to inhomogenous noise).

The estimator of \eqq{fnl_esti} is unbiased and includes information
from the full bispectrum, i.e.~all triangle configurations and
orientations. It is optimal in the limit of weak non-Gaussianity
since it is derived by Edgeworth-expanding the likelihood around a
Gaussian pdf truncating higher orders of $f_\mathrm{NL}$ and connected
$n$-point functions beyond the bispectrum. The
inverse-variance weighting of the density in
\eqq{fnl_esti} is optimal if higher-order contributions to the
bispectrum covariance are neglected. The estimator can be improved by
relaxing these assumptions, but we leave this to future
work.\footnote{Alternatively, constraints can be tightened by 
pushing the analysis to smaller scales, which could be achieved by improving
the theory modeling of the bispectrum (see
  e.g.~\cite{scocci_fitting_formula,hector1111,hector1407theory,tobias1406EFT,angulo1406EFT}),
  or by applying clipping or logarithmic density transforms 
  \cite{Neyrinck0903,SimpsonClipping1107}.}

\subsection{Estimating separable bispectra by cross-correlating linear and
  quadratic fields}
The bispectra discussed above are given by sums
  of terms that have a product-separable form.\footnote{At leading
    order this is the case because the second order density
    \eq{delta_2ndorder_x} and the halo density \eq{bias_relation}
    depend on products of fields evaluated at the same location $\vx$
    which can be written as convolutions with separable kernels in
    Fourier space.}  For simplicity, let us start with a
 bispectrum given by just one such product-separable term
\begin{equation}
  \label{eq:generic_separable_bisp_fgh}
  B_\delta^\mathrm{theo}(\vk_1,\vk_2,\vk_3) = 
f(\vk_1)g(\vk_2)h(\vk_3)  
\end{equation}
for some functions $f, g$ and $h$. Then the estimator \eq{fnl_esti} becomes
\begin{equation}
  \label{eq:fnl_split_integrals}
   \hat{f}_\mathrm{NL}^{B^\mathrm{theo}} 
=
  \frac{(2\pi)^3}{N_\mathrm{theo}} \int
\frac{\d^3 k}{(2\pi)^3} 
\left[
\int \frac{\d^3 q}{(2\pi)^3} 
\frac{f(\vq)\delta(\vq)}{P_\delta(q)}
\frac{g(\vk-\vq)\delta(\vk-\vq)}{P_\delta(|\vk-\vq|)}
\right]
\frac{h(-\vk)\delta(-\vk)}{P_\delta(k)}.
\end{equation}
The integral over $\vq$ is a convolution of two filtered densities
$f\delta/P_\delta$ and $g\delta/P_\delta$,  which we denote as
\begin{equation}
  \label{eq:fg_quadratic_field}
\left[\frac{f\delta}{P_\delta} * \frac{g\delta}{P_\delta}\right](\vk)
 \equiv \int\frac{\d^3q}{(2\pi)^3}
\frac{
f(\vq)\delta(\vq)}{P_\delta(q)}
\frac{g(\vk-\vq)\delta(\vk-\vq)}{P_\delta(|\vk-\vq|)}.
\end{equation}
Then, the estimator \eq{fnl_split_integrals} simplifies to
\begin{eqnarray}
\nonumber
      \hat{f}_\mathrm{NL}^{B^\mathrm{theo}} 
&=&
  \frac{1}{N_\mathrm{theo}} \int
\d k\frac{ k^2}{P_\delta(k)}\int\d\Omega_{\hat\vk}
\left[\frac{f\delta}{P_\delta} * \frac{g\delta}{P_\delta}\right]\!(\vk)\,
[h\delta](-\vk) 
\\
  \label{eq:fnl_separable_fgh}
&=&
  \frac{4\pi L^3}{N_\mathrm{theo}} \int
\d k \frac{k^2}{P_\delta(k)}
 \hat P_{\frac{f\delta}{P_\delta} *
 \frac{g\delta}{P_\delta},\,h\delta}(k),  
\end{eqnarray}
where we used \eqq{hat_cross_power}.
This is a (weighted) integral over the estimated cross-spectrum
between the
quadratic field \eq{fg_quadratic_field} and the 
 filtered density
\begin{equation}
  \label{eq:54}
[h\delta](\vk)
\equiv h(\vk)\delta(\vk).
\end{equation}
Rather than performing the 
integration of the cross-spectrum over wavenumbers $k$ in
\eqq{fnl_separable_fgh}, it is useful to consider the cross-spectrum
as the fundamental observable. This can simplify comparisons of theory,
simulations and observations as a function of scale $k$, and offers
the possibility to incorporate covariances between cross-spectra.

The convolution in \eqq{fg_quadratic_field} can be calculated efficiently by filtering the
density in $k$-space with two filters $f/P$ and $g/P$, Fourier
transforming to configuration space, multiplying the fields with each other in configuration space, and
Fourier transforming back to $k$-space
(e.g.~for $f=g=P_\delta$, \eqq{fg_quadratic_field} is the Fourier transform of
$\delta^2(\vx)$).

If the theoretical bispectrum consists of a sum of separable terms, each
term gives rise to a cross-spectrum of accordingly filtered
densities.  Each cross-spectrum can be measured to analyse
individual contributions to the bispectrum, or they can be combined to
constrain the overall bispectrum amplitude.

\subsection{Gravitational bispectrum and  bias estimators}
Eqs.~\eq{BDM_l}, \eq{Bmmh_l0}-\eq{Bmmh_l2} and \eq{Bhhh_l0}-\eq{Bhhh_l2} show that
matter-matter-matter, matter-matter-halo and halo-halo-halo bispectra
can be constructed from contributions of the form\footnote{In all
  equations $l$ is fixed to $l=0,1,2$, unless we explicitly write
  $\sum_l$ to sum over $l$. We never use Einstein summation notation
  for $l$.}
\begin{equation}
  \label{eq:Blunsym}
  B^{(l)}_\mathrm{unsym}(\vk_1,\vk_2,\vk_3) \equiv
  2P^\mathrm{lin}_\mathrm{mm}(k_1)P^\mathrm{lin}_\mathrm{mm}(k_2)F_2^{l}(k_1,k_2)\mathsf{P}_l(\hat\vk_1\cdot\hat\vk_2), 
\end{equation}
which are not symmetric in the $k_i$.  Note that every $F_2^{l}\mathsf{P}_l$
kernel is separable (or a sum of separable terms), because squaring
the triangle condition $\vk_1+\vk_2=-\vk_3$ implies $\vk_1\cdot\vk_2 =
\tfrac{1}{2} (k_3^2-k_1^2-k_2^2)$. Consequently,
the maximum likelihood estimator for the amplitude of a bispectrum
contribution \eq{Blunsym} can be expressed as an integral over a
cross-spectrum between a quadratic field and the density. Explicitly, the
maximum likelihood estimator \eq{fnl_esti} for the amplitude of a
bispectrum \eq{Blunsym} involving the densities $\delta_a$, $\delta_a$ and $\delta_b$
(for $a,b\in\{\mathrm{m},\mathrm{h}\}$ labeling DM or halo densities)
gives\footnote{Note that the estimators for the $l=0,1,2$ contributions to the full
symmetric gravitational bispectrum $B^{(l)}$ contain an additional factor of
$3$ because  \eqq{Blunsym} picked one out of three permutations in the
$k_i$.}  
\begin{equation}
  \label{eq:fnl_Bgrav_unsym_general}
\hat{f}^{B^{(l)}_\mathrm{unsym}}_\mathrm{NL}  
=
\frac{8\pi L^3}{N_{B^{(l)}_\mathrm{unsym}}}
\int\d k \frac{ k^2}{P_{bb}(k)}
\hat P_{\tilde F_2^{l}\mathsf{P}_l[\delta_a], \,\delta_b}(k)
\end{equation}
where we  defined
\begin{equation}
  \label{eq:general_f2tildeell_applied_to_delta_a}
  \tilde F_2^{l}\mathsf{P}_l[\delta_a](\vk) \equiv
  \int\frac{\d^3q}{(2\pi)^3}
\frac{P^\mathrm{lin}_\mathrm{mm}(q)P^\mathrm{lin}_\mathrm{mm}(|\vk-\vq|)}{P_{aa}(q)P_{aa}(|\vk-\vq|)}
F_2^{l}(q,|\vk-\vq|)\mathsf{P}_l(\mu)
\delta_a(\vq)\delta_a(\vk-\vq),
\end{equation}
where $\mu$ is the cosine \eq{mu_def} between $\vq$ and $\vk-\vq$. 
  The power spectrum ratio in the integrand
serves as a weight which becomes unity  on large scales for dark
matter.  To gain intuition, we choose a
slightly less optimal weight by setting the power spectrum ratio to
unity on all scales\footnote{If $\delta_a$ is the halo
  density this implies a sub-optimal weighting at high $k$, but
  results are not biased because we treat simulations and theory
  consistently. Also, high $k$ modes are suppressed by smoothing,
  which we apply before squaring any fields to suppress nonlinear mode
  coupling, similarly to cutting off bispectrum analyses at some
  maximum wavenumber $k_\mathrm{max}$. We do this because we do not have a 
reliable way to model these high $k$ modes. In our simulations, the power
  ratio $P^\mathrm{lin}_\mathrm{mm}/P_\mathrm{hh}$ drops by $30\%$ or
  less between $k\rightarrow 0$ and $k=0.3h/\mathrm{Mpc}$. All
  smoothing kernels we use asymptote to $0$ much faster for increasing
  $k$, so that corrections due to the power spectrum ratio weight are
  likely small.} (dropping the tilde on $F_2$),
\begin{equation}
  \label{eq:general_f2ell_applied_to_delta_a}
 F_2^{l}\mathsf{P}_l[\delta_a](\vk) \equiv
  \int\frac{\d^3q}{(2\pi)^3}F_2^{l}(q,|\vk-\vq|)\mathsf{P}_l(\mu)
\delta_a(\vq)\delta_a(\vk-\vq).
\end{equation}

In configuration space, the quadratic fields
\eq{general_f2ell_applied_to_delta_a} are proportional to
 \begin{eqnarray}
   \label{eq:55}
   \mathsf{P}_0[\delta](\vx) &=& 
\delta^2(\vx)\\
   -F_2^{1}\mathsf{P}_1[\delta](\vx) &=& 
-\Psi^i(\vx)\partial_i\delta(\vx)\\
\mathsf{P}_2[\delta](\vx) &=& 
s^2(\vx).
 \end{eqnarray}
 According to \eqq{fnl_Bgrav_unsym_general} the (integrated)
 cross-spectra of these three quadratic fields with the density are
 optimal estimators for the amplitude of the $l=0,1,2$ contributions
 to the gravitational bispectrum,
 \begin{eqnarray}
\label{eq:fnl_B0_k}
     \hat{f}^{B^{(0)}_\mathrm{unsym}}_\mathrm{NL}  
&=&
\frac{17}{21}\frac{8\pi L^3}{N_{B^{(0)}_\mathrm{unsym}}}
\int\d k \frac{ k^2}{P_{bb}(k)}
\hat P_{\delta_a^2, \,\delta_b}(k)
\\
\label{eq:fnl_B1_k}
  \hat{f}^{B^{(1)}_\mathrm{unsym}}_\mathrm{NL}  
&=&
-\frac{8\pi L^3}{N_{B^{(1)}_\mathrm{unsym}}}
\int\d k \frac{ k^2}{P_{bb}(k)}
\hat P_{-\Psi^i_a\partial_i\delta_a, \,\delta_b}(k)
\\
\label{eq:fnl_B2_k}
  \hat{f}^{B^{(2)}_\mathrm{unsym}}_\mathrm{NL}  
&=&
\frac{4}{21}
\frac{8\pi L^3}{N_{B^{(2)}_\mathrm{unsym}}}
\int\d k \frac{ k^2}{P_{bb}(k)}
\hat P_{s_a^2, \,\delta_b}(k).
 \end{eqnarray}

These nonlinearity amplitudes $\hat f_\mathrm{NL}$ depend on bias
parameters, e.g.~on $b_1^3$, $b_1^2b_2$ and $b_1^2b_{s^2}$ for
halo-halo-halo cross-spectra. Bias parameters can therefore be
estimated by combining the measured nonlinearity amplitudes
appropriately. Similarly, the bias parameters can be obtained by
jointly fitting them to the three measured cross-spectra $\hat
P_{D[\delta_a]\delta_b}(k)$ for $D\in\{\mathsf{P}_0, -F_2^1\mathsf{P}_1, \mathsf{P}_2\}$.

These cross-spectra contain the entire information that a full
bispectrum analysis would yield, if we are only interested in the
amplitudes of fixed shape contributions to the bispectrum, which is
often the case, e.g.~for estimating bias parameters.

As will be discussed later, the $k^2/P_{bb}(k)$ weighting of the
cross-spectra corresponds to inverse-variance weighting in the limit
of $k\rightarrow 0$ assuming Gaussian bispectrum covariance. This
weighting is improved if cross-spectrum variances and covariances
obtained from $N$-body simulations or mock catalogues are used to fit
cross-spectrum models to measurements. Estimating covariances from
simulations is computationally much less expensive for cross-spectra
than for bispectra because the three cross-spectra depend only on a single
argument $k$ whereas bispectra depend on triangle configurations
$(k_1,k_2,k_3)$, of which there can be many thousands, especially if
fine $k$-binning is used to distinguish different bispectrum shapes.

\subsection{Configuration space estimators}
We mainly worked in Fourier space so far which has the
  advantage that different modes are uncorrelated at leading
  order. Sometimes it is more convenient to work in configuration
  space because it may be easier to include effects that are localized in
  configuration space (e.g.~the survey selection function).
It turns out that the optimal
bispectrum estimator \eq{fnl_separable_fgh} for the amplitude of a
generic separable bispectrum \eq{generic_separable_bisp_fgh} can be
rewritten in configuration space instead of Fourier space as\footnote{To see
  this, express the quadratic field in \eq{fnl_separable_fgh} as $\int
  \d^3 x e^{-i\vk\vx} \frac{f\delta}{P}(\vx)\frac{g\delta}{P}(\vx)$
  and the linear field as $\int \d^3 y
  e^{i\vk\vy}\frac{h\delta}{P}(\vy)$. Then, integrating over $\vk$
  gives $\vy=\vx$ and the result \eq{fnlBtheoRealSpace}
  follows. Alternatively, \eqq{fnlBtheoRealSpace} can be derived by
  introducing $\int \d^3 q' \delta_D(\vq'-\vk+\vq)$ in
  \eqq{fnl_split_integrals} and using $(2\pi)^3\delta_D(\vk)=\int \d^3
  x e^{i\vk\vx}$.}
\begin{equation}
  \label{eq:fnlBtheoRealSpace}
\hat{f}_\mathrm{NL}^{B^\mathrm{theo}} 
=
  \frac{(2\pi)^3}{N_\mathrm{theo}} 
\int \d^3 x\,
\frac{f\delta}{P_\delta}(\vx)
\frac{g\delta}{P_\delta}(\vx)
\frac{h\delta}{P_\delta}(\vx),
\end{equation}
where the filtered densities are defined as
\begin{equation}
  \label{eq:53}
  \frac{f\delta}{P_\delta}(\vx) =
\int \frac{\d ^3k}{(2\pi)^3}e^{i\vk\vx}\frac{f(\vk)\delta(\vk)}{P_\delta(k)},
\end{equation}
and similarly if $f$ is replaced by $g$ and $h$. If one wants to avoid
Fourier space entirely, the filtering can be performed with
convolutions in configuration space.

For the gravitational bispectrum contributions
$B^{(l)}_\mathrm{unsym}$ we get\footnote{Again, we assume
  $P_\delta=P^\mathrm{lin}_\mathrm{mm}$ to get a simple weighting, but
  it would be straightforward to include the optimal
  $P_\mathrm{mm}^\mathrm{lin}/P_\delta$ weighting.  }
\begin{eqnarray}
  \label{eq:fnl_B0_x}
  \hat{f}^{B^{(0)}_\mathrm{unsym}}_\mathrm{NL}  
&=&
\frac{34}{21}\frac{(2\pi)^3}{N_{B^{(0)}_\mathrm{unsym}}}\int\d^3 x\,
\delta_a^2(\vx)\frac{\delta_b}{P_{bb}}(\vx),
\\
\label{eq:fnl_B1_x}
  \hat{f}^{B^{(1)}_\mathrm{unsym}}_\mathrm{NL}  
&=&
2\frac{(2\pi)^3}{N_{B^{(1)}_\mathrm{unsym}}}\int\d^3 x\,
\Psi^i_a(\vx)[\partial_i\delta_a(\vx)]\frac{\delta_b}{P_{bb}}(\vx),
\\
\label{eq:fnl_B2_x}
  \hat{f}^{B^{(2)}_\mathrm{unsym}}_\mathrm{NL}  
&=&
\frac{8}{21} \frac{(2\pi)^3}{N_{B^{(2)}_\mathrm{unsym}}}\int\d^3 x\,
s_a^2(\vx)\frac{\delta_b}{P_{bb}}(\vx).
\end{eqnarray}

\section{Theory cross-spectra}
\label{se:TheoryCrossSpectra}

\subsection{Smoothing}

To suppress small-scale modes, we apply a smoothing filter $W_R(k)$ to
the nonlinear (DM or halo)
density,
\begin{equation}
  \label{eq:smoothing}
  \delta^R(\vk) \equiv W_R(k)\delta(\vk), 
\end{equation}
where $R$ is the smoothing radius. 
For Gaussian smoothing,
\begin{equation}
  \label{eq:W_R_Gauss}
  W_R^\mathrm{Gauss}(k) = e^{-\frac{1}{2}k^2R^2}.
\end{equation}
The power spectrum and bispectrum of the smoothed field $\delta^R$ are
\begin{eqnarray}
  \label{eq:30}
P_{\delta^R\delta^R}(k) &=& P^R_{\delta\delta}(k)= W^2_R(k)P_{\delta\delta}(k),
\\
  B_{\delta^R\delta^R\delta^R}(k_1,k_2,k_3) &=& W_R(k_1)W_R(k_2)W_R(k_3)B_{\delta\delta\delta}(k_1,k_2,k_3).
\end{eqnarray}

More small-scale modes can be included by choosing smaller smoothing
scale $R$. This will generally increase the signal to noise in the
observables at the expense of worse agreement with models, so in
practice some trade-off smoothing scale $R$ should be chosen. Note
that the window function is applied to every density field, so an
isotropic survey window function could be included in the model
predictions below by simply modifying the smoothing kernel $W_R$ appropriately.

Haloes are finite size objects and thus the correlators of halo centers should not have structure on scales below the halo scale. This is equivalent to modelling the halo density field in terms of the density field smoothed on the halo scale. There is indeed evidence for a cutoff of the protohalo power spectrum in Lagrangian space \cite{Tobias14}. Gravitational evolution, in particular the collapse, shrinks this scale and generates non-linear contributions to the clustering statistics. While a self consistent theoretical understanding of the combined effects of Lagrangian smoothing and non-linear gravitational evolution does not exist, there are hints from simulation for the existence of a finite smoothing scale in the halo density field in Eulerian space \cite{Pollack1309}.
Such a smoothing scale would also affect our perturbation theory calculations of the correlators of the squared field. In particular, the integral over the smoothed power spectrum would decrease the amplitude of the low-$k$ limit of the $I_{DE}^R$ terms (Eq.~\eq{I_DE_R} below) by a few percent for realistic smoothing scales and thus increase the inferred bias parameters by this amount. At higher wavenumbers, the smoothing scale enters more explicitly, also in the $I_{DE}^{\text{bare},R}$ (Eq.~\eq{I_DE_R_bare} below) and can thus lead to larger changes. We have calculated these effects and found that the additional fitting parameter did not improve our fits of halo-halo-halo statistics. With no apparent improvements and the lack of a theoretical model, we decide to neglect the smoothing corrections in this study but remark that they should be understood and included in the future.

\subsection{General bispectrum}

To compute general theoretical predictions for cross-spectra
 we work with two smoothed fields
$\delta^R_a$ and $\delta^R_b$ that can be dark matter or halo densities,
$a,b\in\{\mathrm{m}, \mathrm{h}\}$. 
The expectation value of the cross-spectrum of a quadratic field $D[\delta^R_a]$
for $D\in\{\mathsf{P}_0,-F_2^1\mathsf{P}_1, \mathsf{P}_2\}$ with the density $\delta^R_b$ 
 is given by an integral over the
bispectrum $B$,
\begin{eqnarray}
  \label{eq:48}
  \la D[\delta_a^R](\vk)\delta_b^R(\vk')\ra
&=&
\int\frac{\d^3q}{(2\pi)^3} D(\vq,\vk-\vq) \la
\delta_a^R(\vq)\delta_a^R(\vk-\vq)
\delta^R_b(\vk')\ra
\\
&=&
(2\pi)^3\delta_D(\vk+\vk')
\int\frac{\d^3q}{(2\pi)^3} D(\vq,\vk-\vq)
B_{\delta_a^R\delta_a^R\delta^R_b}(\vq,\vk-\vq,-\vk).
\end{eqnarray}
For $a\ne b$, the bispectrum $B$ is not symmetric in its arguments and
the last argument $-\vk$ is associated with $\delta_b$.
Writing the smoothing kernels explicitly, we have 
\begin{equation}
  \label{eq:2}
  P_{D[\delta^R_a],\delta^R_b}(k) = W_R(k)\int\frac{\d^3q}{(2\pi)^3}W_R(q)W_R(|\vk-\vq|) D(\vq,\vk-\vq)
B_{\delta_a\delta_a\delta_b}(\vq,\vk-\vq,-\vk).
\end{equation}

\subsection{Matter-matter-matter cross-spectra}
For cross-spectra of smoothed dark matter fields ($a=b=\mathrm{m}$),
the DM bispectrum \eq{BDM} gives
\begin{eqnarray}
  \label{eq:Pcross_mmm_D_in_terms_of_B}
  P_{D[\delta^R_\mathrm{m}],\delta^R_\mathrm{m}}(k) &=& \int\frac{\d^3 q}{(2\pi)^3}
D(\vq,\vk-\vq)
B_\mathrm{\delta^R_\mathrm{m}\delta^R_\mathrm{m}\delta^R_\mathrm{m}}(\vq,
\vk-\vq, -\vk)\\ 
\label{eq:Pcross_mmm_D}
&=&2I^R_{DF_2}(k)
+4 I_{DF_2}^{\mathrm{bare},R}(k),
\end{eqnarray}
where we used that the kernel $D(\vq, \vk-\vq)$ is assumed to be symmetric in its
arguments and we defined for kernels $D$ and $E$ 
\begin{equation}
  \label{eq:I_DE_R}
  I_{DE}^R(k)\equiv
W_R(k)
\int\frac{\d^3 q}{(2\pi)^3}W_R(q)W_R(|\vk-\vq|)P_\mathrm{mm}^\mathrm{lin}(q)P_\mathrm{mm}^\mathrm{lin}(|\vk-\vq|)
D(\vq, \vk-\vq)E(\vq, \vk-\vq),
\end{equation}
which is symmetric under $D\leftrightarrow E$,
and
\begin{equation}
  \label{eq:I_DE_R_bare}
I_{DE}^{\mathrm{bare},R}(k)\equiv W_R(k) P_\mathrm{mm}^\mathrm{lin}(k) \int\frac{\d^3 q}{(2\pi)^3}
 W_R(q) W_R(|\vk-\vq|) P_\mathrm{mm}^\mathrm{lin}(q) D(\vq,\vk-\vq)E(\vq,-\vk),
\end{equation}
which is not symmetric under $D\leftrightarrow E$. 
Explicit predictions for the cross-spectra $P_{\delta_\mathrm{m}^2, \delta_\mathrm{m}}$,
$P_{-\Psi_\mathrm{m}^i\partial_i\delta_\mathrm{m},\delta_\mathrm{m}}$ and $P_{s_\mathrm{m}^2, \delta_\mathrm{m}}$
can be obtained from \eqq{Pcross_mmm_D} by plugging in $D=\mathsf{P}_0$, 
$D=-F_2^{1}\mathsf{P}_1$ and $D=\mathsf{P}_2$, respectively, and setting $E=F_2$. 
Note that there are three powers
of the smoothing kernel because we smooth the nonlinear rather than
the linear field.

The integrals in Eqs.~\eq{I_DE_R} and \eq{I_DE_R_bare} are similar to
typical 1-loop expressions and can be reduced to two-dimensional
integrals over scale $q$ and cosine $\hat \vq\cdot\hat \vk$, which can
be evaluated numerically with little computational cost (see
e.g.~\cite{MPTBREEZE,taruyaRegPT} for public codes that compute similar
integrals).  The factors
$W_R(|\vk-\vq|)$ and $P_\mathrm{mm}^\mathrm{lin}(|\vk-\vq|)$ introduce
a non-trivial angle dependence so that the angular integration
generally needs to be performed numerically.

The only ingredient for the theory prediction of \eqq{Pcross_mmm_D} is
the model for the DM bispectrum. Improved bispectrum models that have
the same form as \eqq{BDM} could easily be included, e.g.~by replacing
the perturbation theory $F_2$ kernel by an effective $F_2$ kernel
fitted to $N$-body simulations
\cite{hector1111,scocci_fitting_formula,hector1407theory}.

\subsection{Matter-matter-halo cross-spectra}
From the  unsymmetric unsmoothed
matter-matter-halo bispectrum of \eqq{Bmmh_unsymm}
we find for the cross-spectrum of a quadratic matter field with the halo density
\begin{eqnarray}
  \label{eq:Pcross_mmh_D}
  P_{D[\delta_\mathrm{m}^R],\delta_\mathrm{h}^R}(k) &=& \int\frac{\d^3 q}{(2\pi)^3}
D(\vq,\vk-\vq)  B^\mathrm{unsym}_\mathrm{\delta_\mathrm{m}^R\delta_\mathrm{m}^R\delta_\mathrm{h}^R}(\vq, \vk-\vq, -\vk)\\
&=&
2b_1I^R_{DF_2}(k) + 4b_1I_{DF_2}^{\mathrm{bare},R}(k)
+2b_2I^R_{D\mathsf{P}_0}(k)+\frac{4}{3}b_{s^2}I^R_{D\mathsf{P}_2}(k).
\end{eqnarray}
The part depending on $b_1$ can be expressed in terms of the
matter-matter-matter cross-spectrum  so that
\begin{equation}
  \label{eq:Pcross_mmh_D_in_terms_of_mmm}
  P_{D[\delta_\mathrm{m}^R],\delta_\mathrm{h}^R}(k) - b_1P_{D[\delta^R_\mathrm{m}],\delta^R_\mathrm{m}}(k)
  =
 2b_2I^R_{D\mathsf{P}_0}(k)+\frac{4}{3}b_{s^2}I^R_{D\mathsf{P}_2}(k).
\end{equation}

\subsection{Halo-halo-halo cross-spectra}
The halo-halo-halo bispectrum \eq{Bhhh} gives for the halo-halo-halo
cross-spectra
\begin{equation}
  \label{eq:hhh_theory_all_cross_spectra}
  P_{D[\delta_\mathrm{h}^R],\delta^R_\mathrm{h}}(k) 
=2b_1^3\left[I^R_{DF_2}(k)
+2 I_{DF_2}^{\mathrm{bare},R}(k)\right] +
2b_1^2b_2 \left[I^R_{D\mathsf{P}_0}(k) + 
2 I_{D\mathsf{P}_0}^{\mathrm{bare},R}(k)\right]+
\frac{4}{3}b_1^2b_{s^2} \left[I^R_{D\mathsf{P}_2}(k) + 
2 I_{D\mathsf{P}_2}^{\mathrm{bare},R}(k)\right].\quad
\end{equation}
Contributions depending on $b_1^3$ also appear in matter-matter-halo
cross spectra, so that
\begin{equation}
  \label{eq:27}
    P_{D[\delta_\mathrm{h}^R],\delta^R_\mathrm{h}}(k) 
- b_1^2 P_{D[\delta_\mathrm{m}^R],\delta^R_\mathrm{h}}(k) 
=
4b_1^2\left[
b_2 I_{D\mathsf{P}_0}^{\mathrm{bare},R}(k)
+\frac{2}{3}b_{s^2} I_{D\mathsf{P}_2}^{\mathrm{bare},R}(k)
\right].
\end{equation}
Decomposing the
$F_2$ kernel in Legendre polynomials as in \eqq{F2LegendreDecomp},
\eqq{hhh_theory_all_cross_spectra} can be rewritten as
\begin{align}
\nonumber
    P_{D[\delta_\mathrm{h}^R],\delta^R_\mathrm{h}}(k) 
=\,&
\left(\frac{34}{21}b_1^3 + 2b_1^2b_2\right)\left[
I^R_{D\mathsf{P}_0}(k) + 2I^{\mathrm{bare},R}_{D\mathsf{P}_0}(k)
\right]\\
\nonumber
&\,+
2b_1^3\left[I^R_{D,F_2^1\mathsf{P}_1}(k) +2I^{\mathrm{bare},R}_{D,F_2^1\mathsf{P}_1}(k)  \right]\\
&\,+ 
\left(\frac{8}{21}b_1^3 + \frac{4}{3}b_1^2b_{s^2}\right)
\left[I^{R}_{D\mathsf{P}_2}(k) + 2I^{\mathrm{bare},R}_{D\mathsf{P}_2}(k) \right].
\end{align}

\begin{figure}[tb]
\centerline{
\includegraphics[width=0.8\textwidth]{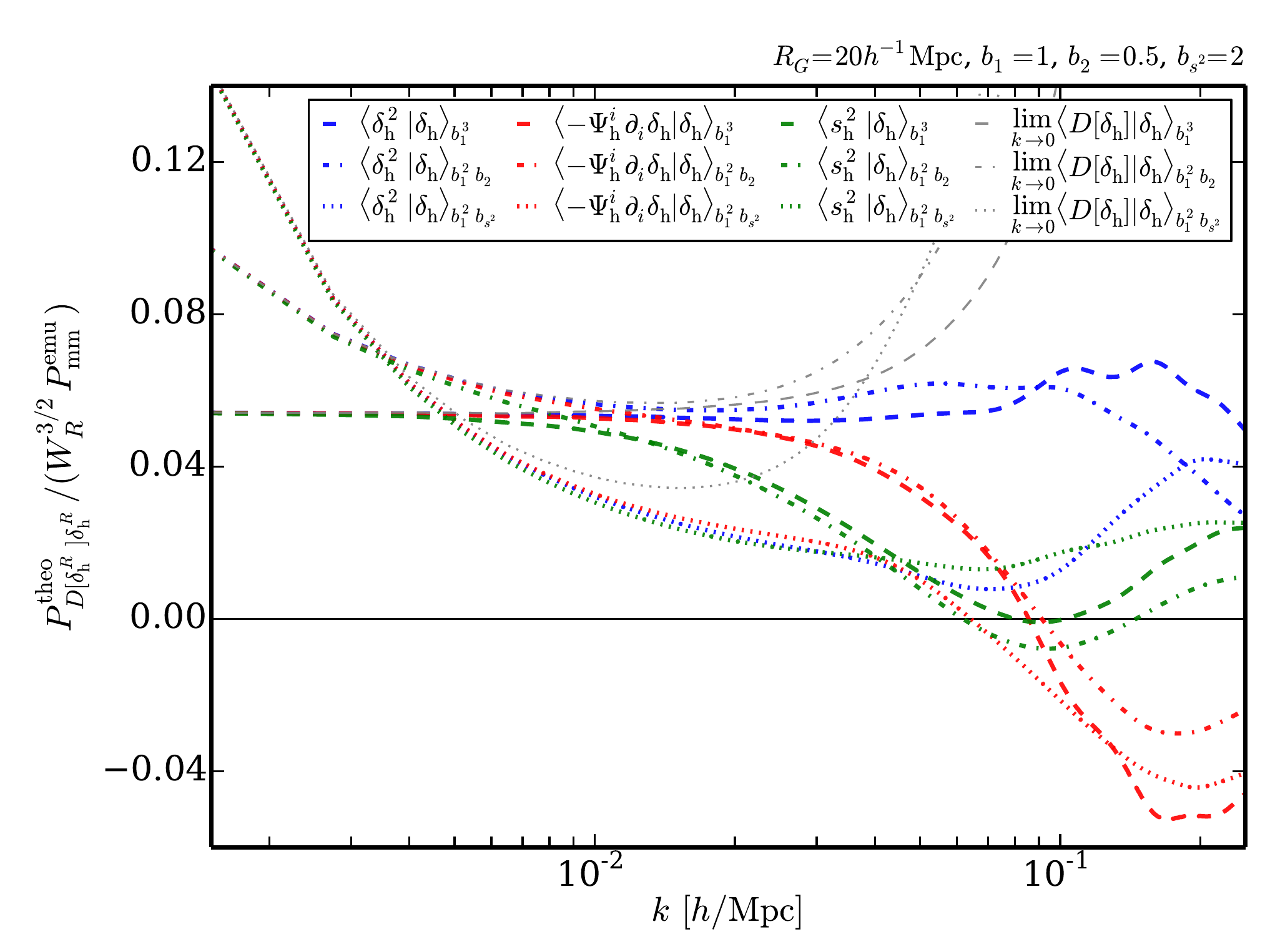}
}
\caption{Theory contributions \eq{hhh_theory_all_cross_spectra} to
  halo-halo-halo cross-spectra scaling like $b_1^3$ (dashed),
  $b_1^2b_2$ (dash-dotted) and $b_1^2 b_{s^2}$ (dotted) for squared
  density $\delta^2_\mathrm{h}(\vx)$ (blue), shift term
  $-\Psi^i_\mathrm{h}(\vx)\partial_i \delta_\mathrm{h}(\vx)$
  (red) and tidal term $s_\mathrm{h}^2(\vx)$ (green), evaluated
  for fixed bias
  parameters $b_1=1$, $b_2=0.5$ and $b_{s^2}=2$, 
  Gaussian smoothing with $R_G=20h^{-1}\mathrm{Mpc}$,
at $z=0.55$,
  with linear matter power spectra in integrands.  Thin gray lines show the large-scale
(low $k$)  limit given by \eqq{lowk_hhh_theory}. The cross-spectra are
divided by the partially smoothed FrankenEmu emulator matter power spectrum
$W_R^{3/2}P_{\mathrm{mm}}^\mathrm{emu}$ \cite{FrankenEmuExt,Emu1,Emu2,Emu3} for plotting convenience.
}
\label{fig:hhhtheory_RGauss20}
\end{figure}

The contributions to the theory expression of \eqq{hhh_theory_all_cross_spectra} are shown in
Fig.~\ref{fig:hhhtheory_RGauss20}
for Gaussian smoothing with $R=20h^{-1}\mathrm{Mpc}$ (see
Fig.~\ref{fig:hhhtheory_RGauss10} in the appendix for
$R=10h^{-1}\mathrm{Mpc}$). Different colors describe
different cross-spectra, $D\in\{\mathsf{P}_0,-F_2^1\mathsf{P}_1,\mathsf{P}_2\}$, while different
line styles correspond to the contributions with different
dependencies on bias parameters, scaling like $b_1^3$, $b_1^2b_2$ or
$b_1^2b_{s^2}$.  The characteristic $k$-dependencies of the different
contributions can be exploited to fit $b_1$, $b_2$ and $b_{s^2}$ to
the three cross-spectra at the same time. In practice, the fitted bias
parameters can still be degenerate because sampling variance at low
$k$ and modeling uncertainty at high $k$ limit the usable $k$ range.
In particular, in the range $0.01h/\mathrm{Mpc}\lesssim k\lesssim
0.1h/\mathrm{Mpc}$, every cross-spectrum depends rather similarly
on $b_1^3$ and $b_1^2b_2$ leading to a degeneracy where larger $b_1$
can be compensated by a smaller $b_2$, which is also present when
considering individual bispectrum triangles rather than
cross-spectra. Consequently, models that extend leading-order PT to
higher $k$ are expected to improve bias constraints
significantly.

In the large-scale limit, $k\ll q$, the three halo-halo-halo
cross-spectra of \eqq{hhh_theory_all_cross_spectra} equal each other
(see Appendix \ref{se:lowk_integral_limits} for 
details),
\begin{align}
\nonumber
  \lim_{k\rightarrow 0} P_{D[\delta^R_\mathrm{h}],\delta^R_\mathrm{h}}(k) 
= W_R(k)\bigg[
b_1^3 P_\mathrm{mm}^\mathrm{lin}(k)\left(
\frac{68}{21}\sigma_{R}^2 - \frac{1}{3}
\sigma_{R,P'}^2
\right)
+
2 b_1^2 b_2 \left(\tau^4_R
+
2 P_\mathrm{mm}^\mathrm{lin}(k)\sigma_R^2 \right)
+
&
\frac{4}{3}b_1^2b_{s^2}\tau^4_R\bigg],&\\
\label{eq:lowk_hhh_theory}
& D\in\big\{\mathsf{P}_0, -F_2^{1}\mathsf{P}_1,
  \mathsf{P}_2\big\},
\end{align}
where $W_R(k)\rightarrow 1$ for $k\rightarrow 0$, and
$\sigma^2_{R,P'}$ and $\tau^4_R$ are defined in
Eqs.~\eq{sigma2_R_Pprime} and \eq{tau_def}, respectively.  In this
limit, the $b_1^3$ term is proportional to the linear matter power
spectrum, whereas the term involving $b_2$ scales like the linear
matter power spectrum plus a $k$-independent correction and the term
involving $b_{s^2}$ is entirely $k$-independent (see thin gray lines
in Figs.~\ref{fig:hhhtheory_RGauss20} and \ref{fig:hhhtheory_RGauss10}
for the the ratio of these limits over the matter power spectrum).
Due to large sampling variance at low $k$, this scale-dependence is
expected to be less powerful in distinguishing bias parameters than
constraints obtained from the different scale-dependencies at high
$k$.

\subsection{Shot noise}
The bispectrum of the smoothed halo density has an additional stochasticity contribution,
\begin{equation}
  \label{eq:43}
\la \hat B_\mathrm{hhh}^R(k_1,k_2,k_3)\ra =
B_\mathrm{hhh}^R(k_1,k_2,k_3)+
 B_\mathrm{hhh}^{R,\mathrm{shot}}(k_1,k_2,k_3),
\end{equation}
whose Poissonian prediction is (see e.g.~\cite{jeongThesis})
\begin{equation}
  \label{eq:Bhhhshot}
B_\mathrm{hhh}^{R,\mathrm{shot}}(k_1,k_2,k_3)
=
W_R(k_1)W_R(k_2)W_R(k_3)\left\{
\frac{1}{\bar n_\mathrm{h}}\left[
P_\mathrm{hh}(k_1)
+ 2\mbox{ perms}
\right]
+ \frac{1}{\bar{n}^2_\mathrm{h}}
\right\}.
\end{equation}
Here, $\bar n_\mathrm{h}$ is the mean halo number density, and
$P_\mathrm{hh}$ is the power spectrum of the unsmoothed continuous
halo density field, which we approximate by the ensemble-averaged,
CIC- and shot-noise-corrected power spectrum of the
unsmoothed halo density measured in the simulations.  The stochasticity
bispectrum contributes to halo cross-spectra as
\begin{equation}
  \label{eq:hhh_cross_spectra_shotnoise}
  P^\mathrm{shot}_{D[\delta^R_\mathrm{h}], \delta^R_\mathrm{h}}(k)
=
\left[\frac{1}{\bar n^2_\mathrm{h}} + \frac{P_\mathrm{hh}(k)}{\bar
    n_\mathrm{h}}\right]J^R_D(k)
+ \frac{2}{\bar n_\mathrm{h}}\tilde J^R_D(k),
\end{equation}
where we defined
\begin{equation}
  \label{eq:50}
  J^R_D(k) \equiv W_R(k)\int\frac{\d^3
    q}{(2\pi)^3}W_R(q)W_R(|\vk-\vq|)D(\vq, \vk-\vq)
\end{equation}
and
\begin{equation}
  \label{eq:51}
  \tilde J^R_D(k) \equiv  W_R(k)\int\frac{\d^3
    q}{(2\pi)^3}W_R(q)W_R(|\vk-\vq|)D(\vq, \vk-\vq) P_\mathrm{hh}(q),
\end{equation}
which depends on the mass bin through $P_\mathrm{hh}$.   
The full model is
\begin{equation}
  \label{eq:Pcross_shotnoisecorrection}
  \la \hat P_{D[\delta^R_\mathrm{h}],\delta^R_\mathrm{h}}(k)\ra = 
 P_{D[\delta^R_\mathrm{h}],\delta^R_\mathrm{h}}(k) +  P^\mathrm{shot}_{D[\delta^R_\mathrm{h}],\delta^R_\mathrm{h}}(k),
\end{equation}
where the first term on the r.h.s.~depends on bias parameters to be
fitted from data, while the stochasticity (shot noise) term does not explicitly
depend on bias parameters because we use the measured
ensemble-averaged halo power spectrum there.

The Poisson stochasticity should be corrected for exclusion and
clustering effects, similarly to the power spectrum results of
\cite{tobias1305}. Phenomenologically, we can model these shot noise
corrections with two scale-independent parameters, $\Delta_1$ and
$\Delta_2$, by adding $\Delta_1[P_\mathrm{hh}(k_1)+2\mbox{perms}]$ to
$\bar{n}_\mathrm{h}^{-1}[P_\mathrm{hh}(k_1)+2\mbox{perms}]$ and
$\Delta_2$ to $\bar{n}_\mathrm{h}^{-2}$ in \eqq{Bhhhshot}, so that
\begin{equation}
  \label{eq:Pcross_shotnoisecorrection_D1_D2}
  P^{\mathrm{shot}}_{D[\delta^R_\mathrm{h}],\delta^R_\mathrm{h}}(k) = 
\left[\left(\bar n^{-2}_\mathrm{h}+\Delta_2\right) +
   \left(\bar
    n^{-1}_\mathrm{h} + \Delta_1\right)P_\mathrm{hh}(k)\right]J^R_D(k)
+ 2\left(\bar n^{-1}_\mathrm{h}+\Delta_1\right)\tilde J^R_D(k).
\end{equation}
For $\Delta_2=\Delta_1/\bar{n}_\mathrm{h}$, this is
equivalent to rescaling the Poisson shot noise by an overall
scale-independent amplitude 
 as done in e.g.~\cite{hector1407theory}. The Poisson prediction is
 recovered for $\Delta_1=\Delta_2=0$.

\subsection{Covariances}
\label{se:TheoryCovariances}
To estimate bias and cosmological parameters from cross-spectra we
need to know their noise and covariance properties. Leading-order
perturbation theory predicts for the covariance between two
cross-spectra at the same wavenumber (see Appendix
\ref{se:TheoryCovariancesAppendix})
\begin{equation}
  \label{eq:cov_cross_spectra_diagonal}
  \mathrm{cov}(\hat P_{D[\delta^R_a],\delta^R_b}(k),
\hat P_{E[\delta^R_a],\delta^R_b}(k)) =
  \frac{2}{N_\mathrm{modes}(k)}P^R_{bb}(k)I^{P^R_\mathit{aa}P^R_\mathit{aa}}_{DE}(k),
\end{equation}
where $I^{P^R_\mathit{aa}P^R_\mathit{aa}}_{DE}$ is defined in \eqq{I_DE_PR_PR_cov}. 
The correlation between two cross-spectra is therefore
\begin{equation}
  \label{eq:correl_cross_spectra_diagonal}
  \mathrm{correl}(\hat P_{D[\delta^R_a],\delta^R_b}(k),
\hat P_{E[\delta^R_a],\delta^R_b}(k)) =
\frac{I^{P^R_\mathit{aa}P^R_\mathit{aa}}_{DE}(k)}
{\sqrt{I^{P^R_\mathit{aa}P^R_\mathit{aa}}_{DD}(k) I^{P^R_\mathit{aa}P^R_\mathit{aa}}_{EE}(k)}},
\end{equation}
which does not depend on the type $b$ of the linear field.
 Note that the perturbative
calculation in Appendix \ref{se:TheoryCovariancesAppendix} predicts
additional covariances between cross-spectra at different wavenumbers
$k\ne k'$, but we neglect them here for simplicity.
Eqs.~\eq{cov_cross_spectra_diagonal} and
\eq{correl_cross_spectra_diagonal} will be compared against
simulations in Section \ref{se:EstimatedCovs} and
Fig.~\ref{fig:correl_mmm_RGauss20}. 

For sufficiently large smoothing scale $R$ and low $k$,  we can approximate
$P^R_\mathrm{hh}\approx b_1^2P^R_\mathrm{mm}$ in the integrand of
\eqq{I_DE_PR_PR_cov}, so that the halo correlation 
($a=\mathrm{h}$ in \eqq{correl_cross_spectra_diagonal}) approaches the dark matter correlation
($a=\mathrm{m}$ in \eqq{correl_cross_spectra_diagonal}). 
In the large scale limit $k\rightarrow 0$ the kernels $D,
E\in\{\mathsf{P}_0,-F_2^1\mathsf{P}_1,\mathsf{P}_2\}$ become unity (see Appendix
\ref{se:lowk_integral_limits}), so that
\begin{equation}
  \label{eq:42}
  \lim_{k\rightarrow 0} \mathrm{correl}(\hat P_{D[\delta^R_a],\delta^R_b}(k),
\hat P_{E[\delta^R_a],\delta^R_b}(k)) =1, \qquad\qquad D, E\in\{\mathsf{P}_0,
-F_2^{1}\mathsf{P}_1, \mathsf{P}_2\}
\end{equation}
i.e.~the three cross-spectra are perfectly correlated on large scales.
Therefore all information is already contained in any one of the three
cross-spectra.  On intermediate scales (higher $k$) the cross-spectra
are less correlated (and have different expectation values), so that
constraints are expected to improve if more than just one
cross-spectrum is considered.  On
small scales, smoothing destroys clustering information and the
cross-spectra become perfectly correlated or anti-correlated (see
Fig.~\ref{fig:correl_mmm_RGauss20} below; this happens at higher $k$ if smaller
smoothing scale $R$ is chosen).

On large scales, $k\rightarrow 0$, the variance of the cross-spectra
scales like $P^R_{bb}(k)/k^2$, because the large-scale limit of
\eqq{I_DE_PR_PR_cov} is independent of $k$ and
$N_\mathrm{modes}\propto k^2$. This confirms that the $k^2/P$
weighting in the optimal bispectrum estimators in
Eqs.~\eq{fnl_separable_fgh}, \eq{fnl_Bgrav_unsym_general},
\eq{fnl_B0_k}, \eq{fnl_B1_k} and \eq{fnl_B2_k} corresponds to
inverse-variance weighting on large scales.

\section{Simulations}
\label{se:Simulations}

\subsection{Setup}

We use ten realizations of $N$-body simulations that were also used in
\cite{beth1404,mwhite1408}. The simulations were run with the TreePM
code of \cite{mwhiteTreePM2002}.  Each realization has $2048^3$ DM
particles in a box of side length $L=1380h^{-1}\mathrm{Mpc}$. The
cosmology is flat $\Lambda\mathrm{CDM}$ with $\Omega_bh^2 = 0.022,
\Omega_mh^2 = 0.139, n_s = 0.965, h = 0.69$ and $\sigma_8 = 0.82$, and
we only use the snapshot at $z=0.55$.  Details of the simulations can
be found in \cite{beth1404,mwhite1408}.  

To obtain the DM density, for  each realization the full set of
$2048^3$ DM particles is interpolated to a $N_g^3=512^3$ grid using
the cloud-in-cell (CIC) scheme.  Halos are identified using the FoF
algorithm with linking length $b=0.168$. The halo sample is split into
four mass bins, each spanning a factor of three in mass, and
interpolated to halo density grids using CIC. The CIC window is
deconvolved from DM and halo densities.  The inverse number density
$1/\bar{n}$ in units of $h^{-3}\mathrm{Mpc}^3$ is $0.306$ for dark
matter and $351.5$, $746.6$, $2026.2$ and $6561.3$ for halo bins ordered by
increasing mass. 

Before squaring any fields, we apply a Gaussian smoothing filter \eq{W_R_Gauss} to
the density.
The squared density $\delta^2(\vx)$ in \eqq{delta2_x} is obtained by
squaring this smoothed density in configuration space. To obtain the shift term
$-\Psi^i(\vx)\partial_i\delta(\vx)$ of \eqq{shift_x}, the density is
first Fourier-transformed to $k$ space, where it is multiplied by
$\vk/k^2$ or $\vk$ to get the displacement or density gradient fields in
$k$ space. Then both fields are Fourier-transformed back to configuration
space, where they are multiplied and contracted as in \eqq{shift_x}. A
similar procedure is used to obtain $s_{ij}(\vk)$ and $s^2(\vx)$ as
defined in \eqq{s2_x}.  Finally, the three quadratic fields
$\delta^2(\vx)$, $-\Psi^i(\vx)\partial_i\delta(\vx)$ and $s^2(\vx)$
are Fourier-transformed back to $k$-space, where their cross-spectra with
the density $\delta(\vk)$ are estimated using
Eqs.~\eq{hat_cross_power} and \eq{discrete_sum_over_modes}.

The computational cost is dominated by Fourier transforms which can be
evaluated efficiently as FFTs, requiring only $\mathcal{O}(N_g^3\log
N^3_g)$ operations.  Therefore the cross-spectrum analysis with
quadratic fields has the same complexity as a usual power spectrum
analysis in $k$-space, but it is sensitive to the full bispectrum
information.  Note that brute-force estimation of the bispectrum triangle
by triangle is computationally more expensive by several orders of
magnitude because it requires $\mathcal{O}(N_g^6)$ operations.

Theoretical expressions for expectation values and covariances use the
linear matter-matter power spectrum computed by CAMB \cite{camb} at $z=0.55$
for our fiducial cosmology. If theoretical expressions involve
halo-halo power spectra, we use the estimated ensemble-averaged halo power
spectrum corrected for shot noise and CIC.  For plotting convenience,
the cross-spectrum expectation values are typically divided by the
partially smoothed nonlinear matter power spectrum
$W_R^{3/2}P_\mathrm{mm}^\mathrm{emu}$ which is calculated with the FrankenEmu emulator
\cite{FrankenEmuExt,Emu1,Emu2,Emu3}.
Error bars in all plots of this
paper show the standard error of the mean of the ten realizations,
which is estimated as the sample standard deviation divided by
$\sqrt{10}$. This corresponds to $1\sigma$ errors in the total volume
$26.3h^{-3}\mathrm{Gpc}^3$ of the ten realizations.\footnote{Due to
  the small number of realizations the estimated error bars are rather
  uncertain, see Fig.~\ref{fig:correl_mmm_RGauss20} below for a
  comparison with theoretical error bars.}

\begin{figure}[tp]
\centerline{
\includegraphics[width=0.5\textwidth]{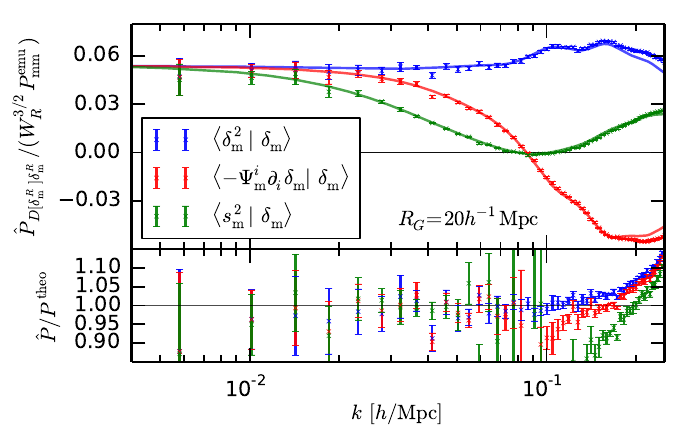}
\includegraphics[width=0.5\textwidth]{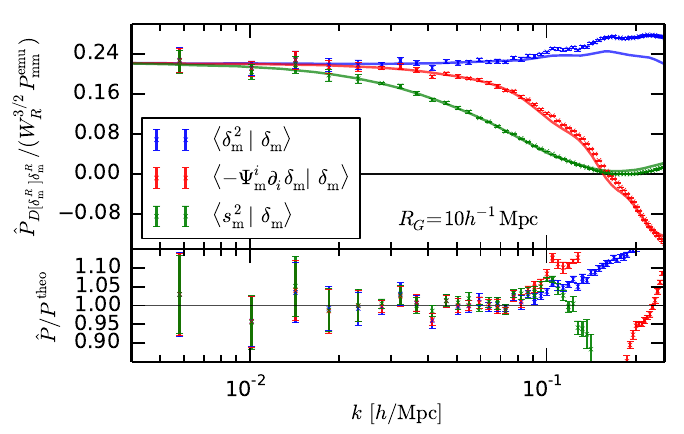}}
\caption{Matter-matter-matter cross-spectra measured from $10$
  realizations at $z=0.55$ (crosses with error bars), compared with leading order theory
  prediction of \eqq{Pcross_mmm_D} (solid lines), neglecting shot noise.  Upper panels show
  cross-spectra divided by the partially smoothed emulator matter
  power spectrum  $W_R^{3/2}P^\mathrm{emu}_\mathrm{mm}$, lower panels
  show the ratio of measured cross-spectra over their theory
  expectation \eq{Pcross_mmm_D}. Gaussian smoothing is applied with
  $R_G=20h^{-1}\mathrm{Mpc}$ (left) and   $R_G=10h^{-1}\mathrm{Mpc}$
  (right). Different colors represent different cross-spectra (squared
  density in blue, shift term in red and tidal term in green). }
\label{fig:mmm_model_vs_data_RGauss10}
\end{figure}

\begin{figure}[tp]
\centerline{
\includegraphics[width=0.5\textwidth]{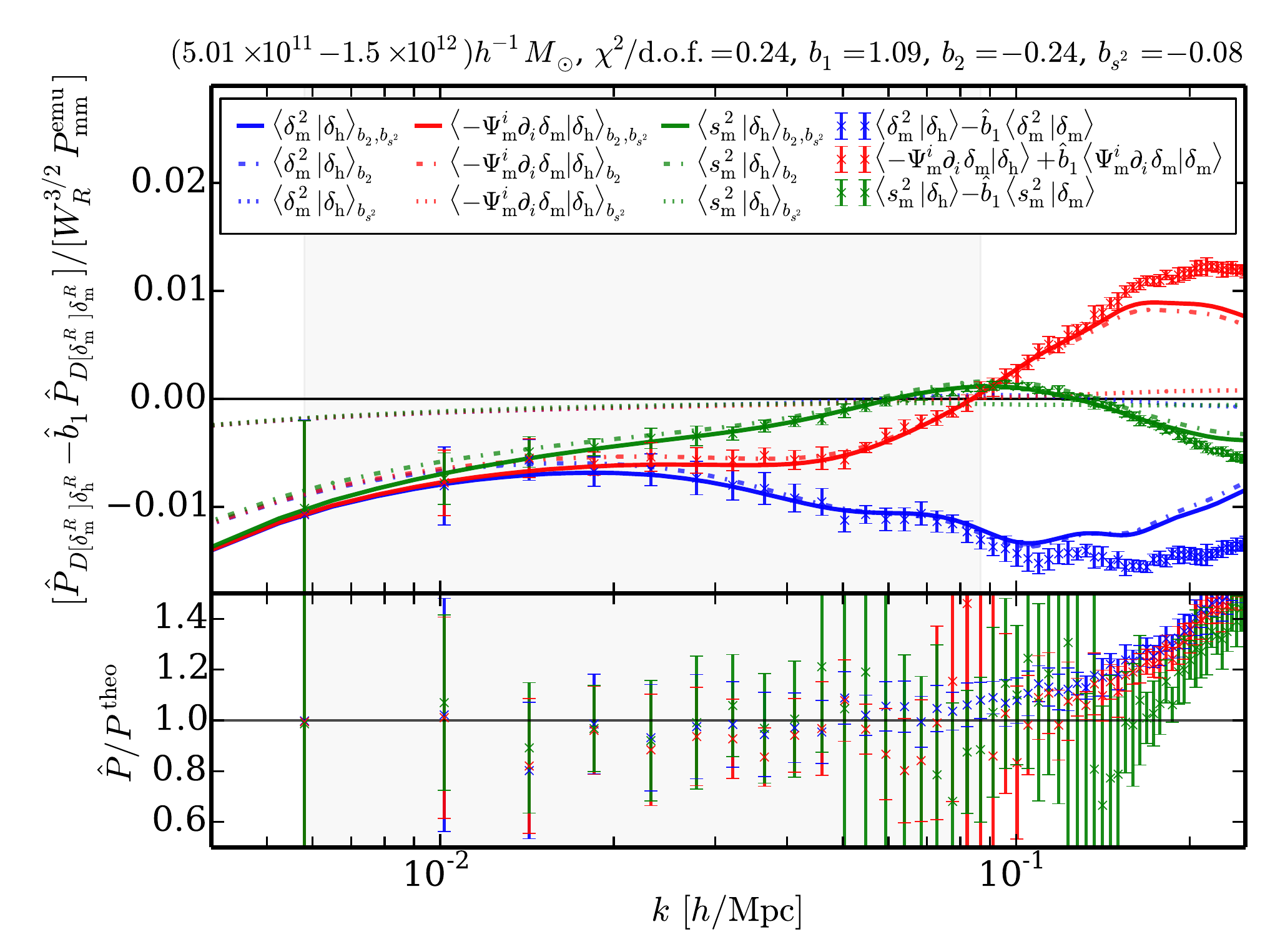}
\includegraphics[width=0.5\textwidth]{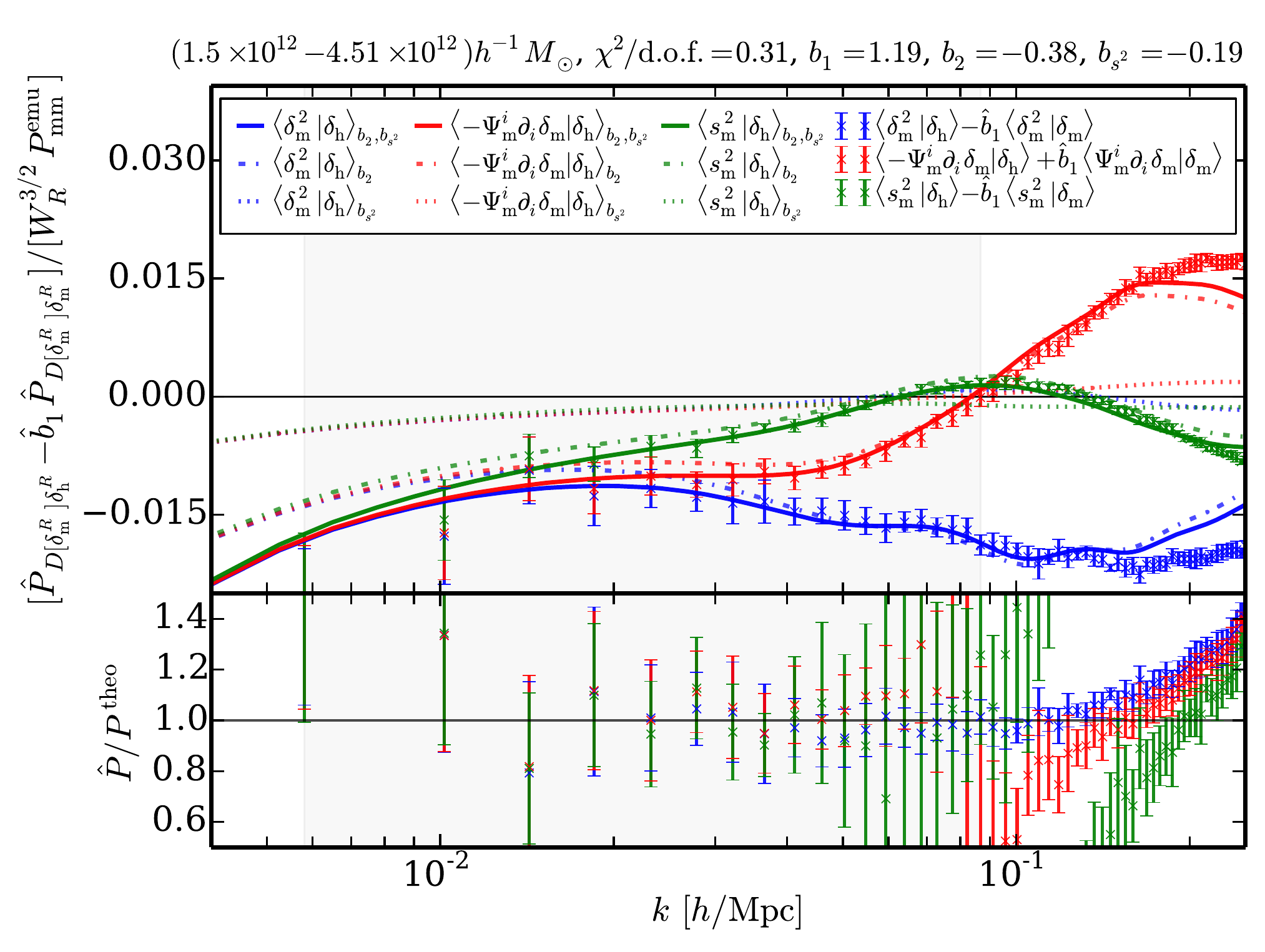}}
\centerline{
\includegraphics[width=0.5\textwidth]{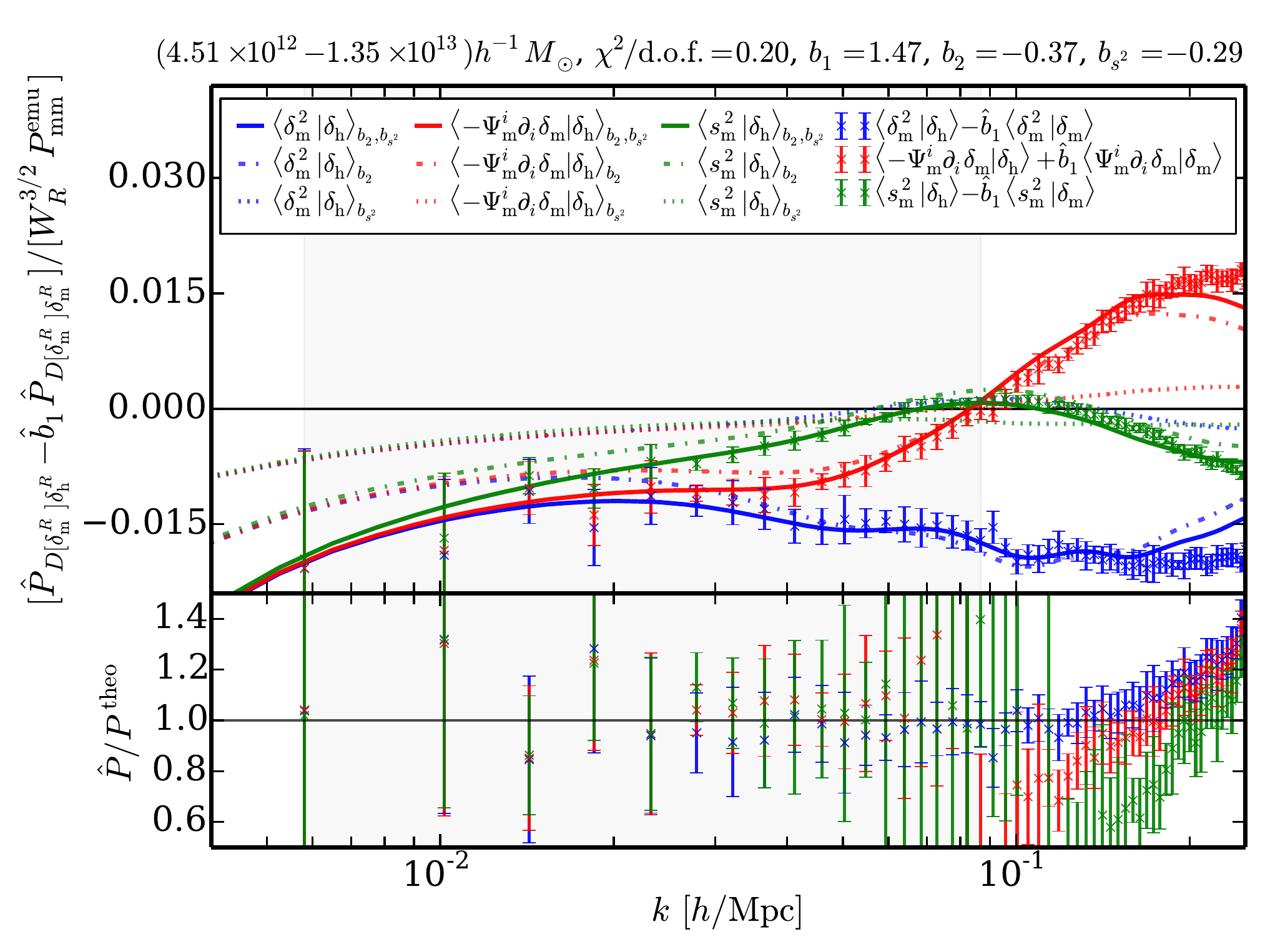}
\includegraphics[width=0.5\textwidth]{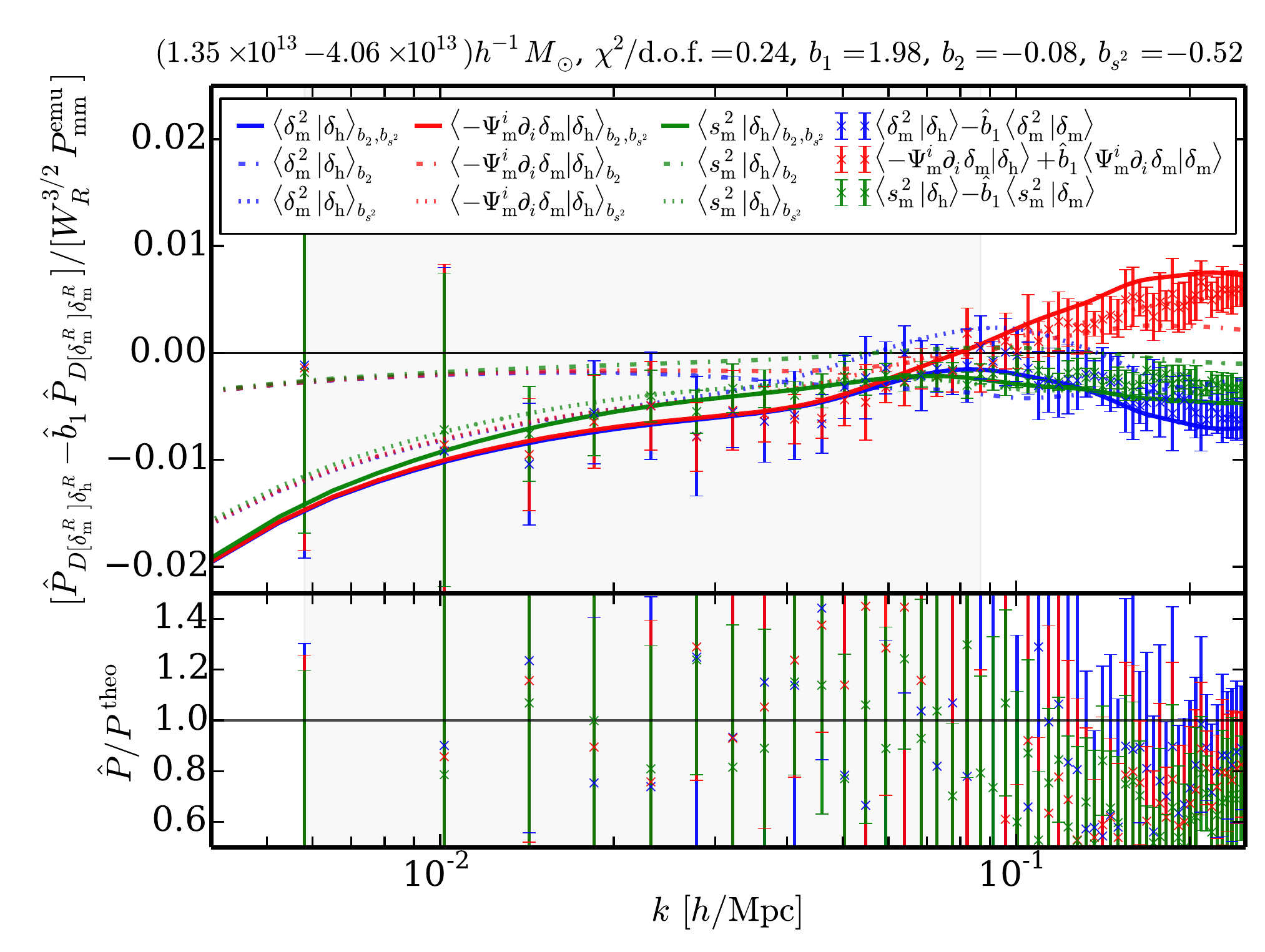}}
\caption{Test of $b_2$ and $b_{s^2}$ contributions to
  matter-matter-halo cross-spectra.  First, $\hat b_1$ is obtained
  from $\hat P_\mathrm{hm}/\hat P_\mathrm{mm}$ at
  $k<0.04h/\mathrm{Mpc}$. Then $b_2$ and $b_{s^2}$ are obtained by
  fitting the model \eq{Pcross_mmh_D_in_terms_of_mmm} to the measured
  excess cross-spectra $\hat
  P_{D[\delta^R_\mathrm{m}]\delta^R_\mathrm{h}} - \hat b_1\hat
  P_{D[\delta^R_\mathrm{m}]\delta^R_\mathrm{m}}$ (crosses). The
  best-fit total model (solid lines) consists of the $b_2$
  contribution (dash-dotted) and the $b_{s^2}$ contribution (dotted),
  while shot noise is neglected.  Different plots show different mass
  bins (increasing from upper left to lower right).  The upper
  sub-panels show excess cross-spectra divided by the partially
  smoothed emulator matter power
  $W_R^{3/2}P^\mathrm{emu}_\mathrm{mm}$, the lower sub-panels show
  measured excess cross-spectra divided by their theory
  expectation. The fit is obtained from the grey shaded region,
  assuming estimated standard errors of the mean without any
  covariances.  Best-fit bias parameters, reduced $\chi^2$ and halo
  mass range  are reported
  at the top of each plot.  Gaussian smoothing with
  $R_G=20h^{-1}\mathrm{Mpc}$ is applied to matter and halo densities.
}
\label{fig:mmh_model_vs_data_RGauss20}
\end{figure}

\subsection{Cross-spectrum expectation values}
We test the consistency of the model by comparing
theory expressions for matter-matter-matter, matter-matter-halo and
halo-halo-halo cross-spectra with simulations. Since the goal of this
section is to test the model, we consider statistics involving the
dark matter field although they cannot directly be observed.

Fig.~\ref{fig:mmm_model_vs_data_RGauss10} compares
matter-matter-matter cross-spectra measured in simulations against
the theory expression of \eqq{Pcross_mmm_D}, finding agreement at the
$5\%$ level for $k\lesssim 0.09h/\mathrm{Mpc}$ for
$R=20h^{-1}\mathrm{Mpc}$ and $R=10h^{-1}\mathrm{Mpc}$.  This demonstrates that
the model for matter-matter-matter cross-spectra works well on large scales.

Figs.~\ref{fig:mmh_model_vs_data_RGauss20} and
\ref{fig:mmh_model_vs_data_RGauss10} test the model for
matter-matter-halo cross-spectra by comparing the excess cross-spectra
\begin{equation}
  \label{eq:69}
\hat P_{D[\delta^R_\mathrm{m}]\delta^R_\mathrm{h}}(k) - \hat b_1\hat
P_{D[\delta^R_\mathrm{m}]\delta^R_\mathrm{m}}(k)
\end{equation}
to the theory expectation of \eqq{Pcross_mmh_D_in_terms_of_mmm}, where
$\hat b_1$ is obtained from large-scale $\hat P_\mathrm{hm}/\hat
P_\mathrm{mm}$, while $b_2$ and $b_{s^2}$ are jointly fitted to the
three excess cross-spectra at $k\le
0.09h/\mathrm{Mpc}$.\footnote{The fits in
  Figs.~\ref{fig:mmh_model_vs_data_RGauss20},
  \ref{fig:mmh_model_vs_data_RGauss10} and
  \ref{fig:hhh_model_vs_data_RGauss20_fitAshot} use estimated
  variances of the cross-spectra in the likelihood and neglect
  covariances for simplicity. In contrast,
  Fig.~\ref{fig:contours_hhh_RGauss20_and_Rgauss10_mass3} uses
  theoretical covariances \eq{cov_cross_spectra_diagonal} between
  different cross-spectra with kernels $D\ne E$ at $k=k'$.  Generally,
  error bars of bias parameters are consistent at the few percent
  level if theoretical instead of estimated variances are
  used. Including theoretical covariances
  \eq{cov_cross_spectra_diagonal} between different cross-spectra with
  kernels $D\ne E$ at $k=k'$ typically leads to fractional changes of
  bias parameter error bars by $\sim 10\%$ or less.
  } The plots show that for all mass bins there is a
combination of $b_2$ and $b_{s^2}$ that describes the simulations
within statistical uncertainties for $k \lesssim
0.09h/\mathrm{Mpc}$.\footnote{In fact, the model seems to overfit the
  data in Figs.~\ref{fig:mmh_model_vs_data_RGauss20} and
  \ref{fig:mmh_model_vs_data_RGauss10} because the reduced $\chi^2$ is
  less than $1$. This might be attributed to the fact that $b_2$ and
  $b_{s^2}$ are degenerate in the excess matter-matter-halo
  cross-spectra, so that the fitting procedure can pick a parameter
  combination along the degeneracy that overfits the data. We do not
  address this issue further because the goal of this section is only
  to show that there are bias parameters for which the model of
  \eqq{Pcross_mmh_D_in_terms_of_mmm} agrees with simulations. } 
The fits are somewhat better for $R_G=20h^{-1}\mathrm{Mpc}$ than for
$R_G=10h^{-1}\mathrm{Mpc}$ because the former excludes nonlinear mode
coupling more efficiently.

\begin{figure}[tp]
\centerline{
\includegraphics[width=0.5\textwidth]{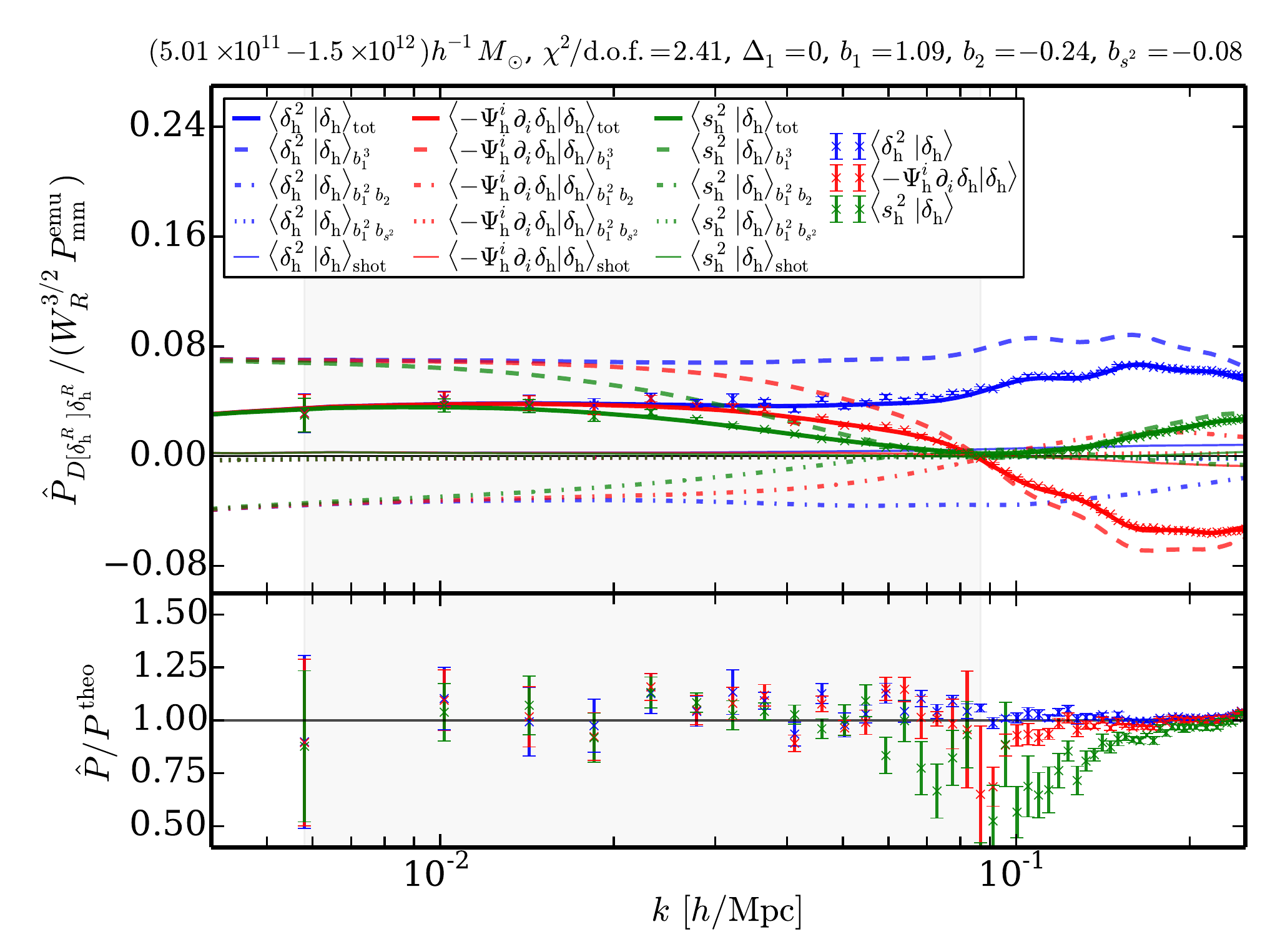}
\includegraphics[width=0.5\textwidth]{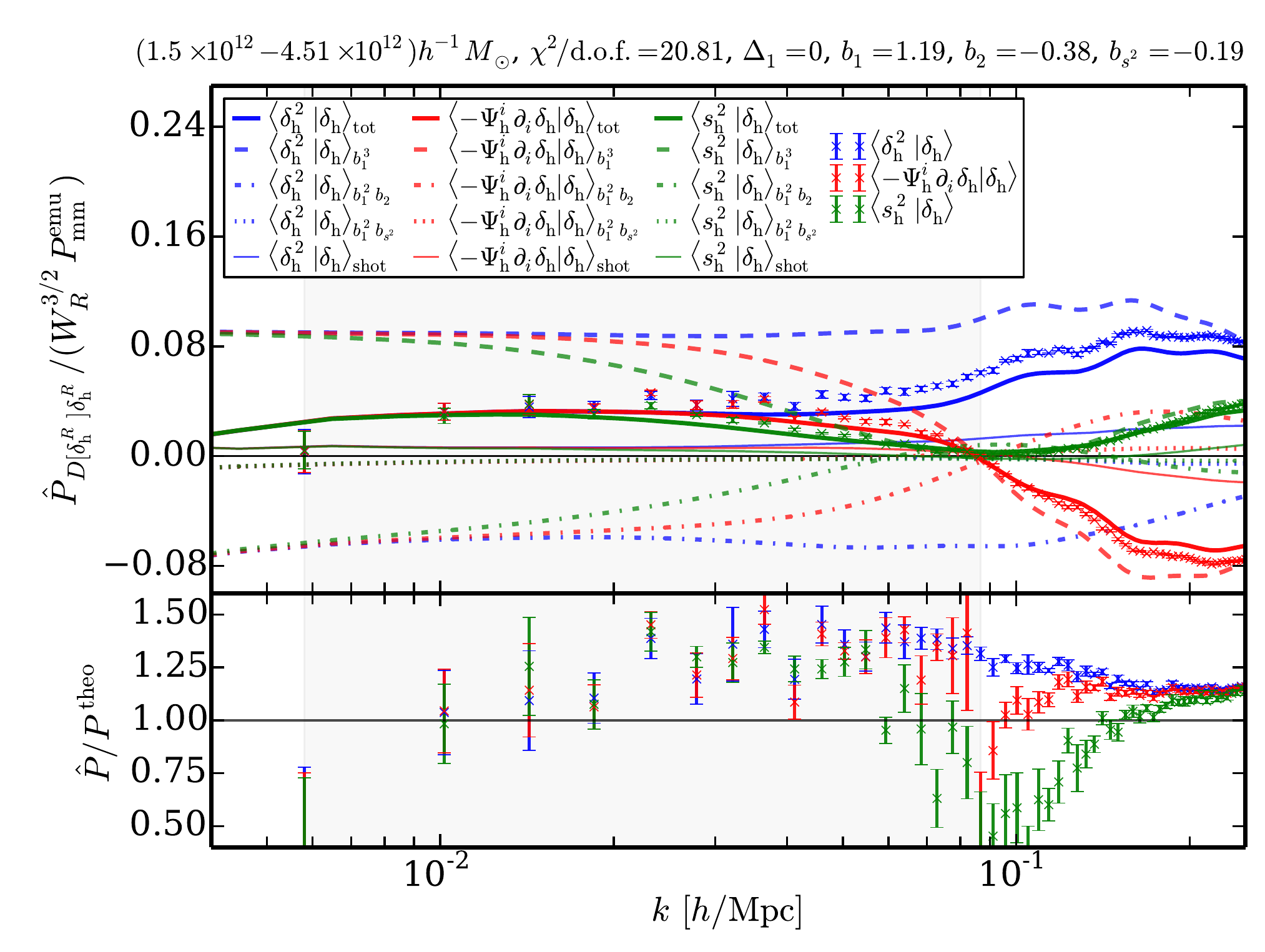}}
\centerline{
\includegraphics[width=0.5\textwidth]{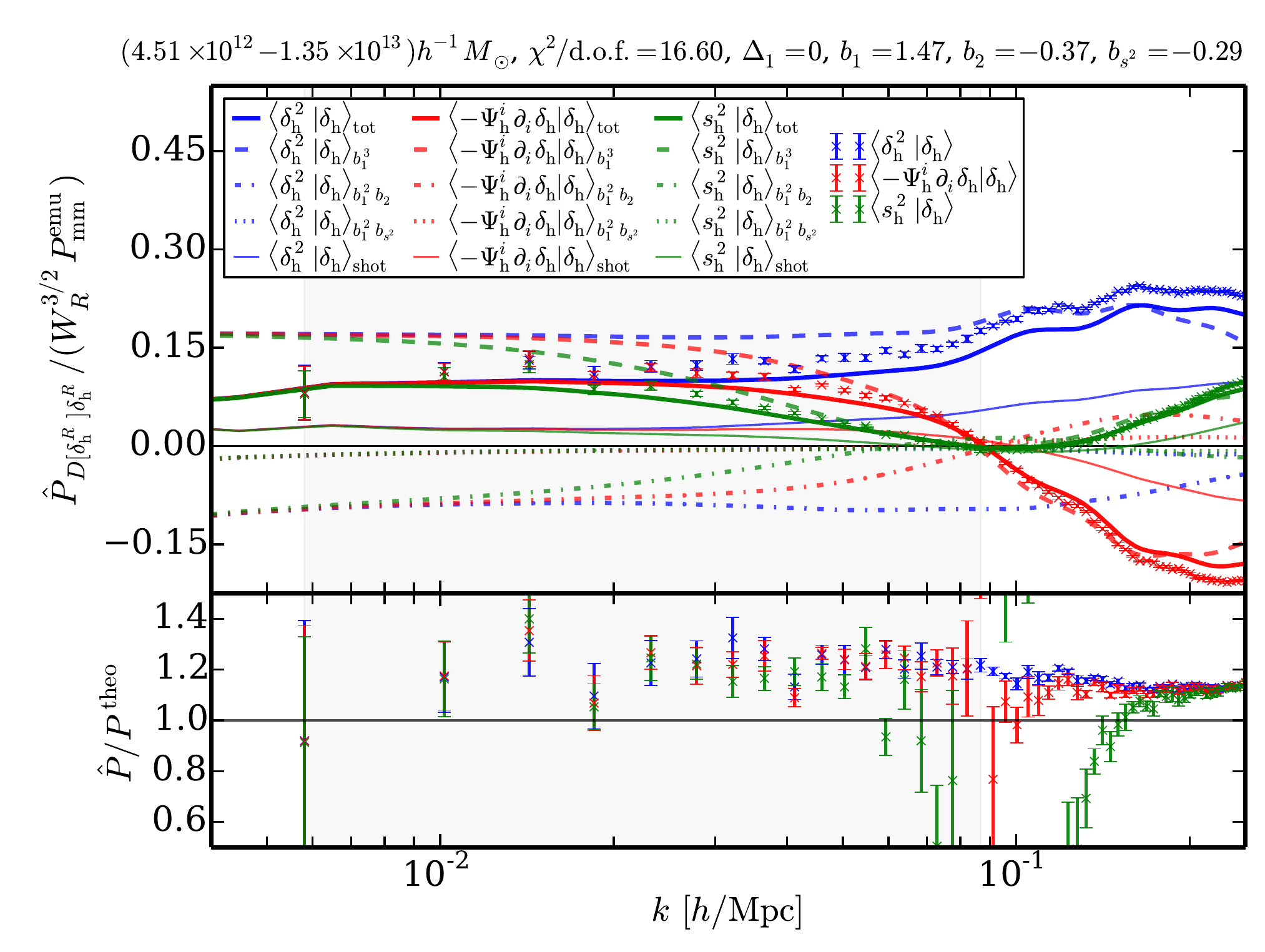}
\includegraphics[width=0.5\textwidth]{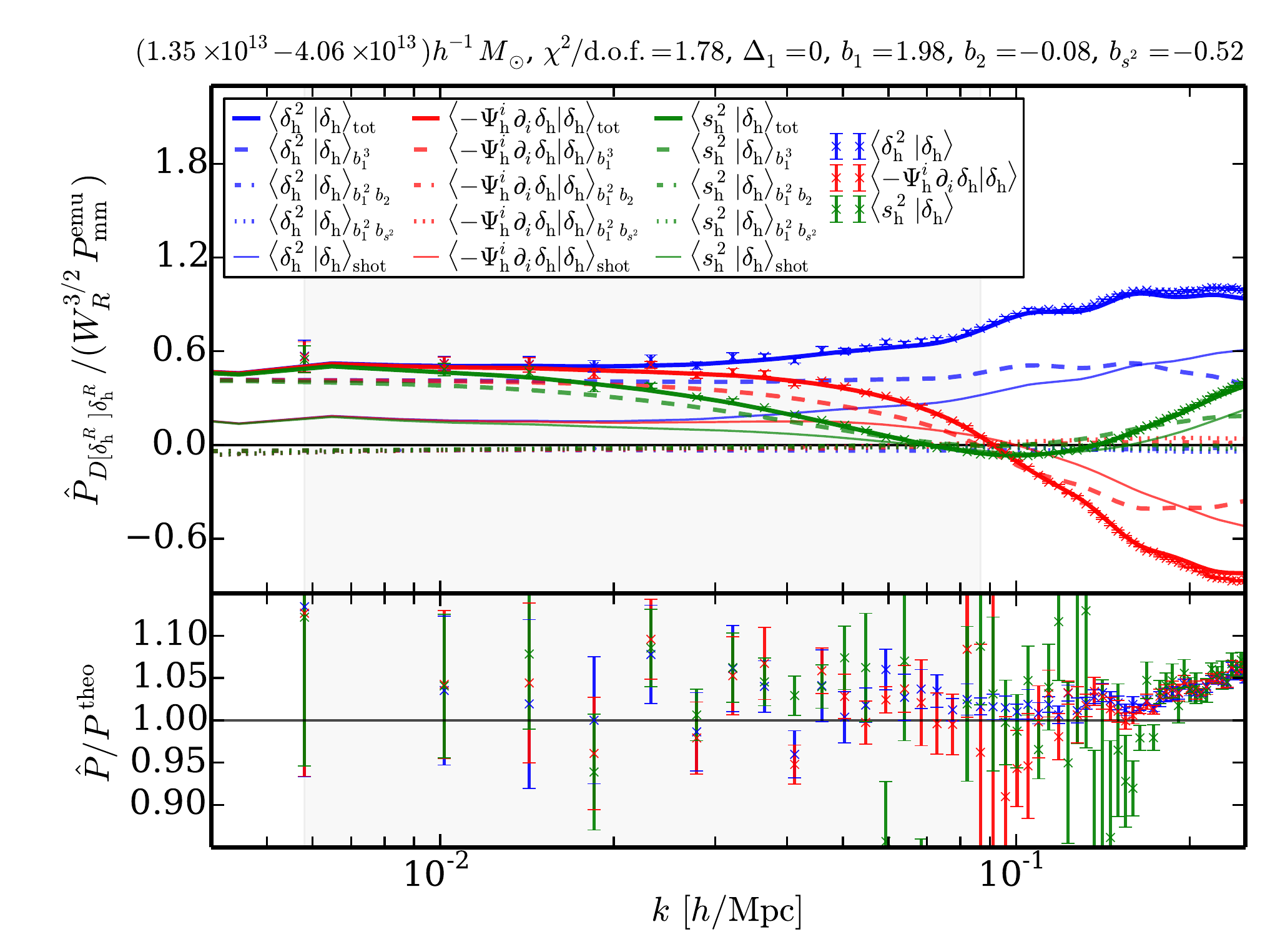}}
\caption{Measured halo-halo-halo cross-spectra (crosses) compared
  against theory (thick solid, \eqq{hhh_theory_all_cross_spectra})
  with bias parameters $\hat b_1$ from $\hat P_\mathrm{hm}/\hat
  P_\mathrm{mm}$ and $b_2$ and $b_{s^2}$ from $\hat
  P_{D[\delta^R_\mathrm{h}]\delta^R_\mathrm{m}}-\hat b_1\hat
  P_{D[\delta^R_\mathrm{m}]\delta^R_\mathrm{m}}$ for
  $R_G=20h^{-1}\mathrm{Mpc}$ smoothing. Upper panels also show theory
  contributions scaling like $b_1^3$ (dashed), $b_1^2b_2$
  (dash-dotted) and $b_1^2b_{s^2}$ (dotted), as well as the
  halo-halo-halo shot noise contribution (thin solid), which is
  assumed to be Poissonian (i.e.~$\Delta_1=\Delta_2=0$ in
  \eqq{Pcross_shotnoisecorrection_D1_D2}). The reduced $\chi^2$ on top of
  the plots quantifies the (dis-)agreement between halo-halo-halo
  cross-spectra measurements and model for the fixed bias
  parameters. It is computed over the gray region, neglecting
  covariances. Note that the shot
  noise contribution fluctuates on very large scales because it is
  computed using the ensemble-averaged estimated halo-halo power spectrum.  }
\label{fig:hhh_model_vs_data_RGauss20_fixAshot0}
\end{figure}

\begin{figure}[tp]
\centerline{
\includegraphics[width=0.5\textwidth]{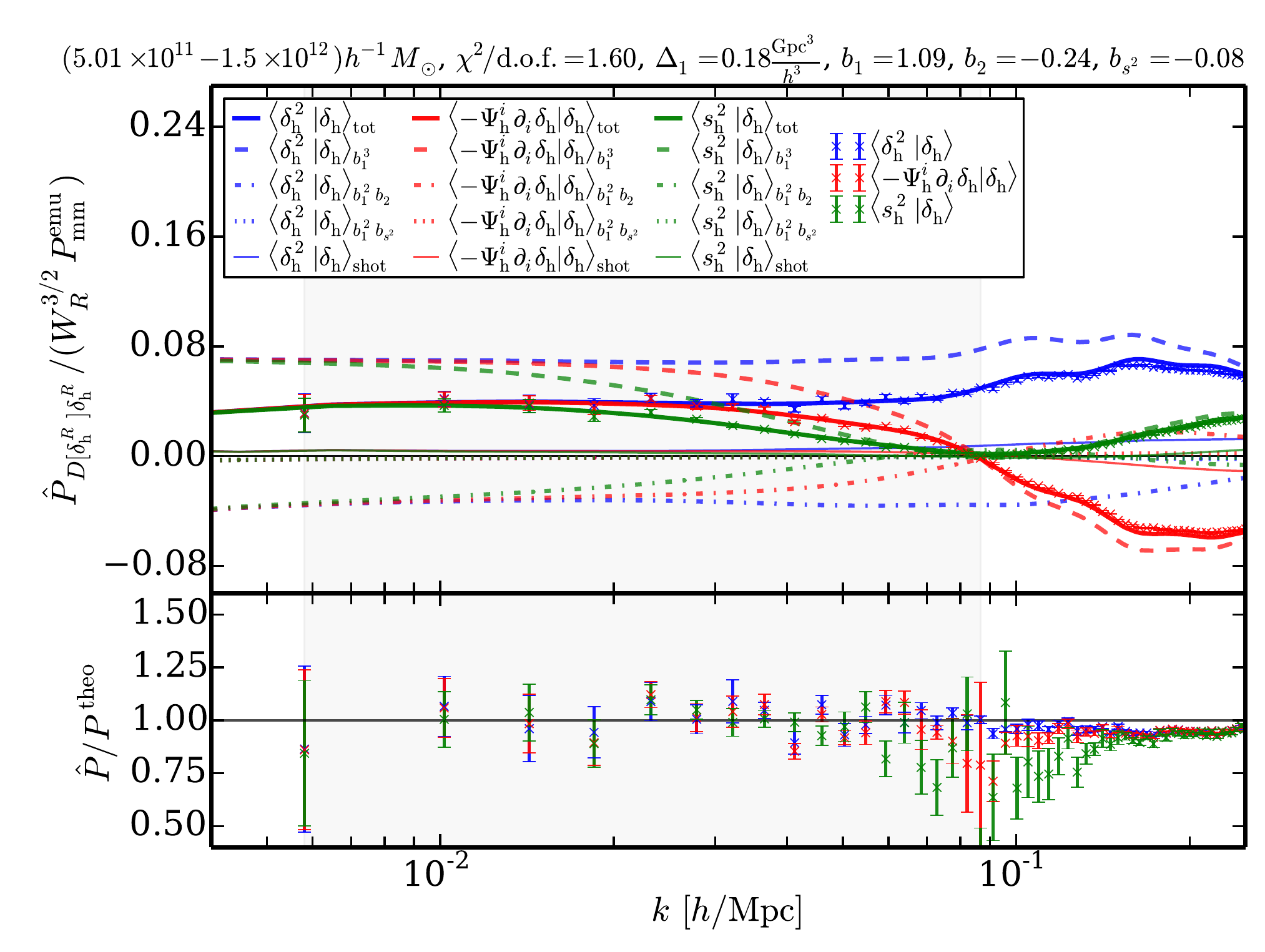}
\includegraphics[width=0.5\textwidth]{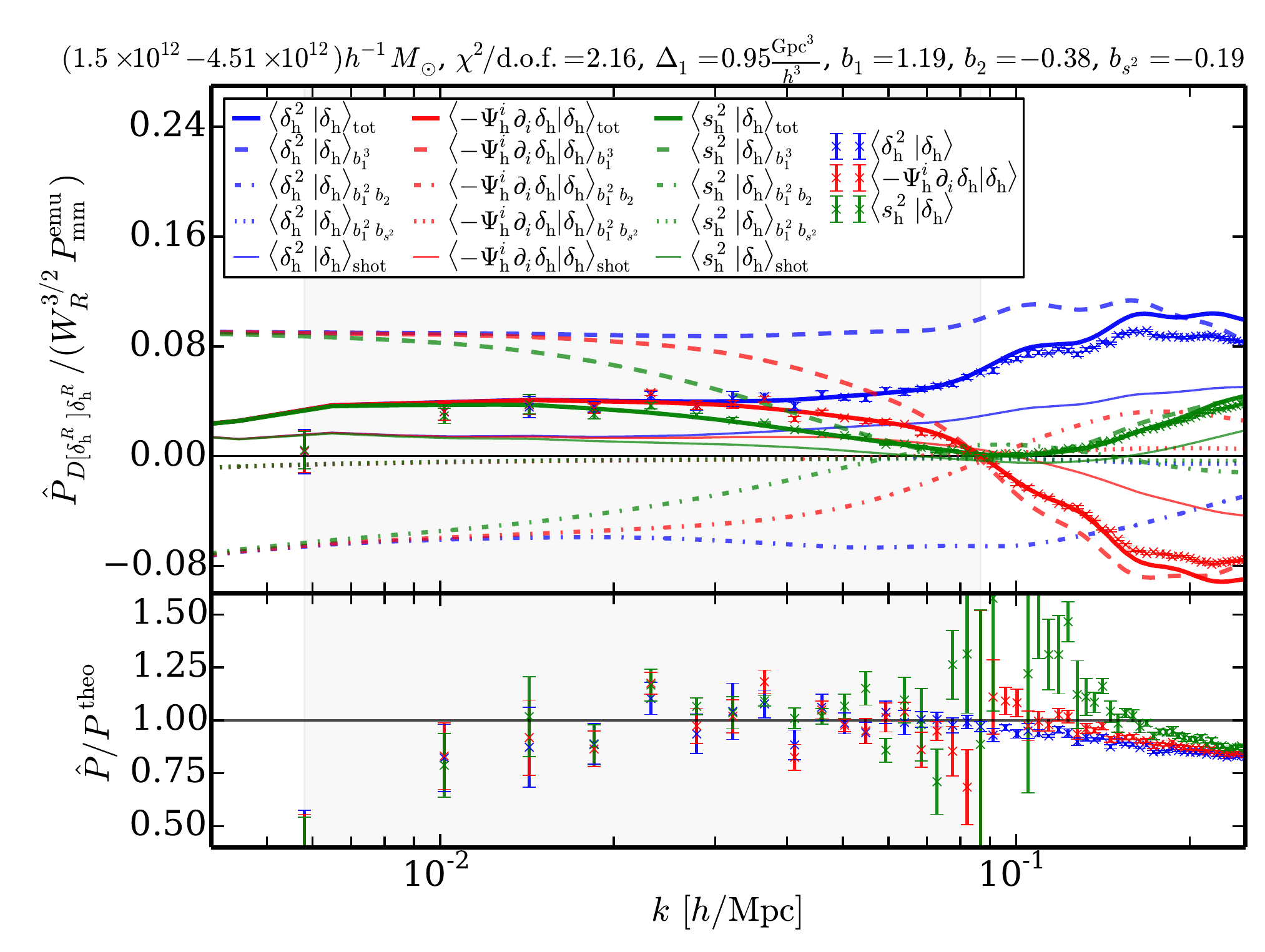}}
\centerline{
\includegraphics[width=0.5\textwidth]{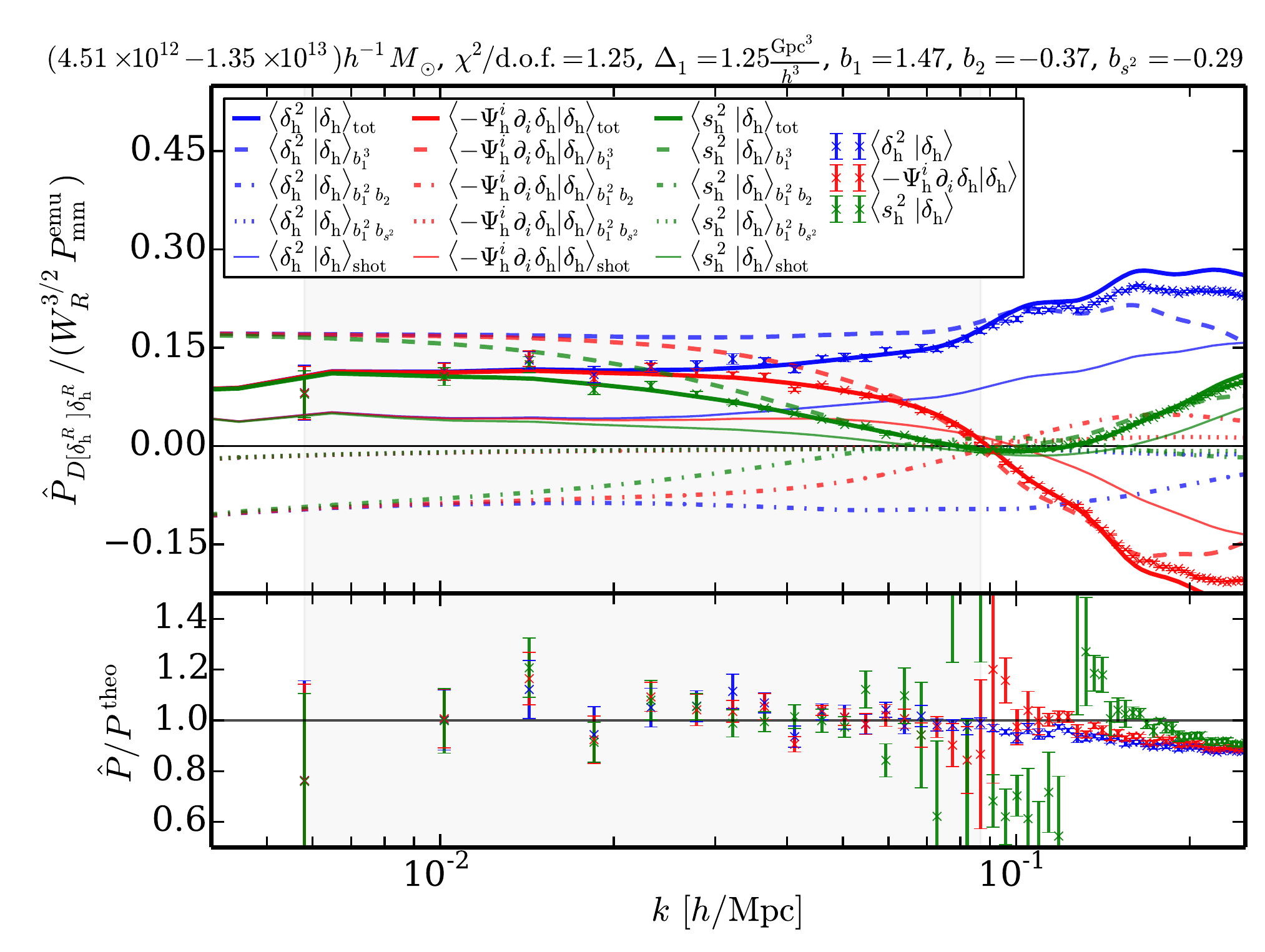}
\includegraphics[width=0.5\textwidth]{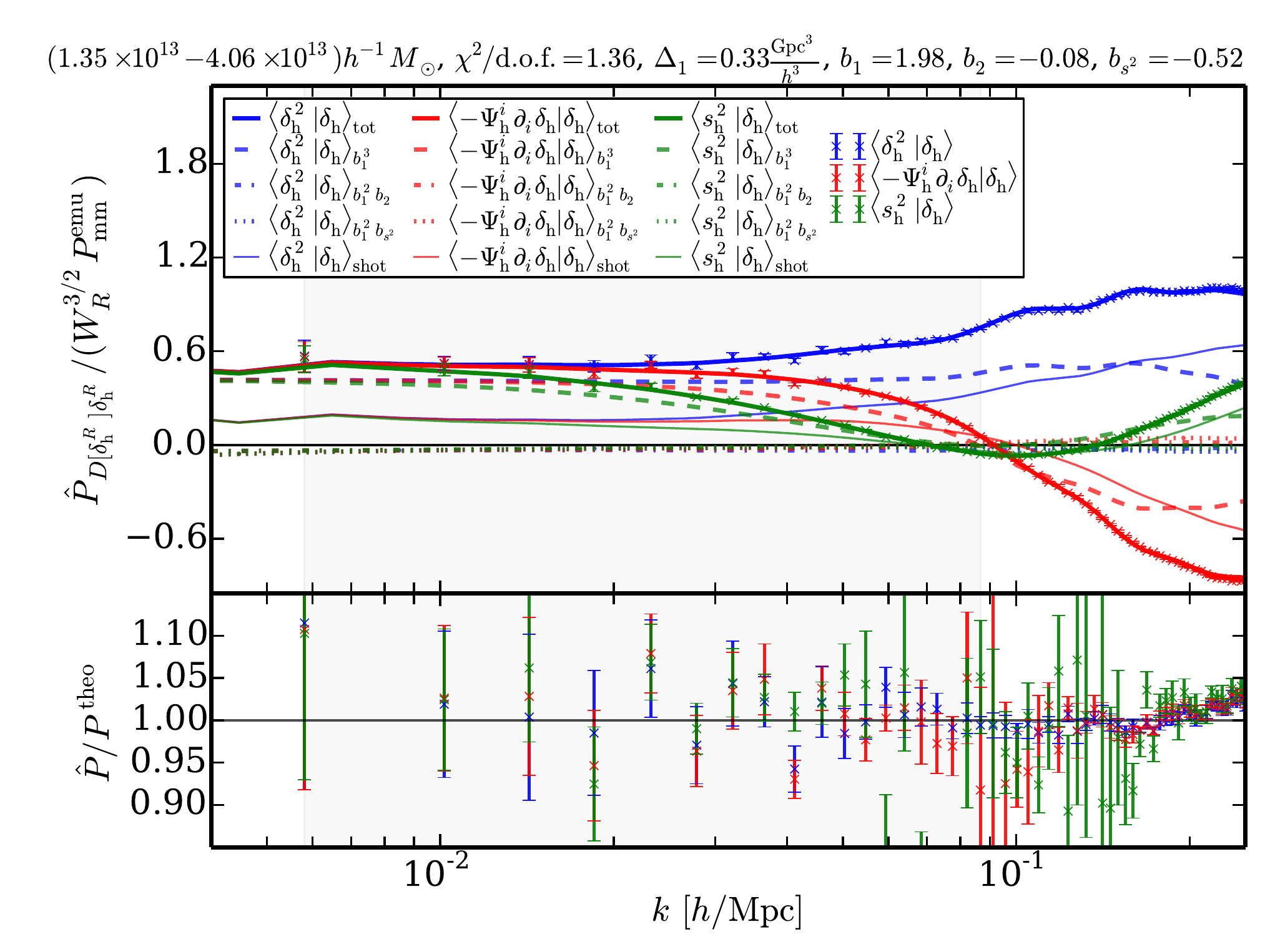}}
\caption{Same as Fig.~\ref{fig:hhh_model_vs_data_RGauss20_fixAshot0}
  if the shot noise correction $\Delta_1$ in
  \eqq{Pcross_shotnoisecorrection_D1_D2} is fitted to measured
  halo-halo-halo cross-spectra, fixing $\Delta_2=\Delta_1/\bar{n}_\mathrm{h}$ (still keeping $b_1$, $b_2$ and
  $b_{s^2}$ fixed to the values obtained from matter-halo and
  matter-matter-halo measurements). }
\label{fig:hhh_model_vs_data_RGauss20_fitAshot}
\end{figure}

Fig.~\ref{fig:hhh_model_vs_data_RGauss20_fixAshot0} tests the
theory prediction of \eqq{hhh_theory_all_cross_spectra} for
halo-halo-halo cross-spectra for $R_G=20h^{-1}\mathrm{Mpc}$. The full measured halo-halo-halo
cross-spectra are compared with their theory prediction
\eq{hhh_theory_all_cross_spectra} for bias parameters fixed to the
values obtained from matter-halo statistics (see caption for details)
and for halo-halo-halo shot noise (stochasticity) fixed to be Poissonian,
corresponding to $\Delta_1=\Delta_2=0$ in
\eqq{Pcross_shotnoisecorrection_D1_D2}.  While theory and simulations do not
differ strongly for the lowest and highest mass bins, the simulations
show a clear excess over theory for the two intermediate mass
bins. Possible reasons for this  could be that the perturbative
treatment breaks down or shot noise is not Poissonian.

To test the latter, Fig.~\ref{fig:hhh_model_vs_data_RGauss20_fitAshot}
shows the same cross-spectra if the shot noise correction $\Delta_1$
in \eqq{Pcross_shotnoisecorrection_D1_D2} is varied as a free
parameter and fitted to the measured halo-halo-halo cross-spectra,
imposing $\Delta_2=\Delta_1/\bar{n}_\mathrm{h}$ and keeping bias
parameters fixed to their values from matter-halo statistics. This
clearly improves the agreement between simulations and theory,
especially for the two intermediate mass bins.\footnote{The worst
  reduced $\chi^2$ is $2.16$, which would improve further if the
  lowest $k$-bin was removed (the theory of this bin is rather noisy
  because the estimated halo-halo power is used to compute the shot
  noise in \eqq{Pcross_shotnoisecorrection_D1_D2}).}  For our fiducial
four mass bins the reduced $\chi^2$ does not improve significantly if
$\Delta_2$ is treated as a free parameter,
i.e.~$\Delta_2=\Delta_1/\bar{n}_\mathrm{h}$ seems to be an acceptable
approximation within the error bars of the simulations.  

\begin{figure}[tp]
\centerline{
\includegraphics[width=0.5\textwidth]{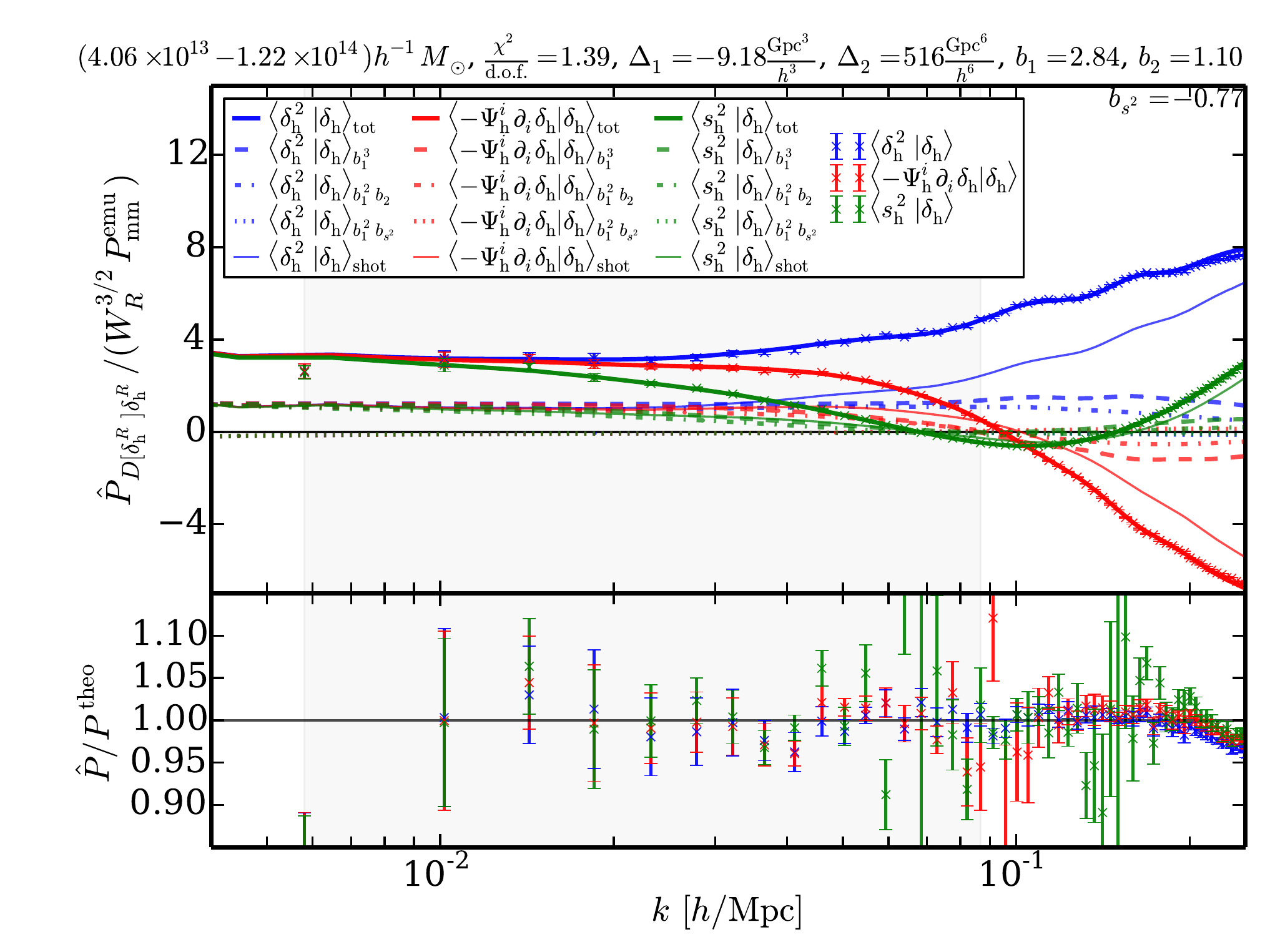}
\includegraphics[width=0.5\textwidth]{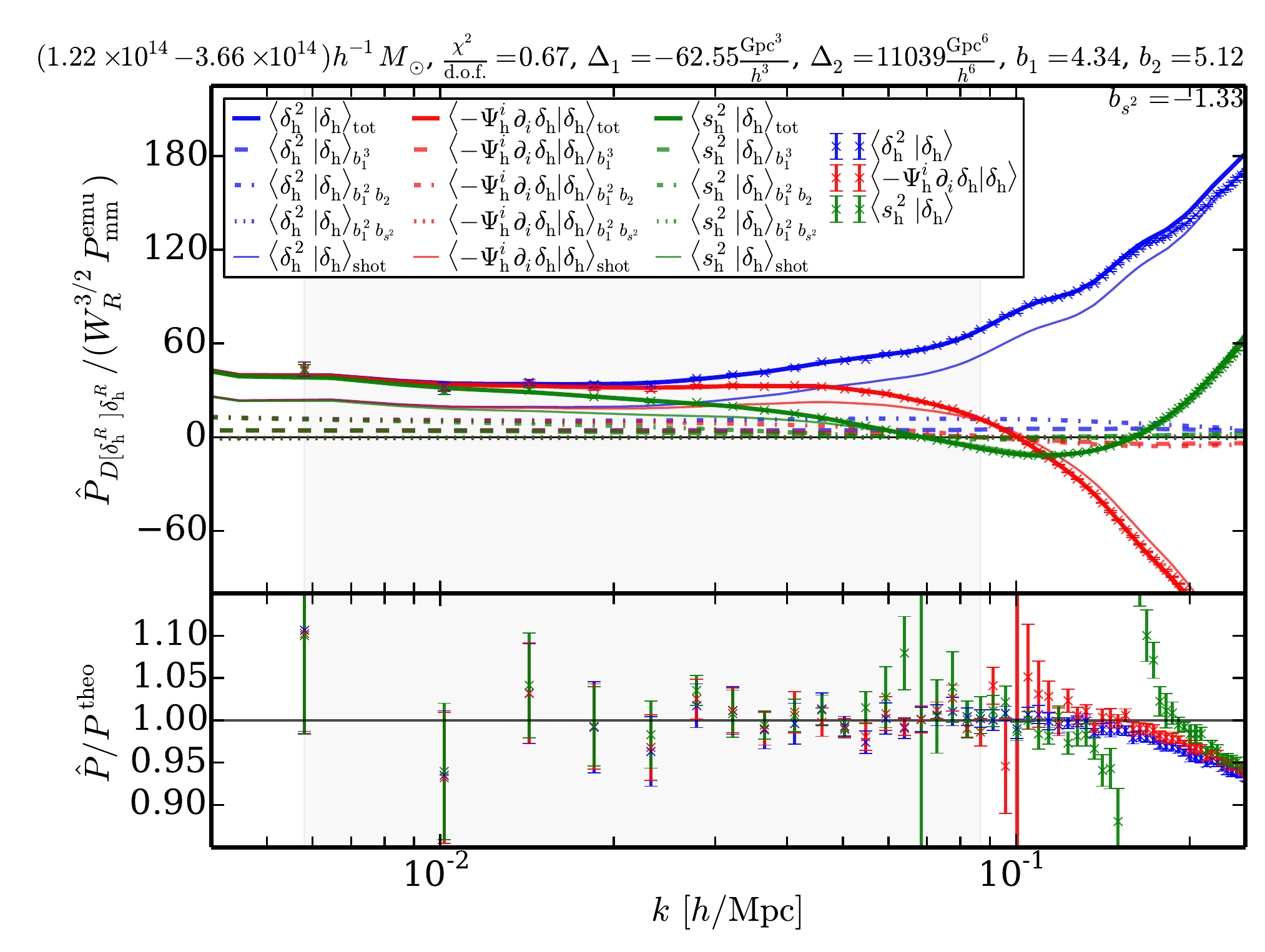}}
\caption{Same as Fig.~\ref{fig:hhh_model_vs_data_RGauss20_fitAshot} for two higher mass bins and treating both 
shot noise corrections  $\Delta_1$ and $\Delta_2$ in
  \eqq{Pcross_shotnoisecorrection_D1_D2} as independent parameters. }
\label{fig:hhh_model_vs_data_RGauss20_fitAshot_mass45}
\end{figure}

\begin{figure}[tp]
\centerline{
\includegraphics[width=0.45\textwidth]{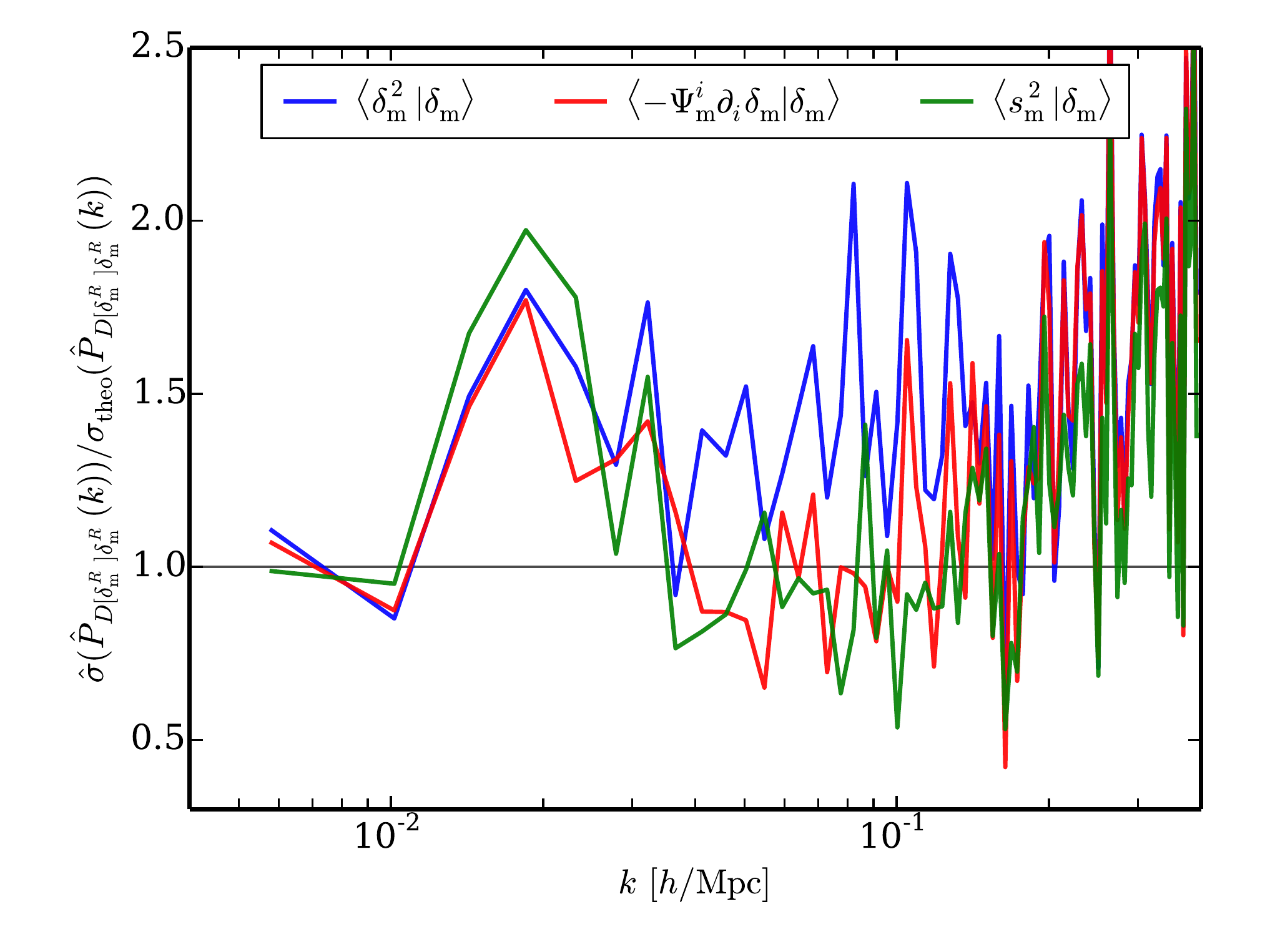}
\includegraphics[width=0.45\textwidth]{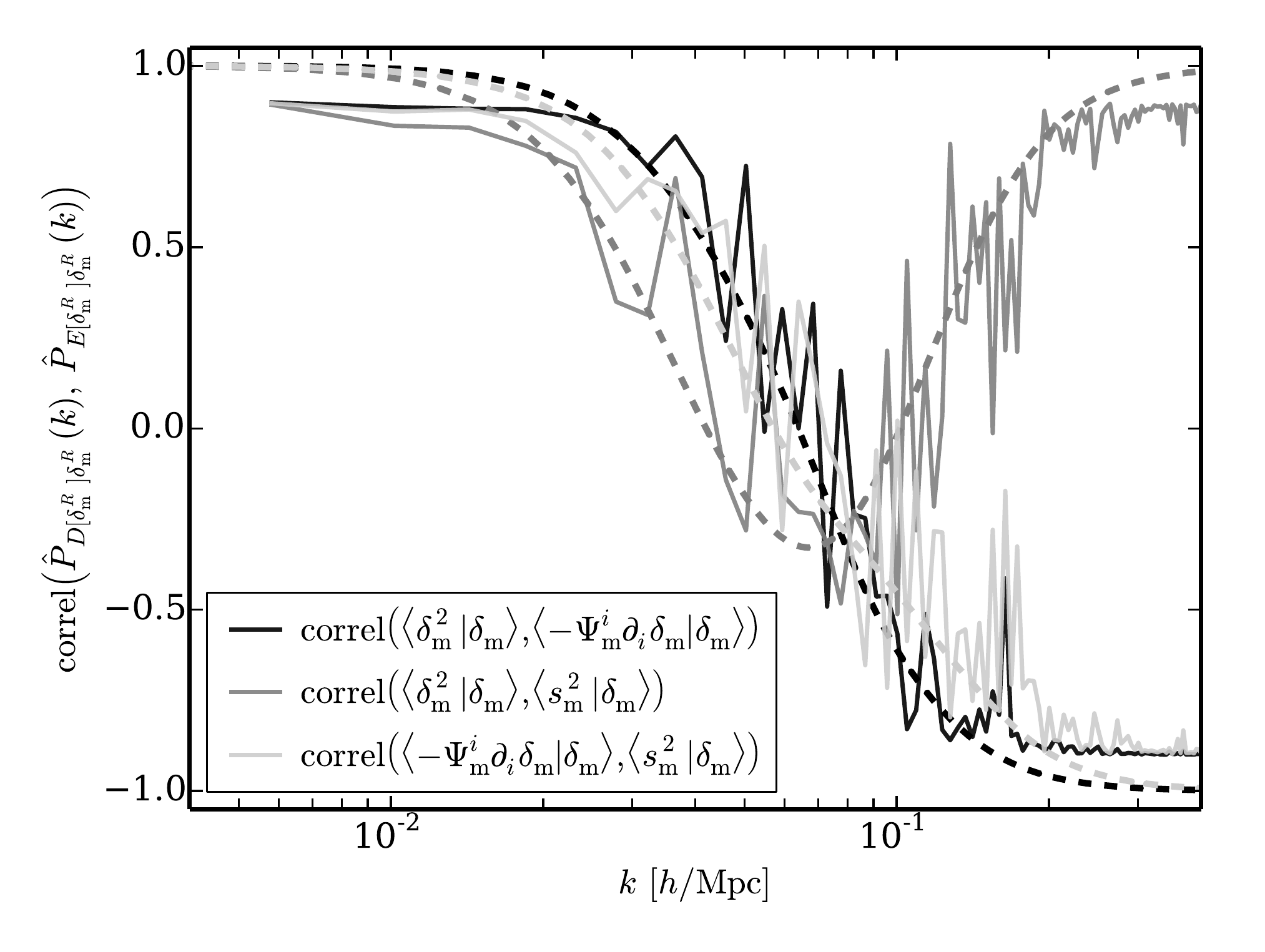}
}
\caption{\emph{Left:}  
Ratio of estimated over theoretical standard deviation of
matter-matter-matter cross-spectra  (using $D=E$ and $a=b=\mathrm{m}$
in \eqq{cov_cross_spectra_diagonal}).
\emph{Right:} Correlations between matter-matter-matter cross-spectra at
  the same scale $k=k'$ predicted by theory 
  (\eqq{correl_cross_spectra_diagonal}, dashed) and estimated from simulations
  (solid).
Both panels assume Gaussian smoothing with
  $R_G=20h^{-1}\mathrm{Mpc}$ and use the linear matter power spectrum in
  theory expressions. For smaller smoothing scale $R$, the zero
  crossing would move to higher $k$.}
\label{fig:correl_mmm_RGauss20}
\end{figure}

To further test the shot noise corrections, we consider two additional
mass bins above the fiducial four mass bins used in the rest of the
paper, with linear bias $b_1=2.8$ and $4.3$. Fitting $\Delta_1$ and
imposing $\Delta_2=\Delta_1/\bar{n}_\mathrm{h}$ gives
$\Delta_1=-4.7h^{-3}\mathrm{Gpc}^3$ with 
$\chi^2/\mathrm{d.o.f.}=1.8$  for the
$b_1=2.8$ mass bin, and $\Delta_1=-16.6h^{-3}\mathrm{Gpc}^3$ with
$\chi^2/\mathrm{d.o.f.}=3.4$ for the $b_1=4.3$ mass bin.  For these
two high-mass bins, the reduced $\chi^2$ of the fits improve to $1.4$ and
$0.67$ if $\Delta_2$ is treated as a free parameter; see
Fig.~\ref{fig:hhh_model_vs_data_RGauss20_fitAshot_mass45}.  

The shot noise correction $\Delta_1$ is positive for the lowest four
mass bins, but becomes negative for the two very high mass bins shown
in Fig.~\ref{fig:hhh_model_vs_data_RGauss20_fitAshot_mass45}.  This
mass dependence is qualitatively consistent with Fig.~11 of
\cite{tobias1305}, where the shot noise correction to the power
spectrum turns negative at around $3\times 10^{13}h^{-1}M_{\odot}$,
because the exclusion effect dominates at high mass.  While
alternative modifications of the model might be able to describe the
simulations similarly well, the qualitative agreement of the mass
dependence of the shot noise correction and the fact that the shot
noise corrections are capable to capture the cross-spectrum
measurements for all mass bins indicate that deviations from Poisson
shot noise caused by exclusion and nonlinear biasing 
are indeed responsible for the disagreement in
Fig.~\ref{fig:hhh_model_vs_data_RGauss20_fixAshot0}.
If so one should be able to model these effects rather than treat them as a 
free parameter (see Fig.~11 of
\cite{tobias1305} for a theoretical model prediction that qualitatively 
agrees with our measurements). We note that a more detailed modeling 
in \cite{tobias1305} predicts this stochasticity term to be constant 
(i.e.~shot noise like) only for low $k$, and is expected to vanish at high $k$ (with 
the transition given by the halo radius scale). A more detailed analysis is needed 
to investigate what the appropriate form is for the bispectrum analysis, and we
expect that the phenomenological approach adopted here can be improved considerably 
with a more detailed modeling. 

\begin{figure}[tp]
\centerline{
\includegraphics[width=0.45\textwidth]{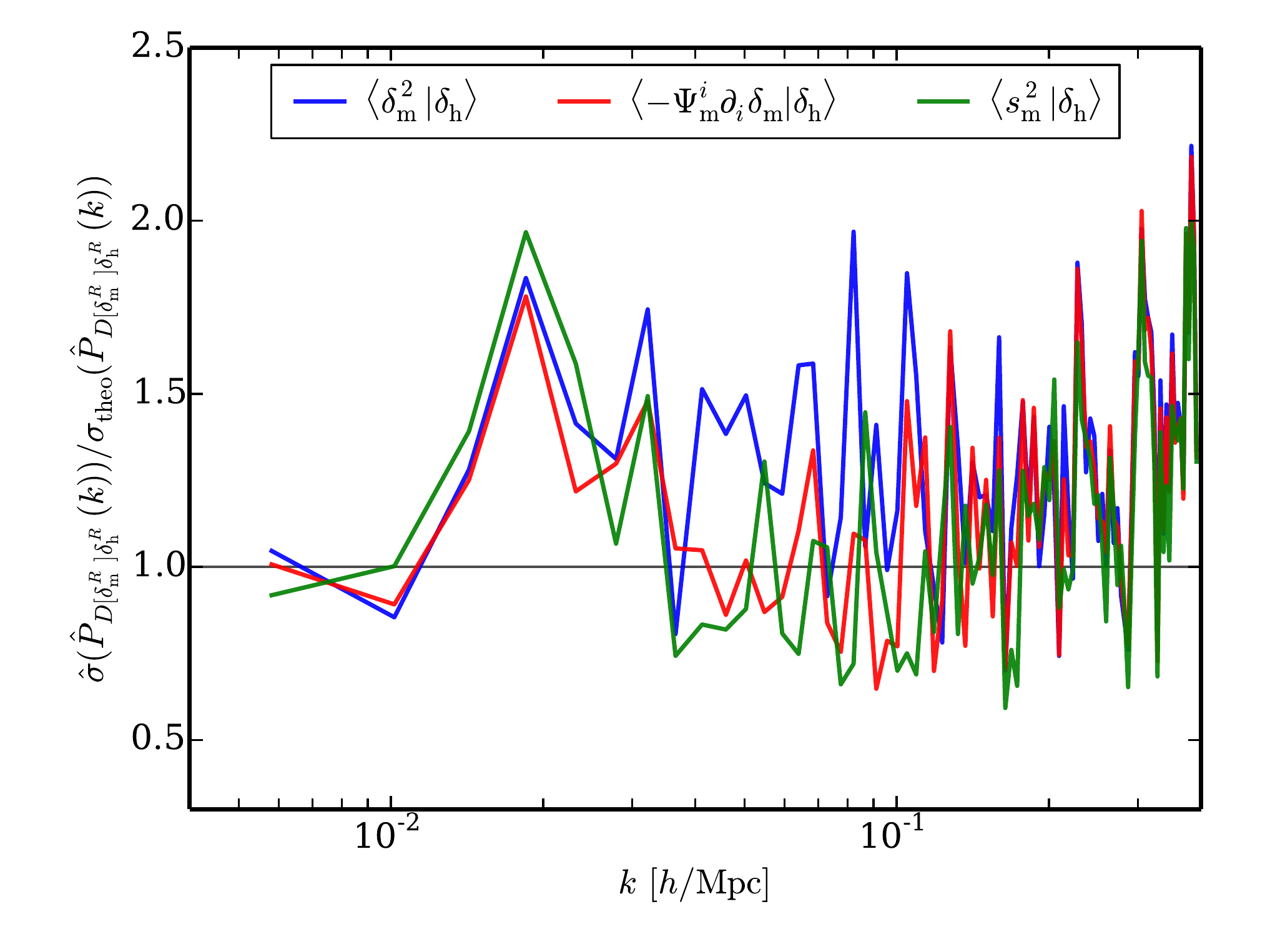}
\includegraphics[width=0.45\textwidth]{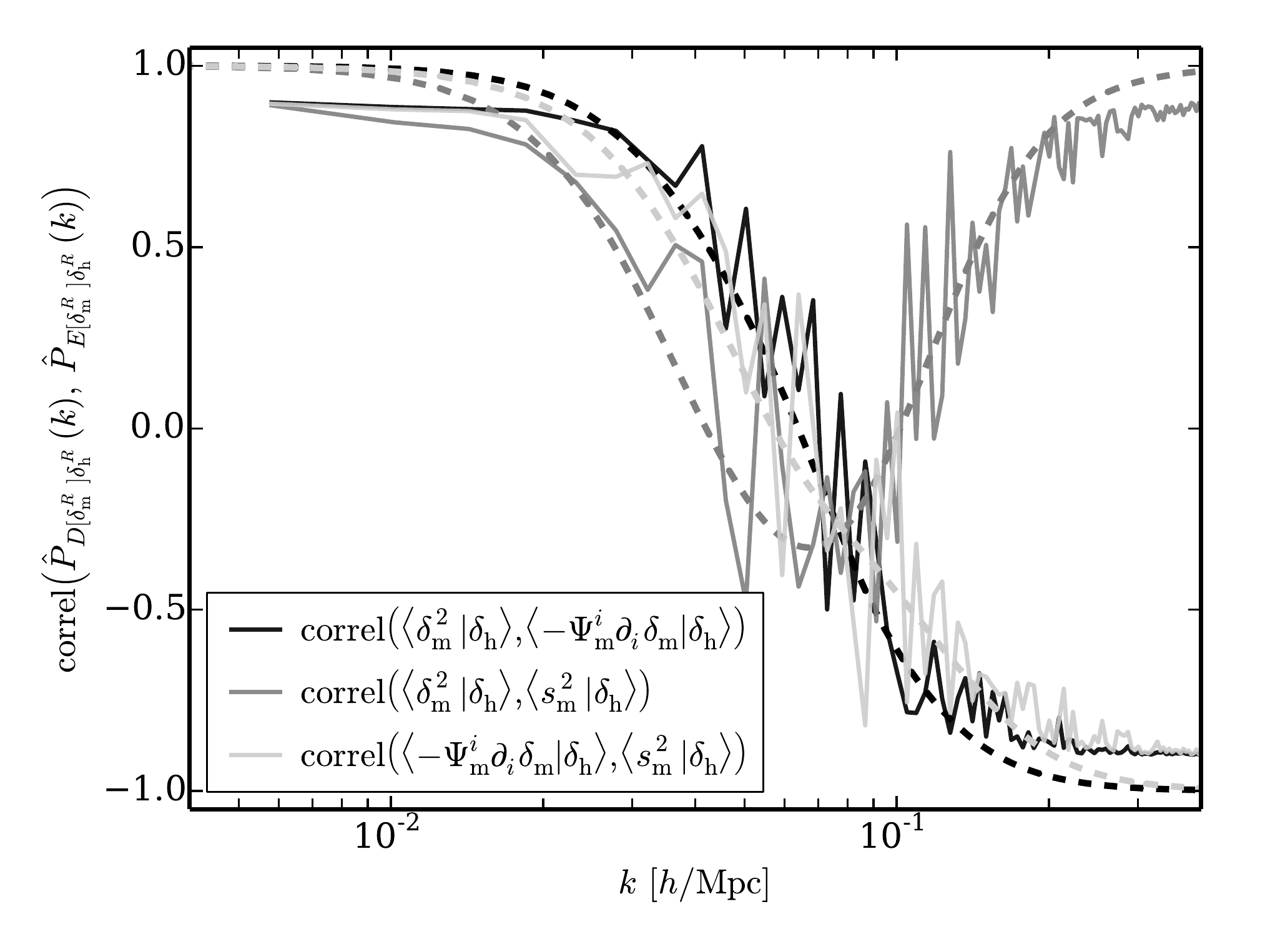}
}
\centerline{
\includegraphics[width=0.45\textwidth]{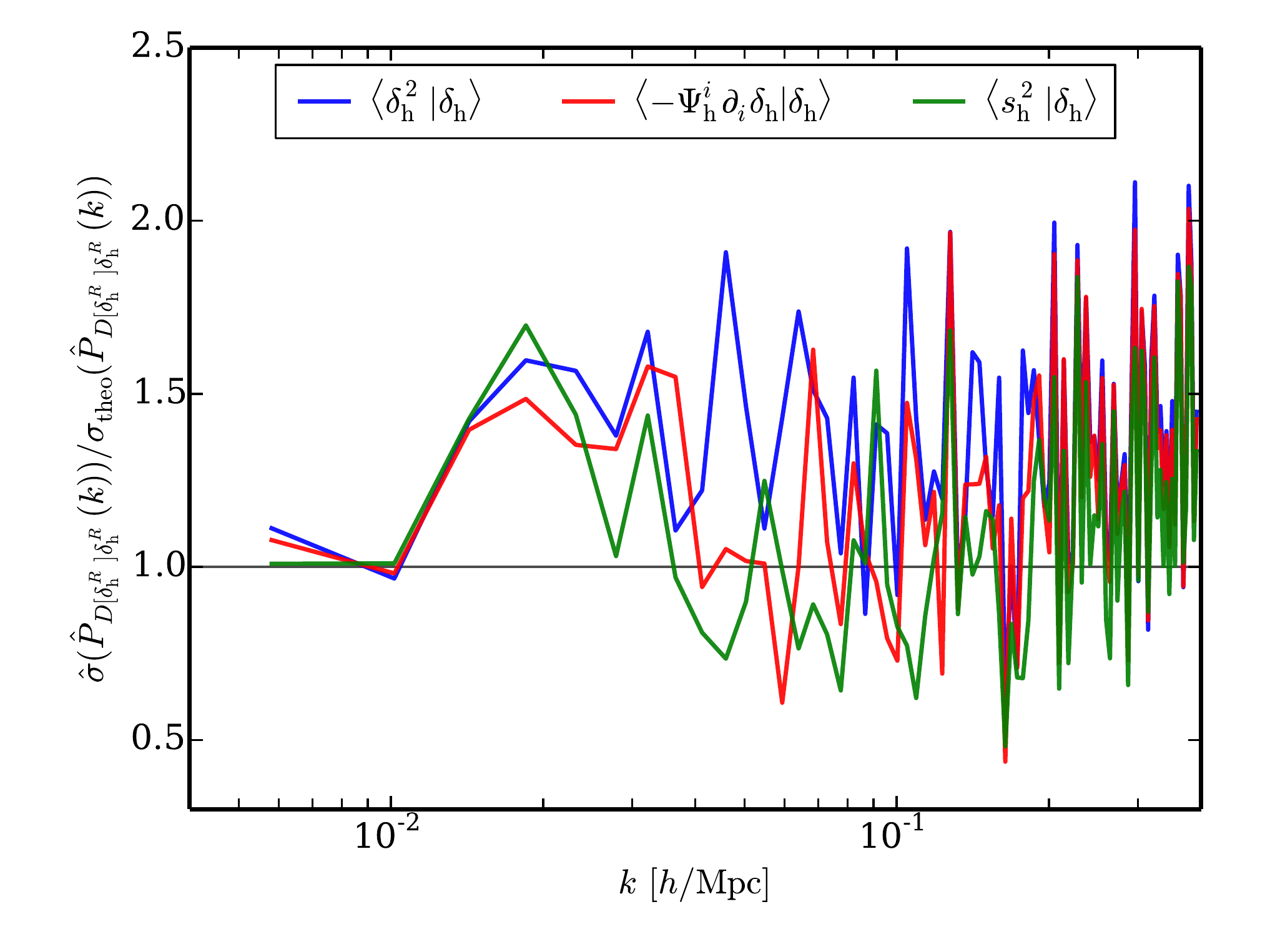}
\includegraphics[width=0.45\textwidth]{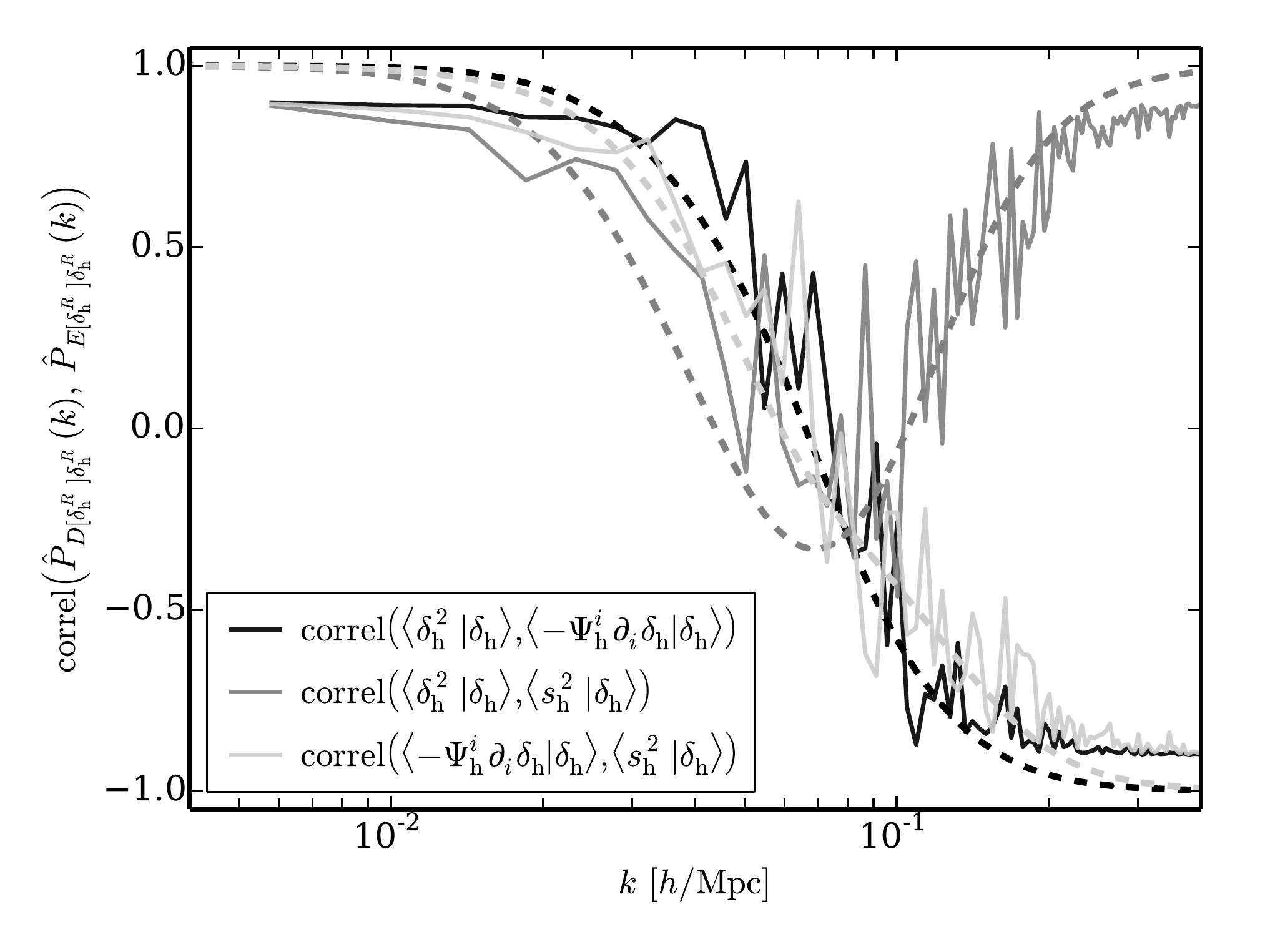}
}
\caption{Same as Fig.~\ref{fig:correl_mmm_RGauss20} but for
  matter-matter-halo cross-spectra (top) and halo-halo-halo
  cross-spectra (bottom), using the halo mass
  bin with $b_1=1.98$. Theoretical standard deviations \eq{cov_cross_spectra_diagonal}
are evaluated using the estimated ensemble-averaged halo-halo power
spectrum including shot noise for $P_\mathrm{hh}$ and the linear
matter power spectrum for $P_\mathrm{mm}$.
Theory correlations of \eqq{correl_cross_spectra_diagonal} are 
  the same as for the matter-matter-matter case. Results for other
  mass bins are similar (not shown). }
\label{fig:correl_mmh_and_hhh_RGauss20_mass3}
\end{figure}

\subsection{Cross-spectrum covariances}
\label{se:EstimatedCovs}
For the case of matter-matter-matter cross-spectra and $R_G=20h^{-1}\mathrm{Mpc}$, 
Fig.~\ref{fig:correl_mmm_RGauss20} tests the predicted cross-spectrum
covariances of \eqq{cov_cross_spectra_diagonal} (counting the number
of modes manually in the code for the given $k$ binning) against 
estimates obtained from ten realizations.  Due to the small number of
realizations, the covariance estimates are rather noisy. Within this
large uncertainty, the standard deviations of the three cross-spectra
(Fig.~\ref{fig:correl_mmm_RGauss20} left) as well as the
cross-correlations between the cross-spectra at the same scale $k=k'$
(Fig.~\ref{fig:correl_mmm_RGauss20} right) are consistent between
simulations and theory at $k\lesssim 0.2h/\mathrm{Mpc}$.

For matter-matter-halo and halo-halo-halo cross-spectra, we find similar agreement; see
Fig.~\ref{fig:correl_mmh_and_hhh_RGauss20_mass3} for the $b_1=1.98$ halo mass
bin. Similar results are obtained for lower mass bins (not shown for
brevity).  In particular, the ratio of measured over theoretical
standard deviations fluctuates between $0.5$ and $2$ for all mass
bins for $k\lesssim 0.2h/\mathrm{Mpc}$.

More work is needed to test the covariances at higher precision,
e.g.~by running more realizations or by dividing simulation boxes into
sub-boxes. We also leave it for future work to test the theoretical
covariances \eq{cov_DE_final} between cross-spectra at different
scales $k\ne k'$.

\subsection{Bias estimation from halo-halo-halo cross-spectra}
\label{se:biasContours}
While the goal of this section thus far has been to test the theory
predictions against simulations, we now aim to get a rough
sense of how well bias parameters could be measured from observable
halo-halo-halo cross spectra alone. This should be regarded mainly as
a motivation to study these observables in more detail in the future
rather than a realistic forecast, because we make a number of
idealistic assumptions that are not valid in practice:  the
covariance between cross-spectra is assumed to be given by the
leading-order theoretical expression of \eqq{correl_cross_spectra_diagonal}, neglecting any
covariance between different wavenumbers $k'\ne k$; observations
are assumed to be in periodic boxes in real space, neglecting
redshift-space distortions; the shot noise correction 
 is treated as a free parameter; and the cosmology is
fixed to the fiducial cosmology of the $N$-body runs. Most of these
assumptions are likely to impact the constraining
power of the cross-spectra and should be addressed before considering
real data.

\begin{figure}[tp]
\centerline{
\includegraphics[width=0.8\textwidth]{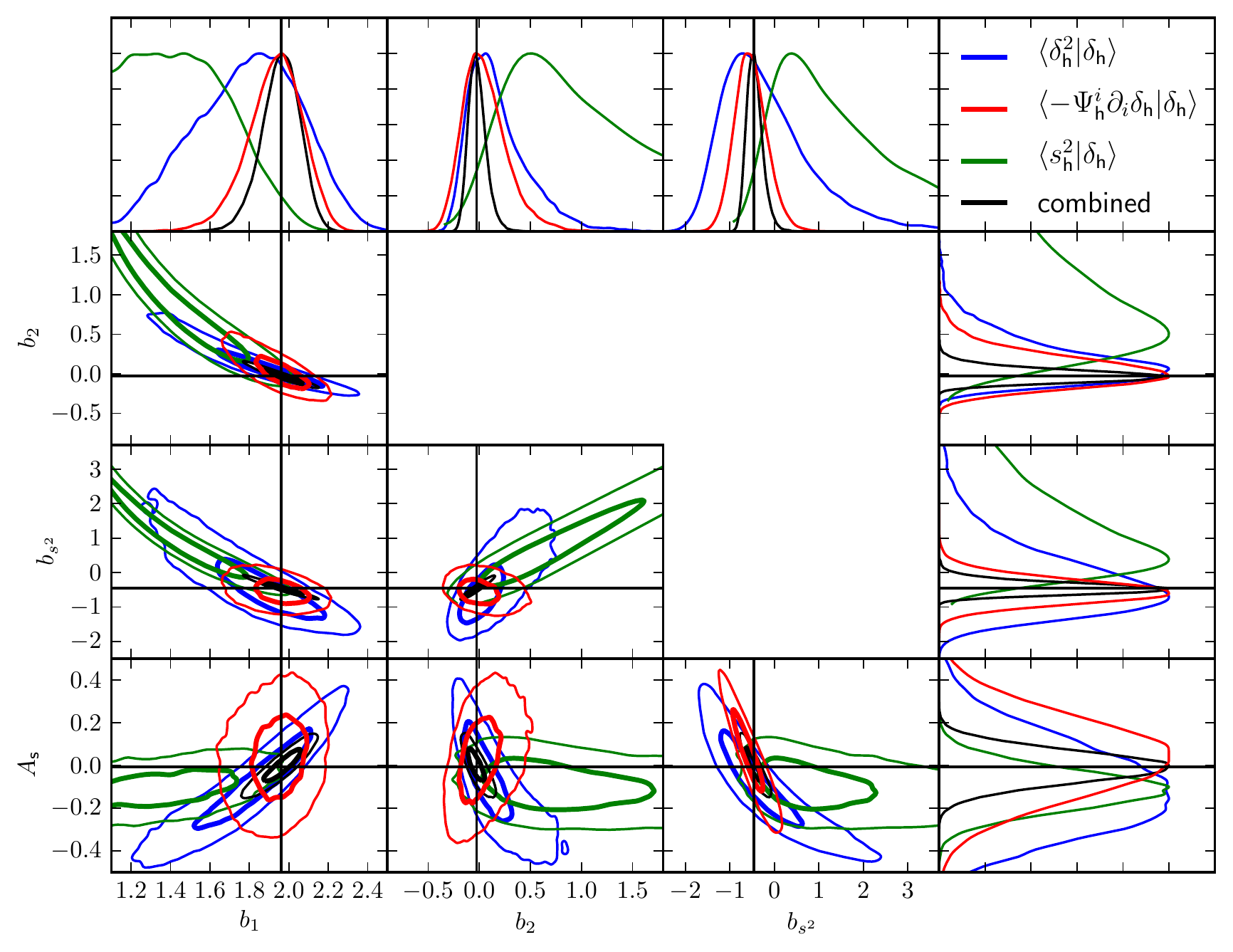}
}
\caption{Results of fitting $b_1$, $b_2$, $b_{s^2}$ and
  $A_\mathrm{shot}\equiv -\bar{n}_\mathrm{h}\Delta_1$ (imposing
  $\Delta_2=\Delta_1/\bar{n}_\mathrm{h}$) to halo-halo-halo
  cross-spectra involving the squared density (blue), shift term (red)
  or tidal term (green), or all three combined (black), choosing Gaussian
  $R=20h^{-1}\mathrm{Mpc}$ smoothing and
  $k_\mathrm{max}=0.09h/\mathrm{Mpc}$ for the halo mass bin with
  $b_1=1.98$.
  The 2d contours of the posterior show $68\%$ and $95\%$
  confidence regions corresponding to the full volume of
  $V=26.3h^{-3}\mathrm{Gpc}^3$ (i.e.~errors in a single realization
  would be larger by a factor of $\sqrt{10}$). 
Thin black lines show the
  maximum-likelihood points corresponding to the black contours.
  The joint likelihood for the
  three halo-halo-halo cross-spectra is assumed to be Gaussian in the
  cross-spectra, with non-zero covariance between cross-spectra at
  $k=k'$ given by the theory expression of
  \eqq{cov_cross_spectra_diagonal}.  The green contours are somewhat
  uncertain because it is not clear how well the MCMC chains sampled
  the elongated degeneracies. } 
\label{fig:contours_hhh_RGauss20_and_Rgauss10_mass3}
\end{figure}

Under the above idealistic assumptions, we use the
  Monte-Carlo sampler emcee \cite{emcee} to fit the bias parameters
  $b_1$, $b_2$ and $b_{s^2}$ as well as the dimensionless shot noise
  amplitude $A_\mathrm{shot}\equiv -\bar{n}_\mathrm{h}\Delta_1$
  (imposing $\Delta_2=\Delta_1/\bar{n}_\mathrm{h}$) to the three
  ensemble-averaged halo-halo-halo cross-spectra
  $P_{D[\delta^R_\mathrm{h}],\delta^R_\mathrm{h}}(k)$,
  $D\in\{\mathsf{P}_0, -F_2^1\mathsf{P}_1, \mathsf{P}_2\}$ for the
  halo mass bin with $b_1=1.98$.  Colored contours in
  Fig.~\ref{fig:contours_hhh_RGauss20_and_Rgauss10_mass3} show results
  if each cross-spectrum is fitted individually, while the black
  contours correspond to the joint fit to all three cross-spectra.
  Focusing on the leftmost column of
  Fig.~\ref{fig:contours_hhh_RGauss20_and_Rgauss10_mass3}, we see that
  the linear bias $b_1$ is best determined by the shift
  cross-spectrum. Adding the other two cross-spectra does not tighten
  the $b_1$ constraint much, i.e.~the shift cross-spectrum contains almost the
  entire bispectrum information on $b_1$.  Once  $b_1$ is known, the
  constraint on $b_2$ is mostly tightened by the squared density
  cross-spectrum (blue degeneracy regions are thinner than green ones in the
  $b_1$ vs $b_2$ panel of
  Fig.~\ref{fig:contours_hhh_RGauss20_and_Rgauss10_mass3}), while the
  constraint on $b_{s^2}$ is mostly 
  tightened by the tidal cross-spectrum (green degeneracy regions are
  thinner than blue ones in the $b_1$ vs $b_{s^2}$ panel).  Intuitively, this can be understood
  from the Legendre decomposition of the halo-halo-halo bispectrum in
  Eqs.~\eq{Bhhh_l0}-\eq{Bhhh_l2}: The  shift term corresponding to $l=1$ is the only
  term that picks up $b_1$ without any contribution from nonlinear
  bias $b_2$ or $b_{s^2}$ (in absence of velocity bias), while the $l=0$ squared
  density term is the only one that gets contributions from $b_2$ and
  the $l=2$ tidal term is the only term depending on
  $b_{s^2}$.\footnote{This intuitive picture is not exact though: In
    practice,   finite $k_\mathrm{max}$ and smoothing 
imply that each cross-spectrum actually picks up dependencies on all
bias parameters, see
Fig.~\ref{fig:contours_hhh_RGauss20_and_Rgauss10_mass3}. 
}  

Overall, the best-fit parameters 
from the combined fit, $(b_1,b_2,b_{s^2},A_\mathrm{shot})=(1.96,
-0.02, -0.44, -0.007)$, agree with the values  obtained from matter-halo
statistics, $(1.98, -0.08, -0.52,
-5\times 10^{-5})$, within the uncertainties shown in
  Fig.~\ref{fig:contours_hhh_RGauss20_and_Rgauss10_mass3},
but the $1\sigma$ error of the mean of $b_1$ is relatively large
($\sim 5\%$, using the full $26.3h^{-3}\mathrm{Gpc}^3$ volume). The
uncertainties decrease significantly for higher $k_\mathrm{max}$ and
smaller smoothing scales $R$, e.g.~$b_1$ could be measured with $\sim
0.4\%$ precision for $k_\mathrm{max}=0.24h/\mathrm{Mpc}$ and
$R=5h^{-1}\mathrm{Mpc}$, but 
 best-fit parameters become inconsistent with the true values, demonstrating that 
modeling needs to be improved or appropriate clipping or
  logarithmic density transforms  \cite{Neyrinck0903,SimpsonClipping1107} need to be applied
before it is possible to push to such
small scales.  This would also make the full advantage of the
  efficient cross-spectrum method over conventional brute-force methods more
  apparent because the latter become computationally unfeasible at high $k$.

The degeneracies of the combined contours in 
Fig.~\ref{fig:contours_hhh_RGauss20_and_Rgauss10_mass3} correspond to the
typical degeneracies between bias parameters estimated from the
halo-halo-halo bispectrum if redshift-space distortions are neglected,
e.g.~higher $b_1$ can be compensated by lower $b_2$ or
$b_{s^2}$. Since the shot noise correction parameterised by $A_\mathrm{shot}$ is also rather degenerate
with the bias parameters, it would be interesting to model this
analytically instead of treating it as a free parameter.

As a consistency check of the assumed likelihood, jack-knife error
bars are obtained by fitting to each of the ten realizations
individually and calculating the scatter of the best-fit values among
realizations. These jack-knife error estimates are larger than the
average uncertainty predicted by the likelihood width for a single
realization by a factor of $\sim 1.8$ for the conservative $k_\mathrm{max}=0.09h/\mathrm{Mpc}$
and by a factor of $\sim 1.25$ for the more ambitious $k_\mathrm{max}=0.24h/\mathrm{Mpc}$. This could
be attributed to additional contributions to the true covariance that
are neglected in our theoretical covariance (e.g.~at $k'\ne k$), or
uncertainty in the determination of the jack-knife errors due to the
small number of realizations, or a departure of the true likelihood
from a Gaussian pdf.  

Summarizing the above, it seems possible to
reach percent-level estimates of the linear bias $b_1$ from
halo-halo-halo cross-spectra in real surveys, but we re-emphasize that
we made a number of unrealistic assumptions, which could easily change
this conclusion (to the better or worse).


\section{Extension to primordial non-Gaussianity}
\label{se:Extensions}

So far we have assumed Gaussian initial conditions.  Multiple-field
inflation models can generate local primordial non-Gaussianity that
induces an additional contribution to the matter-matter-matter
bispectrum of the local form
\begin{equation}
  \label{eq:Bmmmlocal}
  B_\mathrm{mmm}^\mathrm{loc}(k_1,k_2,k_3) =
2  f_\mathrm{NL}^\mathrm{loc} \left[
\frac{M(k_3)}{M(k_1)M(k_2)}P_\mathrm{mm}(k_1)P_\mathrm{mm}(k_2)+2\mbox{
  perms}\right],
\end{equation}
where $M(k)=M(k,z)$ is the linear Poisson conversion factor between the primordial
potential $\Phi$ and the late-time matter density at redshift $z$,
\begin{equation}
  \label{eq:8}
  M(k,z) \equiv \frac{2}{3}\,\frac{k^2T(k)D(z)}{\Omega_\mathrm{m}H_0^2},
\end{equation}
so that $\delta_\mathrm{m}^\mathrm{lin}(\vk,z)=M(k,z)\Phi(\vk)$.
Here, $T(k)$ is the linear transfer function normalized to $T(k)=1$ on
large scales, and the linear growth factor $D(z)$ for
$\Omega_\mathrm{rad}=0$ is normalized to $D(z)=1/(1+z)$ during matter
domination. Note that $M(k)\propto k^2$ for $k\ll k_\mathrm{eq}$ and
 $M(k)\propto k^0$ for $k\gg k_\mathrm{eq}$. The bispectrum
 \eq{Bmmmlocal} is maximal in the squeezed limit (e.g.~$k_1\ll
 k_2\approx k_3$).

Plugging  the bispectrum  \eq{Bmmmlocal} into
\eqq{fnl_separable_fgh}, we get
\begin{equation}
  \label{eq:fnl_local_esti}
  \hat f^\mathrm{loc}_\mathrm{NL} = \frac{24\pi
    L^3}{N_\mathrm{loc}}\int \d k\frac{k^2 M^2(k)}{P_\mathrm{mm}(k)}
\hat P_{[\frac{\delta_\mathrm{m}}{M}]^2, \frac{\delta_\mathrm{m}}{M}}(k),
\end{equation}
where we defined the quadratic field
\begin{equation}
  \label{eq:64}
\left[\frac{\delta_\mathrm{m}}{M}\right]^2(\vk) \equiv \int \frac{\d^3
  q}{(2\pi)^3}\,\frac{\delta_\mathrm{m}(\vq)}{M(q)} \frac{\delta_\mathrm{m}(\vk-\vq)}{M(|\vk-\vq|)}
\end{equation}
and the filtered density
\begin{equation}
  \label{eq:67}
  \frac{\delta_\mathrm{m}}{M}(\vk) \equiv \frac{\delta_\mathrm{m}(\vk)}{M(k)}.
\end{equation}
At leading order, this equals the primordial potential $\Phi$
reconstructed from the DM density $\delta_\mathrm{m}$. The
cross-spectrum in \eqq{fnl_local_esti} then probes the cross-spectrum
of this reconstructed $\Phi$ with $\Phi^2(\vx)$, which corresponds to
the mechanism that generates primordial non-Gaussianity of the local
kind (adding $f_\mathrm{NL}\Phi^2(\vx)$ to $\Phi(\vx)$).

The cross-spectrum appearing in \eqq{fnl_local_esti} could be used to
estimate $f_\mathrm{NL}^\mathrm{loc}$ if the dark matter density was
directly observable.  The extension to observable halo densities is
left for future work.  It would also be
straightforward to extend the cross-spectra to other separable types
of primordial non-Gaussianity generated by other inflation models (e.g.~equilateral or
orthogonal).


\section{Conclusions}
\label{se:Conclusions}
In this paper we explore methods to probe  large-scale
structure bispectrum parameters in a nearly optimal way. 
The tree level bispectrum   receives contributions from 
gravity at second order, which can be Legendre decomposed into 
the squared density $\delta^2(\vx)$, the shift term
  $-\Psi^i(\vx)\partial_i \delta(\vx)$ and the tidal term
  $s^2(\vx)=\tfrac{3}{2}s_{ij}(\vx)s_{ij}(\vx)$.
When applied to galaxies or halos the gravity term is multiplied by the 
appropriate linear bias $b_1$ factor (e.g.~$b_1^3$ when investigating 
the halo bispectrum). In addition, nonlinear biasing can introduce two 
additional terms that contribute at second order, $b_2\delta^2(\vx)$ and $b_{s^2}s^2(\vx)$. 
There is no nonlinear bias associated with the shift term in the absence of velocity bias. 
Since any velocity bias must vanish in the $k \rightarrow 0$ limit as a consequence of 
Galilean invariance, we do not include any such term.\footnote{At a $k^2$ level there could be a velocity bias, 
but we ignore this here since we work at the lowest order in $k$. Indeed, all the biasing terms can
receive $k^2$ type corrections \cite{Tobias14}.  } 

These terms correspond to individual components of the bispectrum in a separable form. In this case, in the 
limit where tree level theory is valid, one can write an optimal bispectrum estimator using these terms. Specifically, 
given a density
$\delta(\vx)$, smoothed on the smallest scale where we 
still trust the theory predictions, the procedure we propose is as
follows: 
\begin{enumerate}
\item Compute the density gradient $\partial_i\delta(\vx)$, the
  displacement field $\Psi_i(\vx)=-\partial_i\partial^{-2} \delta(\vx)$, and
  the tidal tensor $s_{ij}(\vx)=\big[\partial_i\partial_j\partial^{-2}
   - \tfrac{1}{3}\delta_{ij}^{(K)}\big]\delta(\vx)$.
\item Compute the squared density $\delta^2(\vx)$, the shift term
  $-\Psi^i(\vx)\partial_i \delta(\vx)$ and the tidal term
  $s^2(\vx)=\tfrac{3}{2}s_{ij}(\vx)s_{ij}(\vx)$. 
\item Fourier transform
  the three quadratic fields to get $[\delta^2](\vk)$,
  $[-\Psi^i\partial_i\delta](\vk)$ and $[s^2](\vk)$.
\item Compute the cross-spectra between the  quadratic fields and the
  density, i.e.~(suppressing division by the number of modes)
  \begin{eqnarray}
    \label{eq:delta2_conclusion}
\hat P_{\delta^2,\delta}(k)& \sim &
\sum_{\vk,|\vk|=k}
  [\delta^2](\vk)\delta(-\vk),
\\
    \label{eq:shift_conclusion}
\hat P_{-\Psi^i\partial_i\delta,\delta}(k)&\sim &
\sum_{\vk,|\vk|=k}
  [-\Psi^i\partial_i\delta](\vk)\delta(-\vk)
\\
    \label{eq:s2_conclusion}
\hat P_{s^2,\delta}(k)&\sim &
\sum_{\vk,|\vk|=k} [s^2](\vk)\delta(-\vk).
  \end{eqnarray}
\end{enumerate}

As expected from the Legendre decomposition of the
  halo bispectrum, the $l=1$ shift
cross-spectrum \eq{shift_conclusion} contains almost the entire
bispectrum information on the linear bias $b_1$, while the $l=0$ squared density
cross-spectrum \eq{delta2_conclusion} and the $l=2$ tidal
cross-spectrum \eq{s2_conclusion} mostly improve constraints on $b_2$
and $b_{s^2}$ once $b_1$ is known.
Measuring all three cross-spectra and comparing them to their theory
predictions is equivalent to an optimal maximum-likelihood estimation
of the amplitudes of contributions to dark matter or halo bispectra
(under certain regularity conditions; see Section
\ref{se:MaxLikeliBispEsti}). Therefore, these cross-spectra contain the same
constraining power on bias parameters and $\sigma_8$ as a full optimal
bispectrum analysis.
Measuring cross-spectra is both
simpler and computationally cheaper than performing direct bispectrum
measurements for individual triangle configurations. Since they only depend on a single rather than three
wavenumbers, modeling the covariance is also simpler.

We have derived leading-order perturbation theory predictions for the
expectation values and covariances of the three cross-spectra, where
both the quadratic and the single field can be dark matter or halo
fields, and second order bias $b_2$ and tidal tensor bias $b_{s^2}$
are included. The results are given by integrals over
matter-matter-matter, matter-matter-halo or halo-halo-halo bispectra. 

The proposed cross-spectra were measured on a set of ten large $N$-body
simulations. The expectation values
are consistent with perturbation theory at the few percent level for
$k \lesssim 0.09h/\mathrm{Mpc}$ at $z=0.55$ for matter-matter-matter
and matter-matter-halo combinations, if all fields are smoothed by a
Gaussian with smoothing scale $R=20h^{-1}\mathrm{Mpc}$.  For
halo-halo-halo cross-spectra, 
one must include corrections to the Poisson stochasticity. 
 While these corrections are qualitatively similar to corrections to the halo-halo power spectrum due to exclusion and nonlinear biasing \cite{tobias1305}, future work should investigate and model them in more detail. 
The predicted variance of the cross-spectra
and the covariance between any two cross-spectra at the same
wavenumber are found to be consistent with simulations (although the
numerical noise is somewhat large given the small number of
independent realizations). 

The ultimate goal of this is to determine the three bias parameters and dark matter clustering power 
spectrum by combining these three statistics with the measured galaxy power spectrum.  
We have not performed this step in this paper: we plan to explore
potential improvements of the modeling by including higher order
perturbation theory terms as well as improved bias and stochasticity
models to measure the new observables in galaxy surveys in future
work. We also plan to include redshift space distortions
by accordingly modifying and extending the cross-spectra.   Given the 
simplicity of the method and the agreement with
leading-order perturbation theory on large scales, we hope it will become a useful
tool to break degeneracies of bias and cosmological parameters.

\section*{Acknowledgements}
We thank Florian Beutler and Blake Sherwin for many helpful
discussions, and Martin White and Beth Reid for providing the $N$-body
simulations used in this paper.  We are also very grateful to Anatoly
Klypin for sharing a CIC and power spectrum code.  We acknowledge useful
discussions with James Fergusson, Airam Marcos-Caballero, Paul
Shellard and Zvonimir Vlah.  TB gratefully acknowledges support from
the Institute for Advanced Study through the Corning Glass Works
Foundation Fellowship.  This research used resources of the National
Energy Research Scientific Computing Center, a DOE Office of Science
User Facility supported by the Office of Science of the
U.S. Department of Energy under Contract No. DE-AC02-05CH11231.  It
also used the COSMOS Shared Memory system at DAMTP, University of
Cambridge operated on behalf of the STFC DiRAC HPC Facility and funded
by BIS National E-infrastructure capital grant ST/J005673/1 and STFC
grants ST/H008586/1, ST/K00333X/1.  

\appendix

\section{Large-scale limits of integrals}
\label{se:lowk_integral_limits}
The large-scale limits $k\ll q$ of the integrals $I^R_{DE}$ and
$I^{\mathrm{bare},R}_{DE}$ in Eqs.~\eq{I_DE_R} and \eq{I_DE_R_bare}
can be obtained by expanding all integrands consistently in $k/q$ (for
similar calculations see
e.g.~\cite{chiang1403,valageas1311,Kehagias1311,LiHu1401,figueroa1205,SherwinZaldarriaga1202}).
We will only calculate the leading order terms which turn out to be of
order $(k/q)^0$.  The $F_2$ kernel contains a term scaling like $q/k$
which diverges for $k/q\rightarrow 0$. Since this can combine with
contributions of order $k/q$ in other terms to give an overall term of
order $(k/q)^0$, we generally need to keep all terms of order $k/q$.
The orthogonality of Legendre polynomials will be needed for integrals
$I^{\mathrm{bare},R}_{DE}$, but it is not used for integrals $I^R_{DE}$ whose
integrands are angle-independent on large scales.

In detail, to get the low-$k$ limit of
\eqq{I_DE_R},
\begin{equation}
  \label{eq:38}
  \lim_{k\rightarrow 0}    I_{DE}^R(k) =
  \lim_{k\rightarrow 0} W_R(k)
  \int\frac{\d^3 q}{(2\pi)^3}W_R(q)W_R(|\vk-\vq|)P_\mathrm{mm}(q)P_\mathrm{mm}(|\vk-\vq|)
  D(\vq, \vk-\vq)E(\vq, \vk-\vq),
\end{equation}
we expand all factors.
We work with the cosine $\mu$   between $\vq$ and $\vk-\vq$, see \eqq{mu_def},
and the cosine  $\nu$ between $\vq$ and $\vk$
\begin{equation}
  \label{eq:xdef}
  \nu \equiv \frac{\vq\cdot\vk}{qk}.
\end{equation}

From
\begin{equation}
  \label{eq:lowk_k1q_over_q}
 \lim_{k\rightarrow 0}\, \frac{|\vk-\vq|}{q} =  \lim_{k\rightarrow 0}\, \sqrt{1-2\frac{k\nu}{q}+\frac{k^2}{q^2}} =
  1-\frac{k\nu}{q} + \mathcal{O}\left(\frac{k^2}{q^2}\right)
\end{equation}
we get
\begin{equation}
  \label{eq:lowk_mu}
 \lim_{k\rightarrow 0}\,  \mu =  \lim_{k\rightarrow 0}\,
 \frac{-1+\frac{k\nu}{q}}{1-\frac{k\nu}{q}} = -1+\mathcal{O}\left( 
\frac{k^2}{q^2}\right).
\end{equation}
Then
\begin{equation}
  \label{eq:lowk_F2ell1_q_k1q}
 -\lim_{k\rightarrow 0}\, F_2^{1}(q,|\vk-\vq|) \mathsf{P}_1(\mu) = -\lim_{k\rightarrow 0}\,\frac{1}{2}\mu\left(
\frac{|\vk-\vq|}{q} + \frac{q}{|\vk-\vq|} 
\right)
= 1 + \mathcal{O}\left(
\frac{k^2}{q^2}\right)
\end{equation}
and 
\begin{equation}
  \label{eq:lowk_S2_q_k1q}
  \lim_{k\rightarrow 0}\,\mathsf{P}_2(\mu) =   \lim_{k\rightarrow 0}\, \frac{3}{2}\left(\mu^2 - \frac{1}{3}\right) = 1 + \mathcal{O}\left(
\frac{k^2}{q^2}\right).
\end{equation}
This can be summarized as
\begin{equation}
  \label{eq:lowk_D_q_k1q}
  \lim_{k\rightarrow 0} D(\vq, \vk-\vq) = 1 + \mathcal{O}\left(
\frac{k^2}{q^2}\right), \qquad
\qquad D\in\big\{\mathsf{P}_0, -F_2^{1}\mathsf{P}_1,
  \mathsf{P}_2\big\},
\end{equation}
which implies for the full $F_2$ kernel
\begin{equation}
  \label{eq:lowk_F2_q_k1q}
  \lim_{k\rightarrow 0}\, F_2(\vq,\vk-\vq) 
= 0  + \mathcal{O}\left(
\frac{k^2}{q^2}\right).
\end{equation}
Some bare integrals \eq{I_DE_R_bare} contain $E(\vq,
-\vk)$. For $E=-F_2^1\mathsf{P}_1$ this becomes
\begin{equation}
  \label{eq:lowk_F21_q_mk}
  \lim_{k\rightarrow 0} \left[-F_2^{1}(q,k)\mathsf{P}_1(-\nu)\right]
= \frac{\nu}{2}  \lim_{k\rightarrow 0} \left(\frac{q}{k} + \frac{k}{q}\right)
= \frac{q\nu}{2k}  + \mathcal{O}\left(\frac{k}{q}\right),
\end{equation}
which diverges for $k/q\rightarrow 0$.  Note that for $E=\mathsf{P}_2$,
$\mathsf{P}_2(-\nu)=\tfrac{3}{2}(\nu^2-\tfrac{1}{3})$ is independent of $k$, so that
\begin{equation}
  \label{eq:lowk_F2_q_mk}
  \lim_{k\rightarrow 0}F_2(\vq, -\vk) = \frac{17}{21} - \frac{q\nu}{2k}
  + \frac{2}{7}\left(\nu^2 - \frac{1}{3}\right) +
  \mathcal{O}\left(\frac{k}{q}\right). 
\end{equation}
We also expand
\begin{equation}
  \label{eq:lowk_P_k1q}
  P(|\vk-\vq|) = P(q)\left[1 -\frac{k\nu}{q} \frac{\d\ln P(q)}{\d\ln q}
\right]
\end{equation}
and
\begin{equation}
  \label{eq:lowk_WR_k1q}
  W_R(|\vk-\vq|) = W_R(q)\left[1 -\frac{k\nu}{q} \frac{\d\ln W_R(q)}{\d\ln q}
\right].
\end{equation}
For Gaussian smoothing this becomes
\begin{equation}
  \label{eq:lowk_WRGauss_k1q}
  W_{R_G}(|\vk-\vq|) = W_{R_G}(q)\left[1 +\frac{k\nu}{q} (qR)^2
\right].
\end{equation}

Keeping only terms of order $(k/q)^0$ for the final results from now on we get
from
Eqs.~\eq{lowk_D_q_k1q}, \eq{lowk_F2_q_k1q}, \eq{lowk_P_k1q} and \eq{lowk_WR_k1q}
\begin{align}
  \label{eq:59}
 \lim_{k\rightarrow 0} I^R_{DF_2}(k) =0, \qquad D\in\big\{\mathsf{P}_0, -F_2^{1}\mathsf{P}_1,
  \mathsf{P}_2\big\}.
\end{align}
and
\begin{align}
  \label{eq:57}
  \lim_{k\rightarrow 0} I^R_{DE}(k)
= W_R(k)
\tau_R^4, \qquad D\in\big\{\mathsf{P}_0, -F_2^{1}\mathsf{P}_1,
  \mathsf{P}_2\big\},
\;\; E\in\big\{ \mathsf{P}_0, \mathsf{P}_2  \big\},
\end{align}
where
\begin{equation}
  \label{eq:tau_def}
  \tau_R^4\equiv \frac{1}{2\pi^2}\int\d q\, q^2 W_R^2(q)P_\mathrm{mm}^2(q).
\end{equation}

The limits of the bare integrals \eq{I_DE_R_bare} for $E=F_2$ follow
from Eqs.~\eq{lowk_F2_q_mk} and \eq{lowk_WR_k1q} by noting that only
$\mathsf{P}^2_0(\nu)$ and $\mathsf{P}_1^2(\nu)$ survive the $\nu$ integration over the product
of Eqs.~\eq{lowk_F2_q_mk} and \eq{lowk_WR_k1q} because of Legendre
polynomial orthogonality:
\begin{align}
  \label{eq:lowk_Ibare_D_F2_Rgauss}
 \lim_{k\rightarrow 0} I^{\mathrm{bare},R_G}_{DF_2}(k) =
\frac{1}{4}W_{R}(k)P_\mathrm{mm}(k)\left(
\frac{68}{21}\sigma_{R}^2 - \frac{2}{3}
\sigma_{R,W'_R}^2
\right), \qquad
 D\in\big\{\mathsf{P}_0, -F_2^{1}\mathsf{P}_1,
  \mathsf{P}_2\big\},
\end{align}
where
\begin{equation}
  \label{eq:sigma_R}
  \sigma_R^2 \equiv \frac{1}{2\pi^2}\int\d q\,q^2 W_R^2(q)P_\mathrm{mm}(q)
\end{equation}
and
\begin{equation}
  \label{eq:sigma_R_qR}
  \sigma_{R,W'_R}^2 \equiv -\frac{1}{2\pi^2}\int\d q\,q^2 \frac{\d\ln
    W_R(q)}{\d\ln q} W_R^2(q)P_\mathrm{mm}(q) .
\end{equation}
For Gaussian smoothing, $-\d\ln W_R(q)/\d\ln q=(qR)^2$, so no numerical
derivatives are required to evaluate \eqq{sigma_R_qR}.  To gain intuition,
\eqq{sigma_R_qR} can be integrated by parts to get
$\sigma^2_{R,W'_R}=\tfrac{1}{2}\sigma^2_{R,P'}$, where
\begin{equation}
  \label{eq:sigma2_R_Pprime}
\sigma^2_{R,P'} \equiv \frac{1}{2\pi^2}\int\d q\, q^2
W_R^2(q)P_\mathrm{mm}(q)\frac{\d\ln q^3P_\mathrm{mm}(q)}{\d\ln q}. 
\end{equation}
Then,  \eqq{lowk_Ibare_D_F2_Rgauss} becomes for $D\in\{\mathsf{P}_0, -F_2^{1}\mathsf{P}_1,
  \mathsf{P}_2\}$
\begin{eqnarray}
   \lim_{k\rightarrow 0} I^{\mathrm{bare},R_G}_{DF_2}(k) &=&
\frac{1}{4}W_{R}(k)P_\mathrm{mm}(k)\left(
\frac{68}{21}\sigma_R^2 - \frac{1}{3}\sigma^2_{R,P'}
\right)
\\
&=&
\frac{1}{4} W_{R}(k)P_\mathrm{mm}(k)\,
 \frac{1}{2\pi^2}\int\d q\, q^2 W_R^2(q)P_\mathrm{mm}(q)
\left(
\frac{68}{21} - \frac{1}{3}\frac{\d\ln q^3P_\mathrm{mm}(q)}{\d\ln
  q}\right).
\end{eqnarray}
The integrand contains the linear response of the power spectrum to a
long-wavelength overdensity
(e.g.~\cite{SherwinZaldarriaga1202,chiang1403,LiHu1401}). In contrast,
the large-scale limit of the quantity proposed in
\cite{chiang1403} is (by construction) proportional to the
response function on large scales rather than integrating over it.

For $E=\mathsf{P}_0$, only
the $\mathsf{P}_0(\nu)$ part of \eqq{lowk_WR_k1q} survives the $\nu$-integration so that
\begin{align}
  \label{eq:59b}
 \lim_{k\rightarrow 0} I^{\mathrm{bare},R}_{D\mathsf{P}_0}(k)
 =W_R(k)P_\mathrm{mm}(k)\sigma_R^2,
 \qquad D\in\big\{\mathsf{P}_0, -F_2^{1}\mathsf{P}_1,
  \mathsf{P}_2\big\}.
\end{align}
For $E=\mathsf{P}_2$, the integrand of \eqq{I_DE_R_bare} contains $\mathsf{P}_2(\nu)$
multiplied by $\mathsf{P}_0(\nu)$ and $\mathsf{P}_1(\nu)$ from \eqq{lowk_WR_k1q}, which
vanishes upon angular integration because of Legendre polynomial
orthogonality, i.e.
\begin{align}
  \label{eq:59c}
 \lim_{k\rightarrow 0} I^{\mathrm{bare},R}_{D\mathsf{P}_2}(k)
 =0,
 \qquad D\in\big\{\mathsf{P}_0, -F_2^{1}\mathsf{P}_1,
  \mathsf{P}_2\big\}.
\end{align}

Note that for our fiducial cosmology, at $z=0.55$, for
  Gaussian smoothing with $R=20h^{-1}\mathrm{Mpc}$ using the linear
  matter power spectrum in integrands, we get $\sigma_R^2=0.0215$,
  $\sigma_{R,P'}^2 = 0.0464$ and $\tau_R^4=191.9h^{-3}\mathrm{Mpc}^3$.
  Using emulator instead of linear matter power gives
  $\sigma_R^2=0.0213$, $\sigma_{R,P'}^2=0.0460$ and
  $\tau_R^4=189.2h^{-3}\mathrm{Mpc}^3$.

\section{Analytical covariances}
\label{se:TheoryCovariancesAppendix}
This appendix derives the covariance \eq{cov_cross_spectra_diagonal}
between  cross-spectra using leading-order perturbation
theory.  Since the cross-spectra are cubic in the density, their
covariance depends on the density $3$-point and $6$-point functions,
\begin{align}
\nonumber
  \mathrm{cov}(\hat P_{D[\delta^R_a],\delta^R_b}(k),
\hat P_{E[\delta^R_a],\delta^R_b}(k'))
= \,&
\frac{1}{(4\pi L^3)^2}\int\d\Omega_{\hat\vk}\int\d\Omega_{\hat\vk'}
\int\frac{\d^3 q}{(2\pi)^3}\int\frac{\d^3 q'}{(2\pi)^3}
D(\vq,\vk-\vq)E(\vq',\vk'-\vq')\\
\nonumber
&\quad \times \bigg[
\langle
\delta^R_a(\vq)\delta^R_a(\vk-\vq)\delta^R_b(-\vk)
\delta^R_a(\vq')\delta^R_a(\vk'-\vq')\delta^R_b(-\vk')
\rangle
\\
\label{eq:Pcross_cov_general1}
&\qquad\quad
 -\langle
\delta^R_a(\vq)\delta^R_a(\vk-\vq)\delta^R_b(-\vk)\rangle\langle
\delta^R_a(\vq')\delta^R_a(\vk'-\vq')\delta^R_b(-\vk')
\rangle\bigg],
\end{align}
where the kernels $D$ and $E$ are symmetric in their arguments and
$a,b\in\{\mathrm{h},\mathrm{m}\}$ denote halo or dark matter
densities.  The last term in the square brackets of
\eqq{Pcross_cov_general1} can be neglected at leading order. Then, the
fully disconnected part of the $6$-point function results
in\footnote{To arrive at \eqq{disc_6pt_contri} we used that any
  contraction of the $6$-point function that connects two of the three
  leftmost densities with themselves, or two of the three rightmost
  densities with themselves, vanishes, because
  $\delta_{\vk=\mathbf{0}}=0$ (i.e.~$P(0)=0$). We also used
  $E(\vk_1,\vk_2)=E(-\vk_1,-\vk_2)$, which holds for all
  kernels considered in this paper.}  
\begin{align}
\nonumber
\int\frac{\d^3q}{(2\pi)^3}\int\frac{\d^3q'}{(2\pi)^3}
D(\vq,\vk-\vq)&E(\vq',\vk'-\vq')
  \langle
\delta^R_a(\vq) \delta^R_b(\vk-\vq)\delta^R_b(-\vk) \delta^R_a(\vq')\delta^R_b(\vk'-\vq')\delta^R_b(-\vk')
\rangle\\
\nonumber
&=  2 (2\pi)^6\delta^2_D(\vk+\vk')P^R_{bb}(k)
I^{P^R_\mathit{aa}P^R_\mathit{aa}}_{DE}(k)
\\
\label{eq:disc_6pt_contri}
&\quad 
+ 4 (2\pi)^3\delta_D(\mathbf{0}) 
D(\vk',\vk-\vk')E(\vk,\vk'-\vk)
P^R_{ab}(k)P^R_{ab}(k')P^R_{aa}(|\vk-\vk'|),
\end{align}
where
\begin{equation}
  \label{eq:I_DE_PR_PR_cov}
  I_{DE}^{P_\mathit{aa}^RP_\mathit{aa}^R}(k)\equiv
\int\frac{\d^3 q}{(2\pi)^3}P^R_\mathit{aa}(q)P^R_\mathit{aa}(|\vk-\vq|)
D(\vq, \vk-\vq)E(\vq, \vk-\vq).
\end{equation}
All power spectra should be the full power spectra obtained
by contracting two  density fields, including shot
noise.    The covariance
 becomes 
\begin{align}
  \label{eq:cov_DE_final}
\nonumber
    \mathrm{cov}(\hat P_{D[\delta^R_a],\delta^R_b}(k),
\hat P_{E[\delta^R_a],\delta^R_b}(k'))
= \,&
\frac{2 L^6}{(4\pi L^3)^2
  }\delta^\mathrm{(K)}_{k,k'}P^R_{bb}(k)I^{P^R_\mathit{aa}P^R_\mathit{aa}}_{DE}(k)
\int\d\Omega_{\hat \vk}
\\
&\quad 
+ \frac{4L^3}{(4\pi L^3)^2}P^R_{ab}(k)P^R_{ab}(k')\int\d\Omega_{\hat\vk}\int\d\Omega_{\hat\vk'}
D(\vk',\vk-\vk')E(\vk,\vk'-\vk)
P^R_{aa}(|\vk-\vk'|),\quad
\end{align}
where we used  $(2\pi)^3\delta_D(\mathbf{0})=L^3$.
We neglect the $k\ne k'$ covariance in the main text because it is
hard to test its  validity with only ten realizations. This should
 be improved in future analyses.
On a discrete grid, the angular integral is replaced by the sum
over modes as in \eqq{discrete_sum_over_modes}.
 For $k=k'$ this gives 
\begin{equation}
  \label{eq:56}
  \frac{1}{(4\pi)^2}\int\d\Omega_{\hat\vk} \quad\rightarrow \quad
\frac{1}{N^2_\mathrm{modes}(k)}\sum_{\vk, [k-\Delta k/2\le |\vk|\le k+\Delta
k/2]} = \frac{1}{N_\mathrm{modes}(k)},
\end{equation}
which leads to \eqq{cov_cross_spectra_diagonal} in the main text.

\section{Reduced smoothing scale}
\begin{figure}[p]
\centerline{
\includegraphics[width=0.55\textwidth]{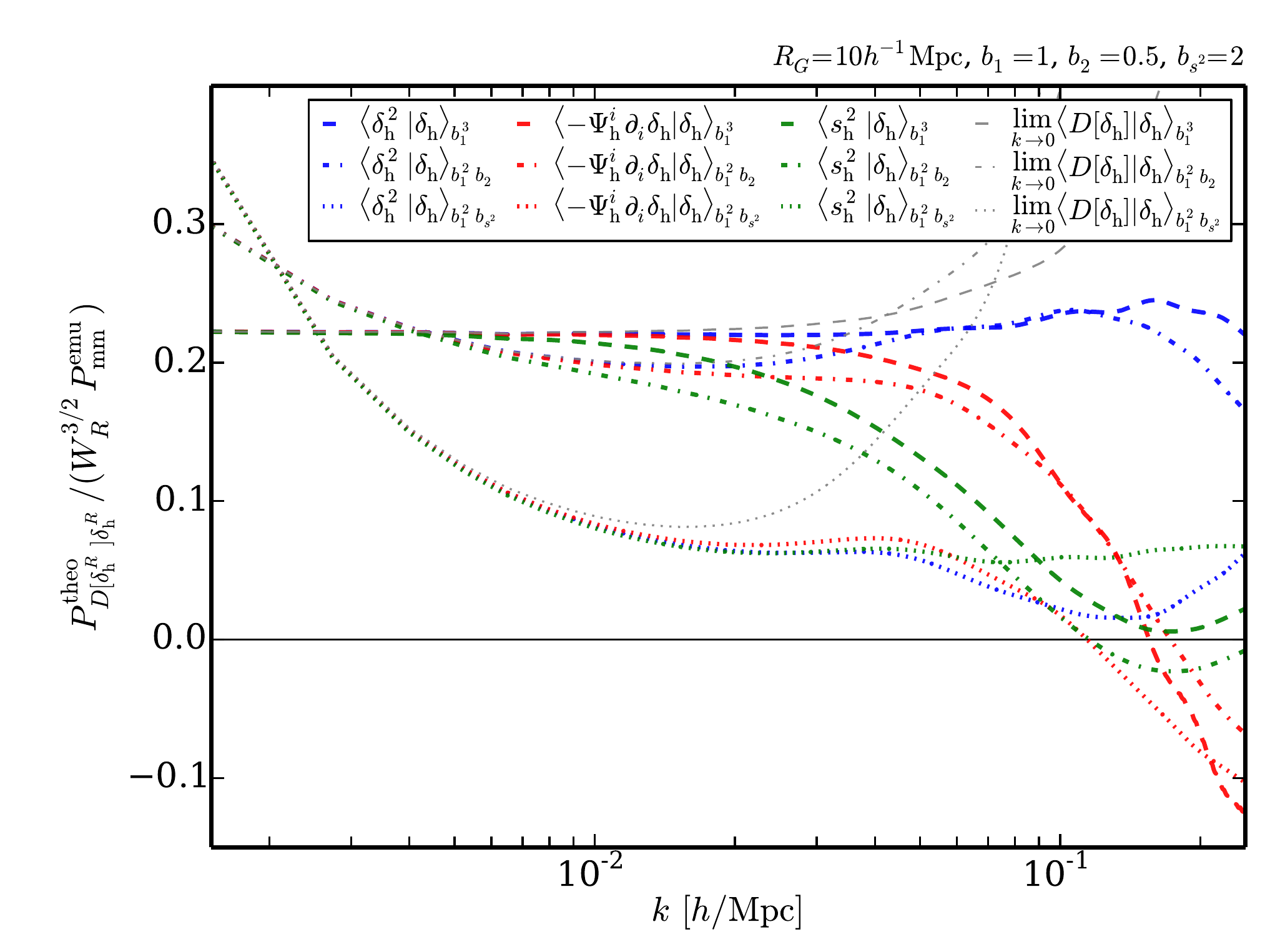}}
\caption{Same as Fig.~\ref{fig:hhhtheory_RGauss20} but for $R_G=10h^{-1}\mathrm{Mpc}$.}
\label{fig:hhhtheory_RGauss10}
\end{figure}

\begin{figure}[p]
\centerline{
\includegraphics[width=0.5\textwidth]{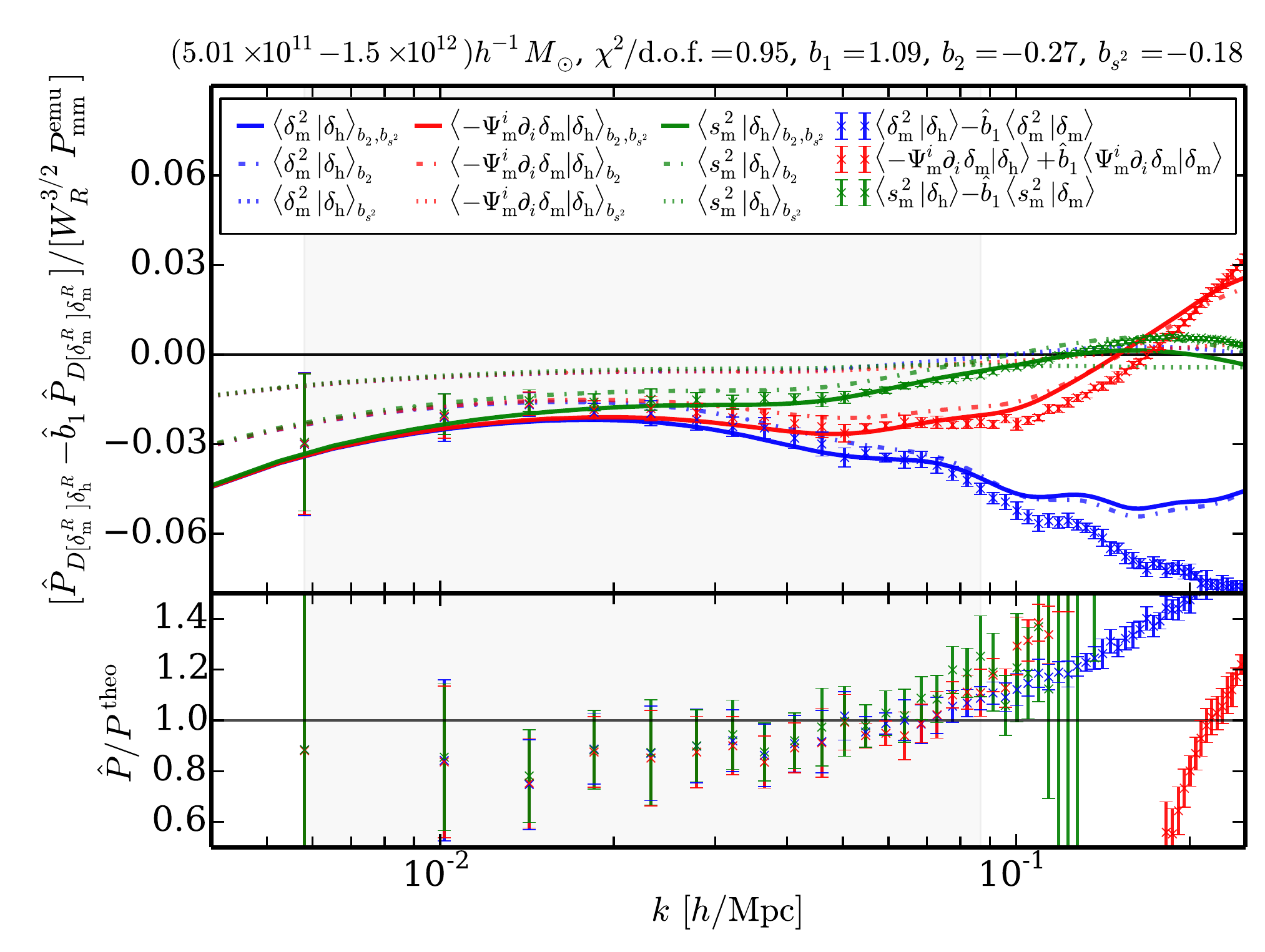}
\includegraphics[width=0.5\textwidth]{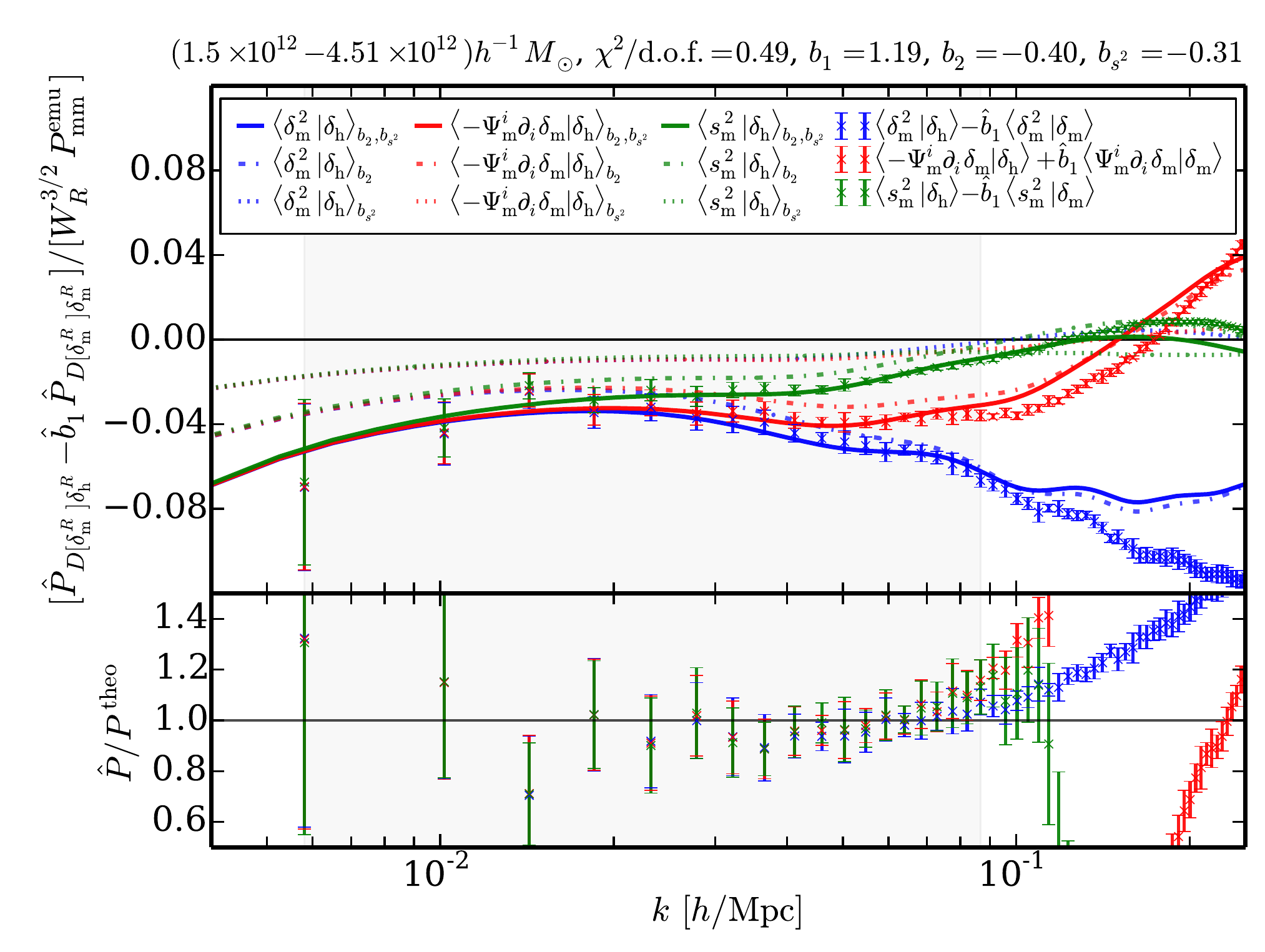}}
\centerline{
\includegraphics[width=0.5\textwidth]{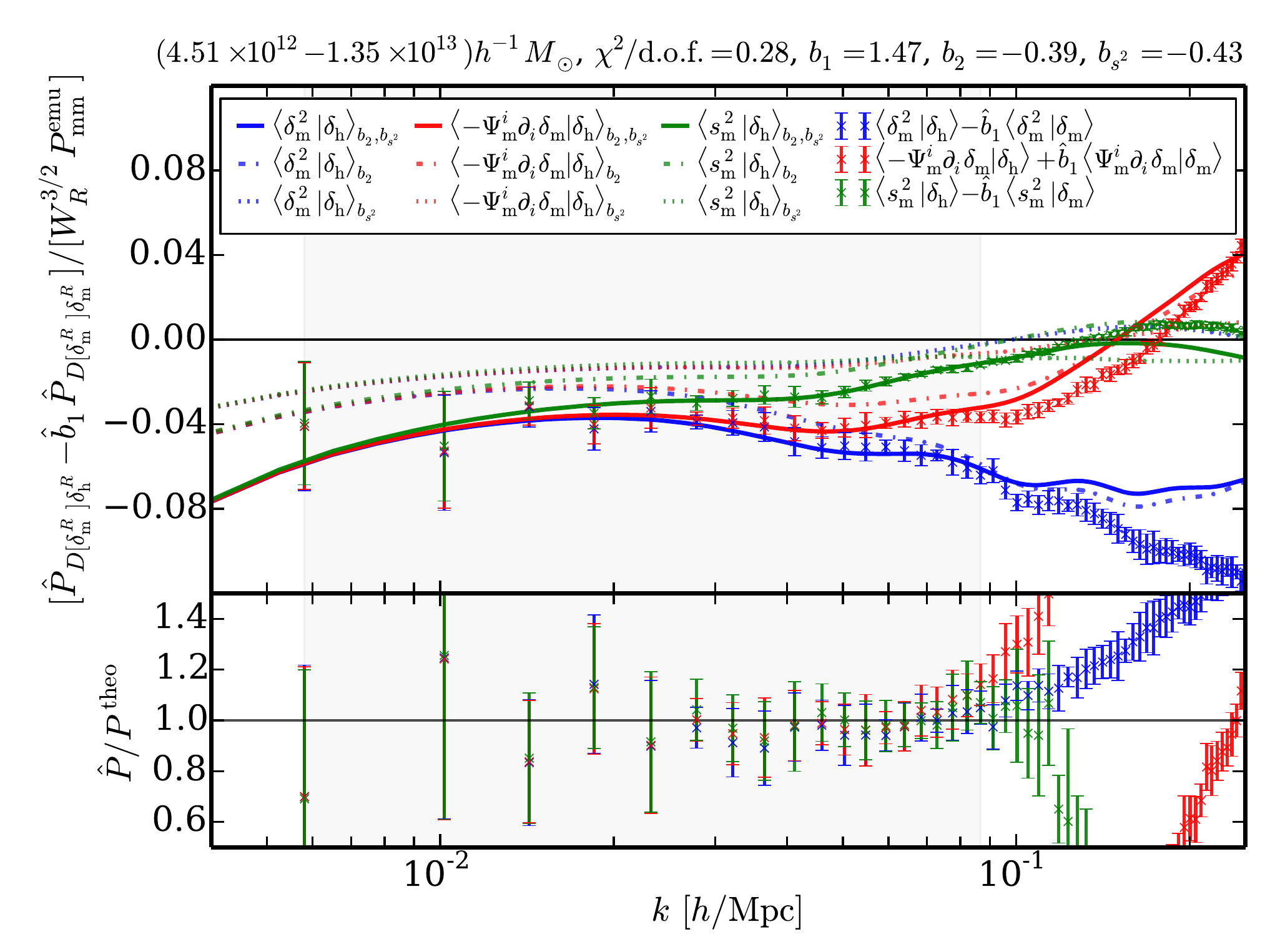}
\includegraphics[width=0.5\textwidth]{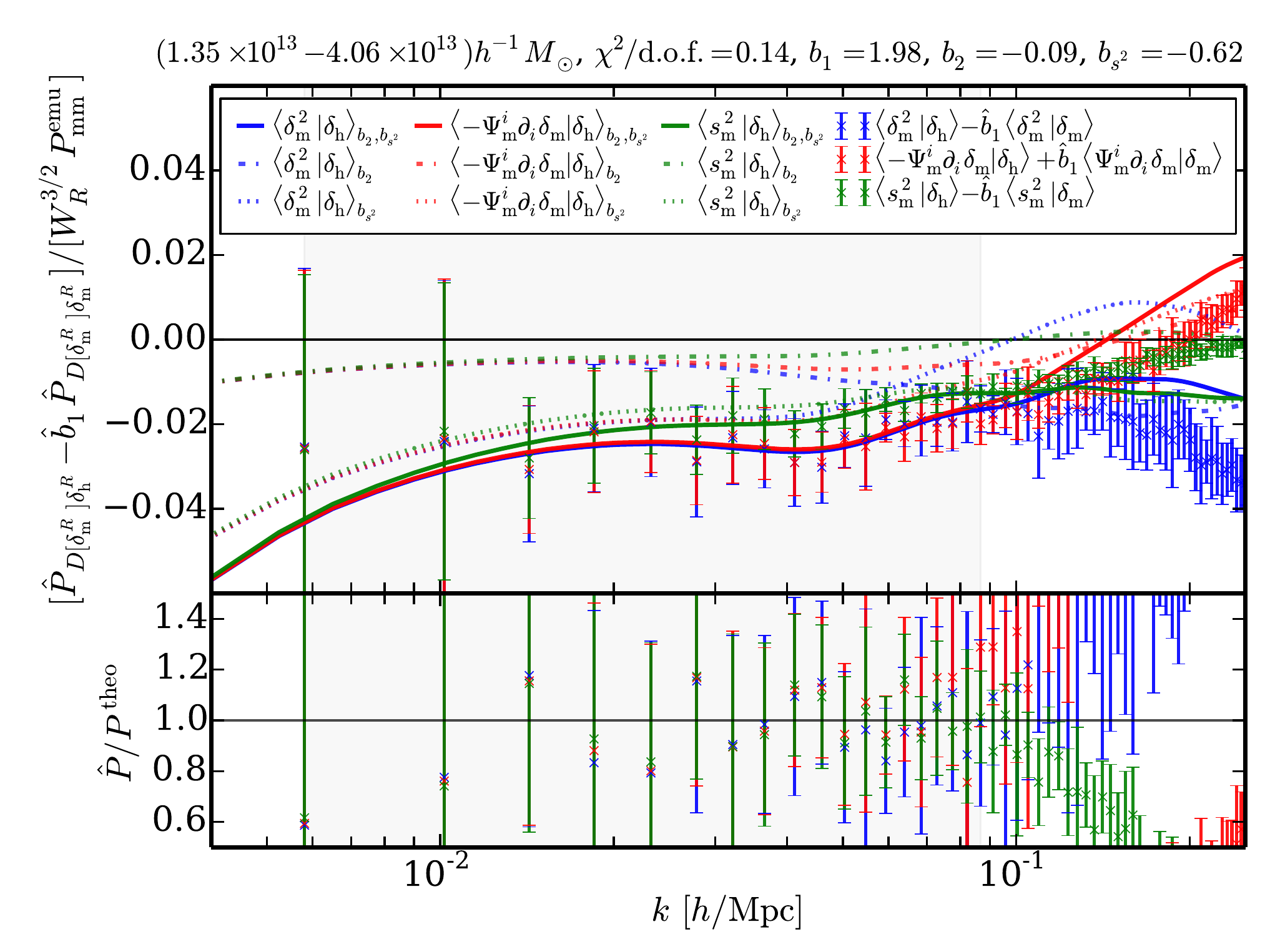}}
\caption{Same as Fig.~\ref{fig:mmh_model_vs_data_RGauss20} but for
  $R_G=10h^{-1}\mathrm{Mpc}$ smoothing.}
\label{fig:mmh_model_vs_data_RGauss10}
\end{figure}

Plots corresponding to Figures \ref{fig:hhhtheory_RGauss20} and
\ref{fig:mmh_model_vs_data_RGauss20} for less aggressive
$R_G=10h^{-1}\mathrm{Mpc}$ 
smoothing are shown in Figures \ref{fig:hhhtheory_RGauss10} and
\ref{fig:mmh_model_vs_data_RGauss10}.
They are discussed in the main body of the paper.

\clearpage
\bibliography{marcel_tobias_bias_estimators}

\begin{thebibliography}{59}%
\makeatletter
\providecommand \@ifxundefined [1]{%
 \@ifx{#1\undefined}
}%
\providecommand \@ifnum [1]{%
 \ifnum #1\expandafter \@firstoftwo
 \else \expandafter \@secondoftwo
 \fi
}%
\providecommand \@ifx [1]{%
 \ifx #1\expandafter \@firstoftwo
 \else \expandafter \@secondoftwo
 \fi
}%
\providecommand \natexlab [1]{#1}%
\providecommand \enquote  [1]{``#1''}%
\providecommand \bibnamefont  [1]{#1}%
\providecommand \bibfnamefont [1]{#1}%
\providecommand \citenamefont [1]{#1}%
\providecommand \href@noop [0]{\@secondoftwo}%
\providecommand \href [0]{\begingroup \@sanitize@url \@href}%
\providecommand \@href[1]{\@@startlink{#1}\@@href}%
\providecommand \@@href[1]{\endgroup#1\@@endlink}%
\providecommand \@sanitize@url [0]{\catcode `\\12\catcode `\$12\catcode
  `\&12\catcode `\#12\catcode `\^12\catcode `\_12\catcode `\%12\relax}%
\providecommand \@@startlink[1]{}%
\providecommand \@@endlink[0]{}%
\providecommand \url  [0]{\begingroup\@sanitize@url \@url }%
\providecommand \@url [1]{\endgroup\@href {#1}{\urlprefix }}%
\providecommand \urlprefix  [0]{URL }%
\providecommand \Eprint [0]{\href }%
\providecommand \doibase [0]{http://dx.doi.org/}%
\providecommand \selectlanguage [0]{\@gobble}%
\providecommand \bibinfo  [0]{\@secondoftwo}%
\providecommand \bibfield  [0]{\@secondoftwo}%
\providecommand \translation [1]{[#1]}%
\providecommand \BibitemOpen [0]{}%
\providecommand \bibitemStop [0]{}%
\providecommand \bibitemNoStop [0]{.\EOS\space}%
\providecommand \EOS [0]{\spacefactor3000\relax}%
\providecommand \BibitemShut  [1]{\csname bibitem#1\endcsname}%
\let\auto@bib@innerbib\@empty
\bibitem [{\citenamefont {{Anderson}}\ \emph {et~al.}(2014)\citenamefont
  {{Anderson}}, \citenamefont {{Aubourg}}, \citenamefont {{Bailey}},
  \citenamefont {{Beutler}}, \citenamefont {{Bhardwaj}}, \citenamefont
  {{Blanton}}, \citenamefont {{Bolton}}, \citenamefont {{Brinkmann}},
  \citenamefont {{Brownstein}}, \citenamefont {{Burden}}, \citenamefont
  {{Chuang}}, \citenamefont {{Cuesta}}, \citenamefont {{Dawson}}, \citenamefont
  {{Eisenstein}}, \citenamefont {{Escoffier}}, \citenamefont {{Gunn}},
  \citenamefont {{Guo}}, \citenamefont {{Ho}}, \citenamefont {{Honscheid}},
  \citenamefont {{Howlett}}, \citenamefont {{Kirkby}}, \citenamefont
  {{Lupton}}, \citenamefont {{Manera}}, \citenamefont {{Maraston}},
  \citenamefont {{McBride}}, \citenamefont {{Mena}}, \citenamefont
  {{Montesano}}, \citenamefont {{Nichol}}, \citenamefont {{Nuza}},
  \citenamefont {{Olmstead}}, \citenamefont {{Padmanabhan}}, \citenamefont
  {{Palanque-Delabrouille}}, \citenamefont {{Parejko}}, \citenamefont
  {{Percival}}, \citenamefont {{Petitjean}}, \citenamefont {{Prada}},
  \citenamefont {{Price-Whelan}}, \citenamefont {{Reid}}, \citenamefont
  {{Roe}}, \citenamefont {{Ross}}, \citenamefont {{Ross}}, \citenamefont
  {{Sabiu}}, \citenamefont {{Saito}}, \citenamefont {{Samushia}}, \citenamefont
  {{S{\'a}nchez}}, \citenamefont {{Schlegel}}, \citenamefont {{Schneider}},
  \citenamefont {{Scoccola}}, \citenamefont {{Seo}}, \citenamefont {{Skibba}},
  \citenamefont {{Strauss}}, \citenamefont {{Swanson}}, \citenamefont
  {{Thomas}}, \citenamefont {{Tinker}}, \citenamefont {{Tojeiro}},
  \citenamefont {{Maga{\~n}a}}, \citenamefont {{Verde}}, \citenamefont
  {{Wake}}, \citenamefont {{Weaver}}, \citenamefont {{Weinberg}}, \citenamefont
  {{White}}, \citenamefont {{Xu}}, \citenamefont {{Y{\`e}che}}, \citenamefont
  {{Zehavi}},\ and\ \citenamefont {{Zhao}}}]{Anderson1312}%
  \BibitemOpen
  \bibfield  {author} {\bibinfo {author} {\bibfnamefont {L.}~\bibnamefont
  {{Anderson}}}, \bibinfo {author} {\bibfnamefont {{\'E}.}~\bibnamefont
  {{Aubourg}}}, \bibinfo {author} {\bibfnamefont {S.}~\bibnamefont {{Bailey}}},
  \bibinfo {author} {\bibfnamefont {F.}~\bibnamefont {{Beutler}}}, \bibinfo
  {author} {\bibfnamefont {V.}~\bibnamefont {{Bhardwaj}}}, \bibinfo {author}
  {\bibfnamefont {M.}~\bibnamefont {{Blanton}}}, \bibinfo {author}
  {\bibfnamefont {A.~S.}\ \bibnamefont {{Bolton}}}, \bibinfo {author}
  {\bibfnamefont {J.}~\bibnamefont {{Brinkmann}}}, \bibinfo {author}
  {\bibfnamefont {J.~R.}\ \bibnamefont {{Brownstein}}}, \bibinfo {author}
  {\bibfnamefont {A.}~\bibnamefont {{Burden}}}, \bibinfo {author}
  {\bibfnamefont {C.-H.}\ \bibnamefont {{Chuang}}}, \bibinfo {author}
  {\bibfnamefont {A.~J.}\ \bibnamefont {{Cuesta}}}, \bibinfo {author}
  {\bibfnamefont {K.~S.}\ \bibnamefont {{Dawson}}}, \bibinfo {author}
  {\bibfnamefont {D.~J.}\ \bibnamefont {{Eisenstein}}}, \bibinfo {author}
  {\bibfnamefont {S.}~\bibnamefont {{Escoffier}}}, \bibinfo {author}
  {\bibfnamefont {J.~E.}\ \bibnamefont {{Gunn}}}, \bibinfo {author}
  {\bibfnamefont {H.}~\bibnamefont {{Guo}}}, \bibinfo {author} {\bibfnamefont
  {S.}~\bibnamefont {{Ho}}}, \bibinfo {author} {\bibfnamefont {K.}~\bibnamefont
  {{Honscheid}}}, \bibinfo {author} {\bibfnamefont {C.}~\bibnamefont
  {{Howlett}}}, \bibinfo {author} {\bibfnamefont {D.}~\bibnamefont {{Kirkby}}},
  \bibinfo {author} {\bibfnamefont {R.~H.}\ \bibnamefont {{Lupton}}}, \bibinfo
  {author} {\bibfnamefont {M.}~\bibnamefont {{Manera}}}, \bibinfo {author}
  {\bibfnamefont {C.}~\bibnamefont {{Maraston}}}, \bibinfo {author}
  {\bibfnamefont {C.~K.}\ \bibnamefont {{McBride}}}, \bibinfo {author}
  {\bibfnamefont {O.}~\bibnamefont {{Mena}}}, \bibinfo {author} {\bibfnamefont
  {F.}~\bibnamefont {{Montesano}}}, \bibinfo {author} {\bibfnamefont {R.~C.}\
  \bibnamefont {{Nichol}}}, \bibinfo {author} {\bibfnamefont {S.~E.}\
  \bibnamefont {{Nuza}}}, \bibinfo {author} {\bibfnamefont {M.~D.}\
  \bibnamefont {{Olmstead}}}, \bibinfo {author} {\bibfnamefont
  {N.}~\bibnamefont {{Padmanabhan}}}, \bibinfo {author} {\bibfnamefont
  {N.}~\bibnamefont {{Palanque-Delabrouille}}}, \bibinfo {author}
  {\bibfnamefont {J.}~\bibnamefont {{Parejko}}}, \bibinfo {author}
  {\bibfnamefont {W.~J.}\ \bibnamefont {{Percival}}}, \bibinfo {author}
  {\bibfnamefont {P.}~\bibnamefont {{Petitjean}}}, \bibinfo {author}
  {\bibfnamefont {F.}~\bibnamefont {{Prada}}}, \bibinfo {author} {\bibfnamefont
  {A.~M.}\ \bibnamefont {{Price-Whelan}}}, \bibinfo {author} {\bibfnamefont
  {B.}~\bibnamefont {{Reid}}}, \bibinfo {author} {\bibfnamefont {N.~A.}\
  \bibnamefont {{Roe}}}, \bibinfo {author} {\bibfnamefont {A.~J.}\ \bibnamefont
  {{Ross}}}, \bibinfo {author} {\bibfnamefont {N.~P.}\ \bibnamefont {{Ross}}},
  \bibinfo {author} {\bibfnamefont {C.~G.}\ \bibnamefont {{Sabiu}}}, \bibinfo
  {author} {\bibfnamefont {S.}~\bibnamefont {{Saito}}}, \bibinfo {author}
  {\bibfnamefont {L.}~\bibnamefont {{Samushia}}}, \bibinfo {author}
  {\bibfnamefont {A.~G.}\ \bibnamefont {{S{\'a}nchez}}}, \bibinfo {author}
  {\bibfnamefont {D.~J.}\ \bibnamefont {{Schlegel}}}, \bibinfo {author}
  {\bibfnamefont {D.~P.}\ \bibnamefont {{Schneider}}}, \bibinfo {author}
  {\bibfnamefont {C.~G.}\ \bibnamefont {{Scoccola}}}, \bibinfo {author}
  {\bibfnamefont {H.-J.}\ \bibnamefont {{Seo}}}, \bibinfo {author}
  {\bibfnamefont {R.~A.}\ \bibnamefont {{Skibba}}}, \bibinfo {author}
  {\bibfnamefont {M.~A.}\ \bibnamefont {{Strauss}}}, \bibinfo {author}
  {\bibfnamefont {M.~E.~C.}\ \bibnamefont {{Swanson}}}, \bibinfo {author}
  {\bibfnamefont {D.}~\bibnamefont {{Thomas}}}, \bibinfo {author}
  {\bibfnamefont {J.~L.}\ \bibnamefont {{Tinker}}}, \bibinfo {author}
  {\bibfnamefont {R.}~\bibnamefont {{Tojeiro}}}, \bibinfo {author}
  {\bibfnamefont {M.~V.}\ \bibnamefont {{Maga{\~n}a}}}, \bibinfo {author}
  {\bibfnamefont {L.}~\bibnamefont {{Verde}}}, \bibinfo {author} {\bibfnamefont
  {D.~A.}\ \bibnamefont {{Wake}}}, \bibinfo {author} {\bibfnamefont {B.~A.}\
  \bibnamefont {{Weaver}}}, \bibinfo {author} {\bibfnamefont {D.~H.}\
  \bibnamefont {{Weinberg}}}, \bibinfo {author} {\bibfnamefont
  {M.}~\bibnamefont {{White}}}, \bibinfo {author} {\bibfnamefont
  {X.}~\bibnamefont {{Xu}}}, \bibinfo {author} {\bibfnamefont {C.}~\bibnamefont
  {{Y{\`e}che}}}, \bibinfo {author} {\bibfnamefont {I.}~\bibnamefont
  {{Zehavi}}}, \ and\ \bibinfo {author} {\bibfnamefont {G.-B.}\ \bibnamefont
  {{Zhao}}},\ }\href {\doibase 10.1093/mnras/stu523} {\bibfield  {journal}
  {\bibinfo  {journal} {\mnras}\ }\textbf {\bibinfo {volume} {441}},\ \bibinfo
  {pages} {24} (\bibinfo {year} {2014})},\ \Eprint
  {http://arxiv.org/abs/1312.4877} {arXiv:1312.4877} \BibitemShut {NoStop}%
\bibitem [{\citenamefont {{Scoccimarro}}\ \emph {et~al.}(2001)\citenamefont
  {{Scoccimarro}}, \citenamefont {{Feldman}}, \citenamefont {{Fry}},\ and\
  \citenamefont {{Frieman}}}]{2001ApJ...546..652S}%
  \BibitemOpen
  \bibfield  {author} {\bibinfo {author} {\bibfnamefont {R.}~\bibnamefont
  {{Scoccimarro}}}, \bibinfo {author} {\bibfnamefont {H.~A.}\ \bibnamefont
  {{Feldman}}}, \bibinfo {author} {\bibfnamefont {J.~N.}\ \bibnamefont
  {{Fry}}}, \ and\ \bibinfo {author} {\bibfnamefont {J.~A.}\ \bibnamefont
  {{Frieman}}},\ }\href {\doibase 10.1086/318284} {\bibfield  {journal}
  {\bibinfo  {journal} {\apj}\ }\textbf {\bibinfo {volume} {546}},\ \bibinfo
  {pages} {652} (\bibinfo {year} {2001})},\ \Eprint
  {http://arxiv.org/abs/astro-ph/0004087} {astro-ph/0004087} \BibitemShut
  {NoStop}%
\bibitem [{\citenamefont {{Feldman}}\ \emph {et~al.}(2001)\citenamefont
  {{Feldman}}, \citenamefont {{Frieman}}, \citenamefont {{Fry}},\ and\
  \citenamefont {{Scoccimarro}}}]{2001PhRvL..86.1434F}%
  \BibitemOpen
  \bibfield  {author} {\bibinfo {author} {\bibfnamefont {H.~A.}\ \bibnamefont
  {{Feldman}}}, \bibinfo {author} {\bibfnamefont {J.~A.}\ \bibnamefont
  {{Frieman}}}, \bibinfo {author} {\bibfnamefont {J.~N.}\ \bibnamefont
  {{Fry}}}, \ and\ \bibinfo {author} {\bibfnamefont {R.}~\bibnamefont
  {{Scoccimarro}}},\ }\href {\doibase 10.1103/PhysRevLett.86.1434} {\bibfield
  {journal} {\bibinfo  {journal} {Physical Review Letters}\ }\textbf {\bibinfo
  {volume} {86}},\ \bibinfo {pages} {1434} (\bibinfo {year} {2001})},\ \Eprint
  {http://arxiv.org/abs/astro-ph/0010205} {astro-ph/0010205} \BibitemShut
  {NoStop}%
\bibitem [{\citenamefont {{Verde}}\ \emph {et~al.}(2002)\citenamefont
  {{Verde}}, \citenamefont {{Heavens}}, \citenamefont {{Percival}},
  \citenamefont {{Matarrese}}, \citenamefont {{Baugh}}, \citenamefont
  {{Bland-Hawthorn}}, \citenamefont {{Bridges}}, \citenamefont {{Cannon}},
  \citenamefont {{Cole}}, \citenamefont {{Colless}}, \citenamefont {{Collins}},
  \citenamefont {{Couch}}, \citenamefont {{Dalton}}, \citenamefont {{De
  Propris}}, \citenamefont {{Driver}}, \citenamefont {{Efstathiou}},
  \citenamefont {{Ellis}}, \citenamefont {{Frenk}}, \citenamefont
  {{Glazebrook}}, \citenamefont {{Jackson}}, \citenamefont {{Lahav}},
  \citenamefont {{Lewis}}, \citenamefont {{Lumsden}}, \citenamefont {{Maddox}},
  \citenamefont {{Madgwick}}, \citenamefont {{Norberg}}, \citenamefont
  {{Peacock}}, \citenamefont {{Peterson}}, \citenamefont {{Sutherland}},\ and\
  \citenamefont {{Taylor}}}]{2002MNRAS.335..432V}%
  \BibitemOpen
  \bibfield  {author} {\bibinfo {author} {\bibfnamefont {L.}~\bibnamefont
  {{Verde}}}, \bibinfo {author} {\bibfnamefont {A.~F.}\ \bibnamefont
  {{Heavens}}}, \bibinfo {author} {\bibfnamefont {W.~J.}\ \bibnamefont
  {{Percival}}}, \bibinfo {author} {\bibfnamefont {S.}~\bibnamefont
  {{Matarrese}}}, \bibinfo {author} {\bibfnamefont {C.~M.}\ \bibnamefont
  {{Baugh}}}, \bibinfo {author} {\bibfnamefont {J.}~\bibnamefont
  {{Bland-Hawthorn}}}, \bibinfo {author} {\bibfnamefont {T.}~\bibnamefont
  {{Bridges}}}, \bibinfo {author} {\bibfnamefont {R.}~\bibnamefont {{Cannon}}},
  \bibinfo {author} {\bibfnamefont {S.}~\bibnamefont {{Cole}}}, \bibinfo
  {author} {\bibfnamefont {M.}~\bibnamefont {{Colless}}}, \bibinfo {author}
  {\bibfnamefont {C.}~\bibnamefont {{Collins}}}, \bibinfo {author}
  {\bibfnamefont {W.}~\bibnamefont {{Couch}}}, \bibinfo {author} {\bibfnamefont
  {G.}~\bibnamefont {{Dalton}}}, \bibinfo {author} {\bibfnamefont
  {R.}~\bibnamefont {{De Propris}}}, \bibinfo {author} {\bibfnamefont {S.~P.}\
  \bibnamefont {{Driver}}}, \bibinfo {author} {\bibfnamefont {G.}~\bibnamefont
  {{Efstathiou}}}, \bibinfo {author} {\bibfnamefont {R.~S.}\ \bibnamefont
  {{Ellis}}}, \bibinfo {author} {\bibfnamefont {C.~S.}\ \bibnamefont
  {{Frenk}}}, \bibinfo {author} {\bibfnamefont {K.}~\bibnamefont
  {{Glazebrook}}}, \bibinfo {author} {\bibfnamefont {C.}~\bibnamefont
  {{Jackson}}}, \bibinfo {author} {\bibfnamefont {O.}~\bibnamefont {{Lahav}}},
  \bibinfo {author} {\bibfnamefont {I.}~\bibnamefont {{Lewis}}}, \bibinfo
  {author} {\bibfnamefont {S.}~\bibnamefont {{Lumsden}}}, \bibinfo {author}
  {\bibfnamefont {S.}~\bibnamefont {{Maddox}}}, \bibinfo {author}
  {\bibfnamefont {D.}~\bibnamefont {{Madgwick}}}, \bibinfo {author}
  {\bibfnamefont {P.}~\bibnamefont {{Norberg}}}, \bibinfo {author}
  {\bibfnamefont {J.~A.}\ \bibnamefont {{Peacock}}}, \bibinfo {author}
  {\bibfnamefont {B.~A.}\ \bibnamefont {{Peterson}}}, \bibinfo {author}
  {\bibfnamefont {W.}~\bibnamefont {{Sutherland}}}, \ and\ \bibinfo {author}
  {\bibfnamefont {K.}~\bibnamefont {{Taylor}}},\ }\href {\doibase
  10.1046/j.1365-8711.2002.05620.x} {\bibfield  {journal} {\bibinfo  {journal}
  {\mnras}\ }\textbf {\bibinfo {volume} {335}},\ \bibinfo {pages} {432}
  (\bibinfo {year} {2002})},\ \Eprint {http://arxiv.org/abs/astro-ph/0112161}
  {astro-ph/0112161} \BibitemShut {NoStop}%
\bibitem [{\citenamefont {{Jing}}\ and\ \citenamefont
  {{B{\"o}rner}}(2004)}]{2004ApJ...607..140J}%
  \BibitemOpen
  \bibfield  {author} {\bibinfo {author} {\bibfnamefont {Y.~P.}\ \bibnamefont
  {{Jing}}}\ and\ \bibinfo {author} {\bibfnamefont {G.}~\bibnamefont
  {{B{\"o}rner}}},\ }\href {\doibase 10.1086/383343} {\bibfield  {journal}
  {\bibinfo  {journal} {\apj}\ }\textbf {\bibinfo {volume} {607}},\ \bibinfo
  {pages} {140} (\bibinfo {year} {2004})},\ \Eprint
  {http://arxiv.org/abs/astro-ph/0311585} {astro-ph/0311585} \BibitemShut
  {NoStop}%
\bibitem [{\citenamefont {{Wang}}\ \emph {et~al.}(2004)\citenamefont {{Wang}},
  \citenamefont {{Yang}}, \citenamefont {{Mo}}, \citenamefont {{van den
  Bosch}},\ and\ \citenamefont {{Chu}}}]{2004MNRAS.353..287W}%
  \BibitemOpen
  \bibfield  {author} {\bibinfo {author} {\bibfnamefont {Y.}~\bibnamefont
  {{Wang}}}, \bibinfo {author} {\bibfnamefont {X.}~\bibnamefont {{Yang}}},
  \bibinfo {author} {\bibfnamefont {H.~J.}\ \bibnamefont {{Mo}}}, \bibinfo
  {author} {\bibfnamefont {F.~C.}\ \bibnamefont {{van den Bosch}}}, \ and\
  \bibinfo {author} {\bibfnamefont {Y.}~\bibnamefont {{Chu}}},\ }\href
  {\doibase 10.1111/j.1365-2966.2004.08141.x} {\bibfield  {journal} {\bibinfo
  {journal} {\mnras}\ }\textbf {\bibinfo {volume} {353}},\ \bibinfo {pages}
  {287} (\bibinfo {year} {2004})},\ \Eprint
  {http://arxiv.org/abs/astro-ph/0404143} {astro-ph/0404143} \BibitemShut
  {NoStop}%
\bibitem [{\citenamefont {{Mar{\'{\i}}n}}\ \emph {et~al.}(2013)\citenamefont
  {{Mar{\'{\i}}n}}, \citenamefont {{Blake}}, \citenamefont {{Poole}},
  \citenamefont {{McBride}}, \citenamefont {{Brough}}, \citenamefont
  {{Colless}}, \citenamefont {{Contreras}}, \citenamefont {{Couch}},
  \citenamefont {{Croton}}, \citenamefont {{Croom}}, \citenamefont {{Davis}},
  \citenamefont {{Drinkwater}}, \citenamefont {{Forster}}, \citenamefont
  {{Gilbank}}, \citenamefont {{Gladders}}, \citenamefont {{Glazebrook}},
  \citenamefont {{Jelliffe}}, \citenamefont {{Jurek}}, \citenamefont {{Li}},
  \citenamefont {{Madore}}, \citenamefont {{Martin}}, \citenamefont
  {{Pimbblet}}, \citenamefont {{Pracy}}, \citenamefont {{Sharp}}, \citenamefont
  {{Wisnioski}}, \citenamefont {{Woods}}, \citenamefont {{Wyder}},\ and\
  \citenamefont {{Yee}}}]{2013MNRAS.432.2654M}%
  \BibitemOpen
  \bibfield  {author} {\bibinfo {author} {\bibfnamefont {F.~A.}\ \bibnamefont
  {{Mar{\'{\i}}n}}}, \bibinfo {author} {\bibfnamefont {C.}~\bibnamefont
  {{Blake}}}, \bibinfo {author} {\bibfnamefont {G.~B.}\ \bibnamefont
  {{Poole}}}, \bibinfo {author} {\bibfnamefont {C.~K.}\ \bibnamefont
  {{McBride}}}, \bibinfo {author} {\bibfnamefont {S.}~\bibnamefont {{Brough}}},
  \bibinfo {author} {\bibfnamefont {M.}~\bibnamefont {{Colless}}}, \bibinfo
  {author} {\bibfnamefont {C.}~\bibnamefont {{Contreras}}}, \bibinfo {author}
  {\bibfnamefont {W.}~\bibnamefont {{Couch}}}, \bibinfo {author} {\bibfnamefont
  {D.~J.}\ \bibnamefont {{Croton}}}, \bibinfo {author} {\bibfnamefont
  {S.}~\bibnamefont {{Croom}}}, \bibinfo {author} {\bibfnamefont
  {T.}~\bibnamefont {{Davis}}}, \bibinfo {author} {\bibfnamefont {M.~J.}\
  \bibnamefont {{Drinkwater}}}, \bibinfo {author} {\bibfnamefont
  {K.}~\bibnamefont {{Forster}}}, \bibinfo {author} {\bibfnamefont
  {D.}~\bibnamefont {{Gilbank}}}, \bibinfo {author} {\bibfnamefont
  {M.}~\bibnamefont {{Gladders}}}, \bibinfo {author} {\bibfnamefont
  {K.}~\bibnamefont {{Glazebrook}}}, \bibinfo {author} {\bibfnamefont
  {B.}~\bibnamefont {{Jelliffe}}}, \bibinfo {author} {\bibfnamefont {R.~J.}\
  \bibnamefont {{Jurek}}}, \bibinfo {author} {\bibfnamefont {I.-h.}\
  \bibnamefont {{Li}}}, \bibinfo {author} {\bibfnamefont {B.}~\bibnamefont
  {{Madore}}}, \bibinfo {author} {\bibfnamefont {D.~C.}\ \bibnamefont
  {{Martin}}}, \bibinfo {author} {\bibfnamefont {K.}~\bibnamefont
  {{Pimbblet}}}, \bibinfo {author} {\bibfnamefont {M.}~\bibnamefont {{Pracy}}},
  \bibinfo {author} {\bibfnamefont {R.}~\bibnamefont {{Sharp}}}, \bibinfo
  {author} {\bibfnamefont {E.}~\bibnamefont {{Wisnioski}}}, \bibinfo {author}
  {\bibfnamefont {D.}~\bibnamefont {{Woods}}}, \bibinfo {author} {\bibfnamefont
  {T.~K.}\ \bibnamefont {{Wyder}}}, \ and\ \bibinfo {author} {\bibfnamefont
  {H.~K.~C.}\ \bibnamefont {{Yee}}},\ }\href {\doibase 10.1093/mnras/stt520}
  {\bibfield  {journal} {\bibinfo  {journal} {\mnras}\ }\textbf {\bibinfo
  {volume} {432}},\ \bibinfo {pages} {2654} (\bibinfo {year} {2013})},\ \Eprint
  {http://arxiv.org/abs/1303.6644} {arXiv:1303.6644 [astro-ph.CO]} \BibitemShut
  {NoStop}%
\bibitem [{\citenamefont {{McBride}}\ \emph
  {et~al.}(2011{\natexlab{a}})\citenamefont {{McBride}}, \citenamefont
  {{Connolly}}, \citenamefont {{Gardner}}, \citenamefont {{Scranton}},
  \citenamefont {{Scoccimarro}}, \citenamefont {{Berlind}}, \citenamefont
  {{Mar{\'{\i}}n}},\ and\ \citenamefont {{Schneider}}}]{2011ApJ...739...85M}%
  \BibitemOpen
  \bibfield  {author} {\bibinfo {author} {\bibfnamefont {C.~K.}\ \bibnamefont
  {{McBride}}}, \bibinfo {author} {\bibfnamefont {A.~J.}\ \bibnamefont
  {{Connolly}}}, \bibinfo {author} {\bibfnamefont {J.~P.}\ \bibnamefont
  {{Gardner}}}, \bibinfo {author} {\bibfnamefont {R.}~\bibnamefont
  {{Scranton}}}, \bibinfo {author} {\bibfnamefont {R.}~\bibnamefont
  {{Scoccimarro}}}, \bibinfo {author} {\bibfnamefont {A.~A.}\ \bibnamefont
  {{Berlind}}}, \bibinfo {author} {\bibfnamefont {F.}~\bibnamefont
  {{Mar{\'{\i}}n}}}, \ and\ \bibinfo {author} {\bibfnamefont {D.~P.}\
  \bibnamefont {{Schneider}}},\ }\href {\doibase 10.1088/0004-637X/739/2/85}
  {\bibfield  {journal} {\bibinfo  {journal} {\apj}\ }\textbf {\bibinfo
  {volume} {739}},\ \bibinfo {eid} {85} (\bibinfo {year}
  {2011}{\natexlab{a}})},\ \Eprint {http://arxiv.org/abs/1012.3462}
  {arXiv:1012.3462 [astro-ph.CO]} \BibitemShut {NoStop}%
\bibitem [{\citenamefont {{McBride}}\ \emph
  {et~al.}(2011{\natexlab{b}})\citenamefont {{McBride}}, \citenamefont
  {{Connolly}}, \citenamefont {{Gardner}}, \citenamefont {{Scranton}},
  \citenamefont {{Newman}}, \citenamefont {{Scoccimarro}}, \citenamefont
  {{Zehavi}},\ and\ \citenamefont {{Schneider}}}]{2011ApJ...726...13M}%
  \BibitemOpen
  \bibfield  {author} {\bibinfo {author} {\bibfnamefont {C.~K.}\ \bibnamefont
  {{McBride}}}, \bibinfo {author} {\bibfnamefont {A.~J.}\ \bibnamefont
  {{Connolly}}}, \bibinfo {author} {\bibfnamefont {J.~P.}\ \bibnamefont
  {{Gardner}}}, \bibinfo {author} {\bibfnamefont {R.}~\bibnamefont
  {{Scranton}}}, \bibinfo {author} {\bibfnamefont {J.~A.}\ \bibnamefont
  {{Newman}}}, \bibinfo {author} {\bibfnamefont {R.}~\bibnamefont
  {{Scoccimarro}}}, \bibinfo {author} {\bibfnamefont {I.}~\bibnamefont
  {{Zehavi}}}, \ and\ \bibinfo {author} {\bibfnamefont {D.~P.}\ \bibnamefont
  {{Schneider}}},\ }\href {\doibase 10.1088/0004-637X/726/1/13} {\bibfield
  {journal} {\bibinfo  {journal} {\apj}\ }\textbf {\bibinfo {volume} {726}},\
  \bibinfo {eid} {13} (\bibinfo {year} {2011}{\natexlab{b}})},\ \Eprint
  {http://arxiv.org/abs/1007.2414} {arXiv:1007.2414 [astro-ph.CO]} \BibitemShut
  {NoStop}%
\bibitem [{\citenamefont {{Gil-Mar{\'{\i}}n}}\ \emph
  {et~al.}(2014{\natexlab{a}})\citenamefont {{Gil-Mar{\'{\i}}n}}, \citenamefont
  {{Nore{\~n}a}}, \citenamefont {{Verde}}, \citenamefont {{Percival}},
  \citenamefont {{Wagner}}, \citenamefont {{Manera}},\ and\ \citenamefont
  {{Schneider}}}]{Hector1407Data1}%
  \BibitemOpen
  \bibfield  {author} {\bibinfo {author} {\bibfnamefont {H.}~\bibnamefont
  {{Gil-Mar{\'{\i}}n}}}, \bibinfo {author} {\bibfnamefont {J.}~\bibnamefont
  {{Nore{\~n}a}}}, \bibinfo {author} {\bibfnamefont {L.}~\bibnamefont
  {{Verde}}}, \bibinfo {author} {\bibfnamefont {W.~J.}\ \bibnamefont
  {{Percival}}}, \bibinfo {author} {\bibfnamefont {C.}~\bibnamefont
  {{Wagner}}}, \bibinfo {author} {\bibfnamefont {M.}~\bibnamefont {{Manera}}},
  \ and\ \bibinfo {author} {\bibfnamefont {D.~P.}\ \bibnamefont
  {{Schneider}}},\ }\href@noop {} {\bibfield  {journal} {\bibinfo  {journal}
  {ArXiv e-prints}\ } (\bibinfo {year} {2014}{\natexlab{a}})},\ \Eprint
  {http://arxiv.org/abs/1407.5668} {arXiv:1407.5668} \BibitemShut {NoStop}%
\bibitem [{\citenamefont {{Gil-Mar{\'{\i}}n}}\ \emph
  {et~al.}(2014{\natexlab{b}})\citenamefont {{Gil-Mar{\'{\i}}n}}, \citenamefont
  {{Verde}}, \citenamefont {{Nore{\~n}a}}, \citenamefont {{Cuesta}},
  \citenamefont {{Samushia}}, \citenamefont {{Percival}}, \citenamefont
  {{Wagner}}, \citenamefont {{Manera}},\ and\ \citenamefont
  {{Schneider}}}]{Hector1408Data2}%
  \BibitemOpen
  \bibfield  {author} {\bibinfo {author} {\bibfnamefont {H.}~\bibnamefont
  {{Gil-Mar{\'{\i}}n}}}, \bibinfo {author} {\bibfnamefont {L.}~\bibnamefont
  {{Verde}}}, \bibinfo {author} {\bibfnamefont {J.}~\bibnamefont
  {{Nore{\~n}a}}}, \bibinfo {author} {\bibfnamefont {A.~J.}\ \bibnamefont
  {{Cuesta}}}, \bibinfo {author} {\bibfnamefont {L.}~\bibnamefont
  {{Samushia}}}, \bibinfo {author} {\bibfnamefont {W.~J.}\ \bibnamefont
  {{Percival}}}, \bibinfo {author} {\bibfnamefont {C.}~\bibnamefont
  {{Wagner}}}, \bibinfo {author} {\bibfnamefont {M.}~\bibnamefont {{Manera}}},
  \ and\ \bibinfo {author} {\bibfnamefont {D.~P.}\ \bibnamefont
  {{Schneider}}},\ }\href@noop {} {\bibfield  {journal} {\bibinfo  {journal}
  {ArXiv e-prints}\ } (\bibinfo {year} {2014}{\natexlab{b}})},\ \Eprint
  {http://arxiv.org/abs/1408.0027} {arXiv:1408.0027} \BibitemShut {NoStop}%
\bibitem [{\citenamefont {{Gil-Mar{\'{\i}}n}}\ \emph
  {et~al.}(2014{\natexlab{c}})\citenamefont {{Gil-Mar{\'{\i}}n}}, \citenamefont
  {{Wagner}}, \citenamefont {{Nore{\~n}a}}, \citenamefont {{Verde}},\ and\
  \citenamefont {{Percival}}}]{hector1407theory}%
  \BibitemOpen
  \bibfield  {author} {\bibinfo {author} {\bibfnamefont {H.}~\bibnamefont
  {{Gil-Mar{\'{\i}}n}}}, \bibinfo {author} {\bibfnamefont {C.}~\bibnamefont
  {{Wagner}}}, \bibinfo {author} {\bibfnamefont {J.}~\bibnamefont
  {{Nore{\~n}a}}}, \bibinfo {author} {\bibfnamefont {L.}~\bibnamefont
  {{Verde}}}, \ and\ \bibinfo {author} {\bibfnamefont {W.}~\bibnamefont
  {{Percival}}},\ }\href@noop {} {\bibfield  {journal} {\bibinfo  {journal}
  {ArXiv e-prints}\ } (\bibinfo {year} {2014}{\natexlab{c}})},\ \Eprint
  {http://arxiv.org/abs/1407.1836} {arXiv:1407.1836} \BibitemShut {NoStop}%
\bibitem [{\citenamefont {{Bernardeau}}(1996)}]{Bernardeau:1996}%
  \BibitemOpen
  \bibfield  {author} {\bibinfo {author} {\bibfnamefont {F.}~\bibnamefont
  {{Bernardeau}}},\ }\href@noop {} {\bibfield  {journal} {\bibinfo  {journal}
  {Astronomy and Astrophysics}\ }\textbf {\bibinfo {volume} {312}},\ \bibinfo
  {pages} {11} (\bibinfo {year} {1996})},\ \Eprint
  {http://arxiv.org/abs/astro-ph/9602072} {astro-ph/9602072} \BibitemShut
  {NoStop}%
\bibitem [{\citenamefont {{Bel}}\ and\ \citenamefont
  {{Marinoni}}(2012)}]{Bel:2012}%
  \BibitemOpen
  \bibfield  {author} {\bibinfo {author} {\bibfnamefont {J.}~\bibnamefont
  {{Bel}}}\ and\ \bibinfo {author} {\bibfnamefont {C.}~\bibnamefont
  {{Marinoni}}},\ }\href {\doibase 10.1111/j.1365-2966.2012.21257.x} {\bibfield
   {journal} {\bibinfo  {journal} {\mnras}\ }\textbf {\bibinfo {volume}
  {424}},\ \bibinfo {pages} {971} (\bibinfo {year} {2012})},\ \Eprint
  {http://arxiv.org/abs/1205.3200} {arXiv:1205.3200 [astro-ph.CO]} \BibitemShut
  {NoStop}%
\bibitem [{\citenamefont {{Hoffmann}}\ \emph {et~al.}(2014)\citenamefont
  {{Hoffmann}}, \citenamefont {{Bel}}, \citenamefont {{Gaztanaga}},
  \citenamefont {{Crocce}}, \citenamefont {{Fosalba}},\ and\ \citenamefont
  {{Castander}}}]{Hoffmann1403}%
  \BibitemOpen
  \bibfield  {author} {\bibinfo {author} {\bibfnamefont {K.}~\bibnamefont
  {{Hoffmann}}}, \bibinfo {author} {\bibfnamefont {J.}~\bibnamefont {{Bel}}},
  \bibinfo {author} {\bibfnamefont {E.}~\bibnamefont {{Gaztanaga}}}, \bibinfo
  {author} {\bibfnamefont {M.}~\bibnamefont {{Crocce}}}, \bibinfo {author}
  {\bibfnamefont {P.}~\bibnamefont {{Fosalba}}}, \ and\ \bibinfo {author}
  {\bibfnamefont {F.~J.}\ \bibnamefont {{Castander}}},\ }\href@noop {}
  {\bibfield  {journal} {\bibinfo  {journal} {ArXiv e-prints}\ } (\bibinfo
  {year} {2014})},\ \Eprint {http://arxiv.org/abs/1403.1259} {arXiv:1403.1259
  [astro-ph.CO]} \BibitemShut {NoStop}%
\bibitem [{\citenamefont {{Pollack}}\ \emph {et~al.}(2014)\citenamefont
  {{Pollack}}, \citenamefont {{Smith}},\ and\ \citenamefont
  {{Porciani}}}]{Pollack1309}%
  \BibitemOpen
  \bibfield  {author} {\bibinfo {author} {\bibfnamefont {J.~E.}\ \bibnamefont
  {{Pollack}}}, \bibinfo {author} {\bibfnamefont {R.~E.}\ \bibnamefont
  {{Smith}}}, \ and\ \bibinfo {author} {\bibfnamefont {C.}~\bibnamefont
  {{Porciani}}},\ }\href {\doibase 10.1093/mnras/stu322} {\bibfield  {journal}
  {\bibinfo  {journal} {\mnras}\ }\textbf {\bibinfo {volume} {440}},\ \bibinfo
  {pages} {555} (\bibinfo {year} {2014})},\ \Eprint
  {http://arxiv.org/abs/1309.0504} {arXiv:1309.0504 [astro-ph.CO]} \BibitemShut
  {NoStop}%
\bibitem [{\citenamefont {{Chiang}}\ \emph {et~al.}(2014)\citenamefont
  {{Chiang}}, \citenamefont {{Wagner}}, \citenamefont {{Schmidt}},\ and\
  \citenamefont {{Komatsu}}}]{chiang1403}%
  \BibitemOpen
  \bibfield  {author} {\bibinfo {author} {\bibfnamefont {C.-T.}\ \bibnamefont
  {{Chiang}}}, \bibinfo {author} {\bibfnamefont {C.}~\bibnamefont {{Wagner}}},
  \bibinfo {author} {\bibfnamefont {F.}~\bibnamefont {{Schmidt}}}, \ and\
  \bibinfo {author} {\bibfnamefont {E.}~\bibnamefont {{Komatsu}}},\ }\href
  {\doibase 10.1088/1475-7516/2014/05/048} {\bibfield  {journal} {\bibinfo
  {journal} {\jcap}\ }\textbf {\bibinfo {volume} {5}},\ \bibinfo {eid} {048}
  (\bibinfo {year} {2014})},\ \Eprint {http://arxiv.org/abs/1403.3411}
  {arXiv:1403.3411} \BibitemShut {NoStop}%
\bibitem [{\citenamefont {{Fergusson}}\ \emph
  {et~al.}(2012{\natexlab{a}})\citenamefont {{Fergusson}}, \citenamefont
  {{Regan}},\ and\ \citenamefont {{Shellard}}}]{shellard1008}%
  \BibitemOpen
  \bibfield  {author} {\bibinfo {author} {\bibfnamefont {J.~R.}\ \bibnamefont
  {{Fergusson}}}, \bibinfo {author} {\bibfnamefont {D.~M.}\ \bibnamefont
  {{Regan}}}, \ and\ \bibinfo {author} {\bibfnamefont {E.~P.~S.}\ \bibnamefont
  {{Shellard}}},\ }\href {\doibase 10.1103/PhysRevD.86.063511} {\bibfield
  {journal} {\bibinfo  {journal} {\prd}\ }\textbf {\bibinfo {volume} {86}},\
  \bibinfo {eid} {063511} (\bibinfo {year} {2012}{\natexlab{a}})},\ \Eprint
  {http://arxiv.org/abs/1008.1730} {arXiv:1008.1730 [astro-ph.CO]} \BibitemShut
  {NoStop}%
\bibitem [{\citenamefont {{Regan}}\ \emph {et~al.}(2012)\citenamefont
  {{Regan}}, \citenamefont {{Schmittfull}}, \citenamefont {{Shellard}},\ and\
  \citenamefont {{Fergusson}}}]{shellard1108}%
  \BibitemOpen
  \bibfield  {author} {\bibinfo {author} {\bibfnamefont {D.~M.}\ \bibnamefont
  {{Regan}}}, \bibinfo {author} {\bibfnamefont {M.~M.}\ \bibnamefont
  {{Schmittfull}}}, \bibinfo {author} {\bibfnamefont {E.~P.~S.}\ \bibnamefont
  {{Shellard}}}, \ and\ \bibinfo {author} {\bibfnamefont {J.~R.}\ \bibnamefont
  {{Fergusson}}},\ }\href {\doibase 10.1103/PhysRevD.86.123524} {\bibfield
  {journal} {\bibinfo  {journal} {\prd}\ }\textbf {\bibinfo {volume} {86}},\
  \bibinfo {eid} {123524} (\bibinfo {year} {2012})},\ \Eprint
  {http://arxiv.org/abs/1108.3813} {arXiv:1108.3813 [astro-ph.CO]} \BibitemShut
  {NoStop}%
\bibitem [{\citenamefont {{Schmittfull}}\ \emph {et~al.}(2013)\citenamefont
  {{Schmittfull}}, \citenamefont {{Regan}},\ and\ \citenamefont
  {{Shellard}}}]{marcel1207}%
  \BibitemOpen
  \bibfield  {author} {\bibinfo {author} {\bibfnamefont {M.~M.}\ \bibnamefont
  {{Schmittfull}}}, \bibinfo {author} {\bibfnamefont {D.~M.}\ \bibnamefont
  {{Regan}}}, \ and\ \bibinfo {author} {\bibfnamefont {E.~P.~S.}\ \bibnamefont
  {{Shellard}}},\ }\href {\doibase 10.1103/PhysRevD.88.063512} {\bibfield
  {journal} {\bibinfo  {journal} {\prd}\ }\textbf {\bibinfo {volume} {88}},\
  \bibinfo {eid} {063512} (\bibinfo {year} {2013})},\ \Eprint
  {http://arxiv.org/abs/1207.5678} {arXiv:1207.5678 [astro-ph.CO]} \BibitemShut
  {NoStop}%
\bibitem [{\citenamefont {{Komatsu}}\ \emph {et~al.}(2005)\citenamefont
  {{Komatsu}}, \citenamefont {{Spergel}},\ and\ \citenamefont
  {{Wandelt}}}]{ksw}%
  \BibitemOpen
  \bibfield  {author} {\bibinfo {author} {\bibfnamefont {E.}~\bibnamefont
  {{Komatsu}}}, \bibinfo {author} {\bibfnamefont {D.~N.}\ \bibnamefont
  {{Spergel}}}, \ and\ \bibinfo {author} {\bibfnamefont {B.~D.}\ \bibnamefont
  {{Wandelt}}},\ }\href {\doibase 10.1086/491724} {\bibfield  {journal}
  {\bibinfo  {journal} {\apj}\ }\textbf {\bibinfo {volume} {634}},\ \bibinfo
  {pages} {14} (\bibinfo {year} {2005})},\ \Eprint
  {http://arxiv.org/abs/astro-ph/0305189} {astro-ph/0305189} \BibitemShut
  {NoStop}%
\bibitem [{\citenamefont {{Lewis}}\ \emph {et~al.}(2011)\citenamefont
  {{Lewis}}, \citenamefont {{Challinor}},\ and\ \citenamefont
  {{Hanson}}}]{lewis1101}%
  \BibitemOpen
  \bibfield  {author} {\bibinfo {author} {\bibfnamefont {A.}~\bibnamefont
  {{Lewis}}}, \bibinfo {author} {\bibfnamefont {A.}~\bibnamefont
  {{Challinor}}}, \ and\ \bibinfo {author} {\bibfnamefont {D.}~\bibnamefont
  {{Hanson}}},\ }\href {\doibase 10.1088/1475-7516/2011/03/018} {\bibfield
  {journal} {\bibinfo  {journal} {\jcap}\ }\textbf {\bibinfo {volume} {3}},\
  \bibinfo {eid} {018} (\bibinfo {year} {2011})},\ \Eprint
  {http://arxiv.org/abs/1101.2234} {arXiv:1101.2234 [astro-ph.CO]} \BibitemShut
  {NoStop}%
\bibitem [{\citenamefont {{Planck Collaboration}}\ \emph
  {et~al.}(2013{\natexlab{a}})\citenamefont {{Planck Collaboration}},
  \citenamefont {{Ade}}, \citenamefont {{Aghanim}}, \citenamefont
  {{Armitage-Caplan}}, \citenamefont {{Arnaud}}, \citenamefont {{Ashdown}},
  \citenamefont {{Atrio-Barandela}}, \citenamefont {{Aumont}}, \citenamefont
  {{Baccigalupi}}, \citenamefont {{Banday}},\ and\ \citenamefont
  {et~al.}}]{Planck1303ISW}%
  \BibitemOpen
  \bibfield  {author} {\bibinfo {author} {\bibnamefont {{Planck
  Collaboration}}}, \bibinfo {author} {\bibfnamefont {P.~A.~R.}\ \bibnamefont
  {{Ade}}}, \bibinfo {author} {\bibfnamefont {N.}~\bibnamefont {{Aghanim}}},
  \bibinfo {author} {\bibfnamefont {C.}~\bibnamefont {{Armitage-Caplan}}},
  \bibinfo {author} {\bibfnamefont {M.}~\bibnamefont {{Arnaud}}}, \bibinfo
  {author} {\bibfnamefont {M.}~\bibnamefont {{Ashdown}}}, \bibinfo {author}
  {\bibfnamefont {F.}~\bibnamefont {{Atrio-Barandela}}}, \bibinfo {author}
  {\bibfnamefont {J.}~\bibnamefont {{Aumont}}}, \bibinfo {author}
  {\bibfnamefont {C.}~\bibnamefont {{Baccigalupi}}}, \bibinfo {author}
  {\bibfnamefont {A.~J.}\ \bibnamefont {{Banday}}}, \ and\ \bibinfo {author}
  {\bibnamefont {et~al.}},\ }\href@noop {} {\bibfield  {journal} {\bibinfo
  {journal} {ArXiv e-prints}\ } (\bibinfo {year} {2013}{\natexlab{a}})},\
  \Eprint {http://arxiv.org/abs/1303.5079} {arXiv:1303.5079 [astro-ph.CO]}
  \BibitemShut {NoStop}%
\bibitem [{\citenamefont {{Fergusson}}\ \emph {et~al.}(2010)\citenamefont
  {{Fergusson}}, \citenamefont {{Liguori}},\ and\ \citenamefont
  {{Shellard}}}]{shellard0912}%
  \BibitemOpen
  \bibfield  {author} {\bibinfo {author} {\bibfnamefont {J.~R.}\ \bibnamefont
  {{Fergusson}}}, \bibinfo {author} {\bibfnamefont {M.}~\bibnamefont
  {{Liguori}}}, \ and\ \bibinfo {author} {\bibfnamefont {E.~P.~S.}\
  \bibnamefont {{Shellard}}},\ }\href {\doibase 10.1103/PhysRevD.82.023502}
  {\bibfield  {journal} {\bibinfo  {journal} {\prd}\ }\textbf {\bibinfo
  {volume} {82}},\ \bibinfo {eid} {023502} (\bibinfo {year} {2010})},\ \Eprint
  {http://arxiv.org/abs/0912.5516} {arXiv:0912.5516 [astro-ph.CO]} \BibitemShut
  {NoStop}%
\bibitem [{\citenamefont {{Fergusson}}\ \emph
  {et~al.}(2012{\natexlab{b}})\citenamefont {{Fergusson}}, \citenamefont
  {{Liguori}},\ and\ \citenamefont {{Shellard}}}]{shellard1006}%
  \BibitemOpen
  \bibfield  {author} {\bibinfo {author} {\bibfnamefont {J.~R.}\ \bibnamefont
  {{Fergusson}}}, \bibinfo {author} {\bibfnamefont {M.}~\bibnamefont
  {{Liguori}}}, \ and\ \bibinfo {author} {\bibfnamefont {E.~P.~S.}\
  \bibnamefont {{Shellard}}},\ }\href {\doibase 10.1088/1475-7516/2012/12/032}
  {\bibfield  {journal} {\bibinfo  {journal} {\jcap}\ }\textbf {\bibinfo
  {volume} {12}},\ \bibinfo {eid} {032} (\bibinfo {year}
  {2012}{\natexlab{b}})},\ \Eprint {http://arxiv.org/abs/1006.1642}
  {arXiv:1006.1642 [astro-ph.CO]} \BibitemShut {NoStop}%
\bibitem [{\citenamefont {{Planck Collaboration}}\ \emph
  {et~al.}(2013{\natexlab{b}})\citenamefont {{Planck Collaboration}},
  \citenamefont {{Ade}}, \citenamefont {{Aghanim}}, \citenamefont
  {{Armitage-Caplan}}, \citenamefont {{Arnaud}}, \citenamefont {{Ashdown}},
  \citenamefont {{Atrio-Barandela}}, \citenamefont {{Aumont}}, \citenamefont
  {{Baccigalupi}}, \citenamefont {{Banday}},\ and\ \citenamefont
  {et~al.}}]{Planck1303NG}%
  \BibitemOpen
  \bibfield  {author} {\bibinfo {author} {\bibnamefont {{Planck
  Collaboration}}}, \bibinfo {author} {\bibfnamefont {P.~A.~R.}\ \bibnamefont
  {{Ade}}}, \bibinfo {author} {\bibfnamefont {N.}~\bibnamefont {{Aghanim}}},
  \bibinfo {author} {\bibfnamefont {C.}~\bibnamefont {{Armitage-Caplan}}},
  \bibinfo {author} {\bibfnamefont {M.}~\bibnamefont {{Arnaud}}}, \bibinfo
  {author} {\bibfnamefont {M.}~\bibnamefont {{Ashdown}}}, \bibinfo {author}
  {\bibfnamefont {F.}~\bibnamefont {{Atrio-Barandela}}}, \bibinfo {author}
  {\bibfnamefont {J.}~\bibnamefont {{Aumont}}}, \bibinfo {author}
  {\bibfnamefont {C.}~\bibnamefont {{Baccigalupi}}}, \bibinfo {author}
  {\bibfnamefont {A.~J.}\ \bibnamefont {{Banday}}}, \ and\ \bibinfo {author}
  {\bibnamefont {et~al.}},\ }\href@noop {} {\bibfield  {journal} {\bibinfo
  {journal} {ArXiv e-prints}\ } (\bibinfo {year} {2013}{\natexlab{b}})},\
  \Eprint {http://arxiv.org/abs/1303.5084} {arXiv:1303.5084 [astro-ph.CO]}
  \BibitemShut {NoStop}%
\bibitem [{\citenamefont {{Zaldarriaga}}\ and\ \citenamefont
  {{Seljak}}(1999)}]{ZaldarriagaSeljak98LensingRec}%
  \BibitemOpen
  \bibfield  {author} {\bibinfo {author} {\bibfnamefont {M.}~\bibnamefont
  {{Zaldarriaga}}}\ and\ \bibinfo {author} {\bibfnamefont {U.}~\bibnamefont
  {{Seljak}}},\ }\href {\doibase 10.1103/PhysRevD.59.123507} {\bibfield
  {journal} {\bibinfo  {journal} {\prd}\ }\textbf {\bibinfo {volume} {59}},\
  \bibinfo {eid} {123507} (\bibinfo {year} {1999})},\ \Eprint
  {http://arxiv.org/abs/astro-ph/9810257} {astro-ph/9810257} \BibitemShut
  {NoStop}%
\bibitem [{\citenamefont {{Okamoto}}\ and\ \citenamefont
  {{Hu}}(2003)}]{okamotoHu0301}%
  \BibitemOpen
  \bibfield  {author} {\bibinfo {author} {\bibfnamefont {T.}~\bibnamefont
  {{Okamoto}}}\ and\ \bibinfo {author} {\bibfnamefont {W.}~\bibnamefont
  {{Hu}}},\ }\href {\doibase 10.1103/PhysRevD.67.083002} {\bibfield  {journal}
  {\bibinfo  {journal} {\prd}\ }\textbf {\bibinfo {volume} {67}},\ \bibinfo
  {eid} {083002} (\bibinfo {year} {2003})},\ \Eprint
  {http://arxiv.org/abs/astro-ph/0301031} {astro-ph/0301031} \BibitemShut
  {NoStop}%
\bibitem [{\citenamefont {{Hanson}}\ \emph {et~al.}(2011)\citenamefont
  {{Hanson}}, \citenamefont {{Challinor}}, \citenamefont {{Efstathiou}},\ and\
  \citenamefont {{Bielewicz}}}]{duncan1008}%
  \BibitemOpen
  \bibfield  {author} {\bibinfo {author} {\bibfnamefont {D.}~\bibnamefont
  {{Hanson}}}, \bibinfo {author} {\bibfnamefont {A.}~\bibnamefont
  {{Challinor}}}, \bibinfo {author} {\bibfnamefont {G.}~\bibnamefont
  {{Efstathiou}}}, \ and\ \bibinfo {author} {\bibfnamefont {P.}~\bibnamefont
  {{Bielewicz}}},\ }\href {\doibase 10.1103/PhysRevD.83.043005} {\bibfield
  {journal} {\bibinfo  {journal} {\prd}\ }\textbf {\bibinfo {volume} {83}},\
  \bibinfo {eid} {043005} (\bibinfo {year} {2011})},\ \Eprint
  {http://arxiv.org/abs/1008.4403} {arXiv:1008.4403 [astro-ph.CO]} \BibitemShut
  {NoStop}%
\bibitem [{\citenamefont {{Baldauf}}\ \emph {et~al.}(2012)\citenamefont
  {{Baldauf}}, \citenamefont {{Seljak}}, \citenamefont {{Desjacques}},\ and\
  \citenamefont {{McDonald}}}]{tobias1201}%
  \BibitemOpen
  \bibfield  {author} {\bibinfo {author} {\bibfnamefont {T.}~\bibnamefont
  {{Baldauf}}}, \bibinfo {author} {\bibfnamefont {U.}~\bibnamefont {{Seljak}}},
  \bibinfo {author} {\bibfnamefont {V.}~\bibnamefont {{Desjacques}}}, \ and\
  \bibinfo {author} {\bibfnamefont {P.}~\bibnamefont {{McDonald}}},\ }\href
  {\doibase 10.1103/PhysRevD.86.083540} {\bibfield  {journal} {\bibinfo
  {journal} {\prd}\ }\textbf {\bibinfo {volume} {86}},\ \bibinfo {eid} {083540}
  (\bibinfo {year} {2012})},\ \Eprint {http://arxiv.org/abs/1201.4827}
  {arXiv:1201.4827 [astro-ph.CO]} \BibitemShut {NoStop}%
\bibitem [{\citenamefont {{Bernardeau}}\ \emph {et~al.}(2002)\citenamefont
  {{Bernardeau}}, \citenamefont {{Colombi}}, \citenamefont {{Gazta{\~n}aga}},\
  and\ \citenamefont {{Scoccimarro}}}]{BernardeauReview}%
  \BibitemOpen
  \bibfield  {author} {\bibinfo {author} {\bibfnamefont {F.}~\bibnamefont
  {{Bernardeau}}}, \bibinfo {author} {\bibfnamefont {S.}~\bibnamefont
  {{Colombi}}}, \bibinfo {author} {\bibfnamefont {E.}~\bibnamefont
  {{Gazta{\~n}aga}}}, \ and\ \bibinfo {author} {\bibfnamefont {R.}~\bibnamefont
  {{Scoccimarro}}},\ }\href {\doibase 10.1016/S0370-1573(02)00135-7} {\bibfield
   {journal} {\bibinfo  {journal} {Physics Reports}\ }\textbf {\bibinfo
  {volume} {367}},\ \bibinfo {pages} {1} (\bibinfo {year} {2002})},\ \Eprint
  {http://arxiv.org/abs/astro-ph/0112551} {astro-ph/0112551} \BibitemShut
  {NoStop}%
\bibitem [{\citenamefont {{Bouchet}}\ \emph {et~al.}(1992)\citenamefont
  {{Bouchet}}, \citenamefont {{Juszkiewicz}}, \citenamefont {{Colombi}},\ and\
  \citenamefont {{Pellat}}}]{Bouchet1992ApJ...394L...5B}%
  \BibitemOpen
  \bibfield  {author} {\bibinfo {author} {\bibfnamefont {F.~R.}\ \bibnamefont
  {{Bouchet}}}, \bibinfo {author} {\bibfnamefont {R.}~\bibnamefont
  {{Juszkiewicz}}}, \bibinfo {author} {\bibfnamefont {S.}~\bibnamefont
  {{Colombi}}}, \ and\ \bibinfo {author} {\bibfnamefont {R.}~\bibnamefont
  {{Pellat}}},\ }\href {\doibase 10.1086/186459} {\bibfield  {journal}
  {\bibinfo  {journal} {\apjl}\ }\textbf {\bibinfo {volume} {394}},\ \bibinfo
  {pages} {L5} (\bibinfo {year} {1992})}\BibitemShut {NoStop}%
\bibitem [{\citenamefont {{Sherwin}}\ and\ \citenamefont
  {{Zaldarriaga}}(2012)}]{SherwinZaldarriaga1202}%
  \BibitemOpen
  \bibfield  {author} {\bibinfo {author} {\bibfnamefont {B.~D.}\ \bibnamefont
  {{Sherwin}}}\ and\ \bibinfo {author} {\bibfnamefont {M.}~\bibnamefont
  {{Zaldarriaga}}},\ }\href {\doibase 10.1103/PhysRevD.85.103523} {\bibfield
  {journal} {\bibinfo  {journal} {\prd}\ }\textbf {\bibinfo {volume} {85}},\
  \bibinfo {eid} {103523} (\bibinfo {year} {2012})},\ \Eprint
  {http://arxiv.org/abs/1202.3998} {arXiv:1202.3998 [astro-ph.CO]} \BibitemShut
  {NoStop}%
\bibitem [{\citenamefont {{McDonald}}\ and\ \citenamefont
  {{Roy}}(2009)}]{McDonaldRoy09}%
  \BibitemOpen
  \bibfield  {author} {\bibinfo {author} {\bibfnamefont {P.}~\bibnamefont
  {{McDonald}}}\ and\ \bibinfo {author} {\bibfnamefont {A.}~\bibnamefont
  {{Roy}}},\ }\href {\doibase 10.1088/1475-7516/2009/08/020} {\bibfield
  {journal} {\bibinfo  {journal} {\jcap}\ }\textbf {\bibinfo {volume} {8}},\
  \bibinfo {eid} {020} (\bibinfo {year} {2009})},\ \Eprint
  {http://arxiv.org/abs/0902.0991} {arXiv:0902.0991 [astro-ph.CO]} \BibitemShut
  {NoStop}%
\bibitem [{\citenamefont {{Chan}}\ \emph {et~al.}(2012)\citenamefont {{Chan}},
  \citenamefont {{Scoccimarro}},\ and\ \citenamefont
  {{Sheth}}}]{KwanScoccimarro1201}%
  \BibitemOpen
  \bibfield  {author} {\bibinfo {author} {\bibfnamefont {K.~C.}\ \bibnamefont
  {{Chan}}}, \bibinfo {author} {\bibfnamefont {R.}~\bibnamefont
  {{Scoccimarro}}}, \ and\ \bibinfo {author} {\bibfnamefont {R.~K.}\
  \bibnamefont {{Sheth}}},\ }\href {\doibase 10.1103/PhysRevD.85.083509}
  {\bibfield  {journal} {\bibinfo  {journal} {\prd}\ }\textbf {\bibinfo
  {volume} {85}},\ \bibinfo {eid} {083509} (\bibinfo {year} {2012})},\ \Eprint
  {http://arxiv.org/abs/1201.3614} {arXiv:1201.3614 [astro-ph.CO]} \BibitemShut
  {NoStop}%
\bibitem [{\citenamefont {{Scoccimarro}}\ and\ \citenamefont
  {{Couchman}}(2001)}]{scocci_fitting_formula}%
  \BibitemOpen
  \bibfield  {author} {\bibinfo {author} {\bibfnamefont {R.}~\bibnamefont
  {{Scoccimarro}}}\ and\ \bibinfo {author} {\bibfnamefont {H.~M.~P.}\
  \bibnamefont {{Couchman}}},\ }\href {\doibase
  10.1046/j.1365-8711.2001.04281.x} {\bibfield  {journal} {\bibinfo  {journal}
  {\mnras}\ }\textbf {\bibinfo {volume} {325}},\ \bibinfo {pages} {1312}
  (\bibinfo {year} {2001})},\ \Eprint {http://arxiv.org/abs/astro-ph/0009427}
  {astro-ph/0009427} \BibitemShut {NoStop}%
\bibitem [{\citenamefont {{Gil-Mar{\'{\i}}n}}\ \emph
  {et~al.}(2012)\citenamefont {{Gil-Mar{\'{\i}}n}}, \citenamefont {{Wagner}},
  \citenamefont {{Fragkoudi}}, \citenamefont {{Jimenez}},\ and\ \citenamefont
  {{Verde}}}]{hector1111}%
  \BibitemOpen
  \bibfield  {author} {\bibinfo {author} {\bibfnamefont {H.}~\bibnamefont
  {{Gil-Mar{\'{\i}}n}}}, \bibinfo {author} {\bibfnamefont {C.}~\bibnamefont
  {{Wagner}}}, \bibinfo {author} {\bibfnamefont {F.}~\bibnamefont
  {{Fragkoudi}}}, \bibinfo {author} {\bibfnamefont {R.}~\bibnamefont
  {{Jimenez}}}, \ and\ \bibinfo {author} {\bibfnamefont {L.}~\bibnamefont
  {{Verde}}},\ }\href {\doibase 10.1088/1475-7516/2012/02/047} {\bibfield
  {journal} {\bibinfo  {journal} {\jcap}\ }\textbf {\bibinfo {volume} {2}},\
  \bibinfo {eid} {047} (\bibinfo {year} {2012})},\ \Eprint
  {http://arxiv.org/abs/1111.4477} {arXiv:1111.4477 [astro-ph.CO]} \BibitemShut
  {NoStop}%
\bibitem [{\citenamefont {{Baldauf}}\ \emph
  {et~al.}(2014{\natexlab{a}})\citenamefont {{Baldauf}}, \citenamefont
  {{Mercolli}}, \citenamefont {{Mirbabayi}},\ and\ \citenamefont
  {{Pajer}}}]{tobias1406EFT}%
  \BibitemOpen
  \bibfield  {author} {\bibinfo {author} {\bibfnamefont {T.}~\bibnamefont
  {{Baldauf}}}, \bibinfo {author} {\bibfnamefont {L.}~\bibnamefont
  {{Mercolli}}}, \bibinfo {author} {\bibfnamefont {M.}~\bibnamefont
  {{Mirbabayi}}}, \ and\ \bibinfo {author} {\bibfnamefont {E.}~\bibnamefont
  {{Pajer}}},\ }\href@noop {} {\bibfield  {journal} {\bibinfo  {journal} {ArXiv
  e-prints}\ } (\bibinfo {year} {2014}{\natexlab{a}})},\ \Eprint
  {http://arxiv.org/abs/1406.4135} {arXiv:1406.4135} \BibitemShut {NoStop}%
\bibitem [{\citenamefont {{Angulo}}\ \emph {et~al.}(2014)\citenamefont
  {{Angulo}}, \citenamefont {{Foreman}}, \citenamefont {{Schmittfull}},\ and\
  \citenamefont {{Senatore}}}]{angulo1406EFT}%
  \BibitemOpen
  \bibfield  {author} {\bibinfo {author} {\bibfnamefont {R.~E.}\ \bibnamefont
  {{Angulo}}}, \bibinfo {author} {\bibfnamefont {S.}~\bibnamefont {{Foreman}}},
  \bibinfo {author} {\bibfnamefont {M.}~\bibnamefont {{Schmittfull}}}, \ and\
  \bibinfo {author} {\bibfnamefont {L.}~\bibnamefont {{Senatore}}},\
  }\href@noop {} {\bibfield  {journal} {\bibinfo  {journal} {ArXiv e-prints}\ }
  (\bibinfo {year} {2014})},\ \Eprint {http://arxiv.org/abs/1406.4143}
  {arXiv:1406.4143} \BibitemShut {NoStop}%
\bibitem [{\citenamefont {{Neyrinck}}\ \emph {et~al.}(2009)\citenamefont
  {{Neyrinck}}, \citenamefont {{Szapudi}},\ and\ \citenamefont
  {{Szalay}}}]{Neyrinck0903}%
  \BibitemOpen
  \bibfield  {author} {\bibinfo {author} {\bibfnamefont {M.~C.}\ \bibnamefont
  {{Neyrinck}}}, \bibinfo {author} {\bibfnamefont {I.}~\bibnamefont
  {{Szapudi}}}, \ and\ \bibinfo {author} {\bibfnamefont {A.~S.}\ \bibnamefont
  {{Szalay}}},\ }\href {\doibase 10.1088/0004-637X/698/2/L90} {\bibfield
  {journal} {\bibinfo  {journal} {\apjl}\ }\textbf {\bibinfo {volume} {698}},\
  \bibinfo {pages} {L90} (\bibinfo {year} {2009})},\ \Eprint
  {http://arxiv.org/abs/0903.4693} {arXiv:0903.4693 [astro-ph.CO]} \BibitemShut
  {NoStop}%
\bibitem [{\citenamefont {{Simpson}}\ \emph {et~al.}(2011)\citenamefont
  {{Simpson}}, \citenamefont {{James}}, \citenamefont {{Heavens}},\ and\
  \citenamefont {{Heymans}}}]{SimpsonClipping1107}%
  \BibitemOpen
  \bibfield  {author} {\bibinfo {author} {\bibfnamefont {F.}~\bibnamefont
  {{Simpson}}}, \bibinfo {author} {\bibfnamefont {J.~B.}\ \bibnamefont
  {{James}}}, \bibinfo {author} {\bibfnamefont {A.~F.}\ \bibnamefont
  {{Heavens}}}, \ and\ \bibinfo {author} {\bibfnamefont {C.}~\bibnamefont
  {{Heymans}}},\ }\href {\doibase 10.1103/PhysRevLett.107.271301} {\bibfield
  {journal} {\bibinfo  {journal} {Physical Review Letters}\ }\textbf {\bibinfo
  {volume} {107}},\ \bibinfo {eid} {271301} (\bibinfo {year} {2011})},\ \Eprint
  {http://arxiv.org/abs/1107.5169} {arXiv:1107.5169 [astro-ph.CO]} \BibitemShut
  {NoStop}%
\bibitem [{\citenamefont {{Baldauf}}\ \emph
  {et~al.}(2014{\natexlab{b}})\citenamefont {{Baldauf}}, \citenamefont
  {{Desjacques}},\ and\ \citenamefont {{Seljak}}}]{Tobias14}%
  \BibitemOpen
  \bibfield  {author} {\bibinfo {author} {\bibfnamefont {T.}~\bibnamefont
  {{Baldauf}}}, \bibinfo {author} {\bibfnamefont {V.}~\bibnamefont
  {{Desjacques}}}, \ and\ \bibinfo {author} {\bibfnamefont {U.}~\bibnamefont
  {{Seljak}}},\ }\href@noop {} {\bibfield  {journal} {\bibinfo  {journal}
  {ArXiv e-prints}\ } (\bibinfo {year} {2014}{\natexlab{b}})},\ \Eprint
  {http://arxiv.org/abs/1405.5885} {arXiv:1405.5885} \BibitemShut {NoStop}%
\bibitem [{\citenamefont {{Crocce}}\ \emph {et~al.}(2012)\citenamefont
  {{Crocce}}, \citenamefont {{Scoccimarro}},\ and\ \citenamefont
  {{Bernardeau}}}]{MPTBREEZE}%
  \BibitemOpen
  \bibfield  {author} {\bibinfo {author} {\bibfnamefont {M.}~\bibnamefont
  {{Crocce}}}, \bibinfo {author} {\bibfnamefont {R.}~\bibnamefont
  {{Scoccimarro}}}, \ and\ \bibinfo {author} {\bibfnamefont {F.}~\bibnamefont
  {{Bernardeau}}},\ }\href {\doibase 10.1111/j.1365-2966.2012.22127.x}
  {\bibfield  {journal} {\bibinfo  {journal} {\mnras}\ }\textbf {\bibinfo
  {volume} {427}},\ \bibinfo {pages} {2537} (\bibinfo {year} {2012})},\ \Eprint
  {http://arxiv.org/abs/1207.1465} {arXiv:1207.1465 [astro-ph.CO]} \BibitemShut
  {NoStop}%
\bibitem [{\citenamefont {{Taruya}}\ \emph {et~al.}(2012)\citenamefont
  {{Taruya}}, \citenamefont {{Bernardeau}}, \citenamefont {{Nishimichi}},\ and\
  \citenamefont {{Codis}}}]{taruyaRegPT}%
  \BibitemOpen
  \bibfield  {author} {\bibinfo {author} {\bibfnamefont {A.}~\bibnamefont
  {{Taruya}}}, \bibinfo {author} {\bibfnamefont {F.}~\bibnamefont
  {{Bernardeau}}}, \bibinfo {author} {\bibfnamefont {T.}~\bibnamefont
  {{Nishimichi}}}, \ and\ \bibinfo {author} {\bibfnamefont {S.}~\bibnamefont
  {{Codis}}},\ }\href {\doibase 10.1103/PhysRevD.86.103528} {\bibfield
  {journal} {\bibinfo  {journal} {\prd}\ }\textbf {\bibinfo {volume} {86}},\
  \bibinfo {eid} {103528} (\bibinfo {year} {2012})},\ \Eprint
  {http://arxiv.org/abs/1208.1191} {arXiv:1208.1191 [astro-ph.CO]} \BibitemShut
  {NoStop}%
\bibitem [{\citenamefont {{Heitmann}}\ \emph {et~al.}(2014)\citenamefont
  {{Heitmann}}, \citenamefont {{Lawrence}}, \citenamefont {{Kwan}},
  \citenamefont {{Habib}},\ and\ \citenamefont {{Higdon}}}]{FrankenEmuExt}%
  \BibitemOpen
  \bibfield  {author} {\bibinfo {author} {\bibfnamefont {K.}~\bibnamefont
  {{Heitmann}}}, \bibinfo {author} {\bibfnamefont {E.}~\bibnamefont
  {{Lawrence}}}, \bibinfo {author} {\bibfnamefont {J.}~\bibnamefont {{Kwan}}},
  \bibinfo {author} {\bibfnamefont {S.}~\bibnamefont {{Habib}}}, \ and\
  \bibinfo {author} {\bibfnamefont {D.}~\bibnamefont {{Higdon}}},\ }\href
  {\doibase 10.1088/0004-637X/780/1/111} {\bibfield  {journal} {\bibinfo
  {journal} {\apj}\ }\textbf {\bibinfo {volume} {780}},\ \bibinfo {eid} {111}
  (\bibinfo {year} {2014})},\ \Eprint {http://arxiv.org/abs/1304.7849}
  {arXiv:1304.7849 [astro-ph.CO]} \BibitemShut {NoStop}%
\bibitem [{\citenamefont {{Heitmann}}\ \emph {et~al.}(2010)\citenamefont
  {{Heitmann}}, \citenamefont {{White}}, \citenamefont {{Wagner}},
  \citenamefont {{Habib}},\ and\ \citenamefont {{Higdon}}}]{Emu1}%
  \BibitemOpen
  \bibfield  {author} {\bibinfo {author} {\bibfnamefont {K.}~\bibnamefont
  {{Heitmann}}}, \bibinfo {author} {\bibfnamefont {M.}~\bibnamefont {{White}}},
  \bibinfo {author} {\bibfnamefont {C.}~\bibnamefont {{Wagner}}}, \bibinfo
  {author} {\bibfnamefont {S.}~\bibnamefont {{Habib}}}, \ and\ \bibinfo
  {author} {\bibfnamefont {D.}~\bibnamefont {{Higdon}}},\ }\href {\doibase
  10.1088/0004-637X/715/1/104} {\bibfield  {journal} {\bibinfo  {journal}
  {\apj}\ }\textbf {\bibinfo {volume} {715}},\ \bibinfo {pages} {104} (\bibinfo
  {year} {2010})},\ \Eprint {http://arxiv.org/abs/0812.1052} {arXiv:0812.1052}
  \BibitemShut {NoStop}%
\bibitem [{\citenamefont {{Heitmann}}\ \emph {et~al.}(2009)\citenamefont
  {{Heitmann}}, \citenamefont {{Higdon}}, \citenamefont {{White}},
  \citenamefont {{Habib}}, \citenamefont {{Williams}}, \citenamefont
  {{Lawrence}},\ and\ \citenamefont {{Wagner}}}]{Emu2}%
  \BibitemOpen
  \bibfield  {author} {\bibinfo {author} {\bibfnamefont {K.}~\bibnamefont
  {{Heitmann}}}, \bibinfo {author} {\bibfnamefont {D.}~\bibnamefont
  {{Higdon}}}, \bibinfo {author} {\bibfnamefont {M.}~\bibnamefont {{White}}},
  \bibinfo {author} {\bibfnamefont {S.}~\bibnamefont {{Habib}}}, \bibinfo
  {author} {\bibfnamefont {B.~J.}\ \bibnamefont {{Williams}}}, \bibinfo
  {author} {\bibfnamefont {E.}~\bibnamefont {{Lawrence}}}, \ and\ \bibinfo
  {author} {\bibfnamefont {C.}~\bibnamefont {{Wagner}}},\ }\href {\doibase
  10.1088/0004-637X/705/1/156} {\bibfield  {journal} {\bibinfo  {journal}
  {\apj}\ }\textbf {\bibinfo {volume} {705}},\ \bibinfo {pages} {156} (\bibinfo
  {year} {2009})},\ \Eprint {http://arxiv.org/abs/0902.0429} {arXiv:0902.0429
  [astro-ph.CO]} \BibitemShut {NoStop}%
\bibitem [{\citenamefont {{Lawrence}}\ \emph {et~al.}(2010)\citenamefont
  {{Lawrence}}, \citenamefont {{Heitmann}}, \citenamefont {{White}},
  \citenamefont {{Higdon}}, \citenamefont {{Wagner}}, \citenamefont {{Habib}},\
  and\ \citenamefont {{Williams}}}]{Emu3}%
  \BibitemOpen
  \bibfield  {author} {\bibinfo {author} {\bibfnamefont {E.}~\bibnamefont
  {{Lawrence}}}, \bibinfo {author} {\bibfnamefont {K.}~\bibnamefont
  {{Heitmann}}}, \bibinfo {author} {\bibfnamefont {M.}~\bibnamefont {{White}}},
  \bibinfo {author} {\bibfnamefont {D.}~\bibnamefont {{Higdon}}}, \bibinfo
  {author} {\bibfnamefont {C.}~\bibnamefont {{Wagner}}}, \bibinfo {author}
  {\bibfnamefont {S.}~\bibnamefont {{Habib}}}, \ and\ \bibinfo {author}
  {\bibfnamefont {B.}~\bibnamefont {{Williams}}},\ }\href {\doibase
  10.1088/0004-637X/713/2/1322} {\bibfield  {journal} {\bibinfo  {journal}
  {\apj}\ }\textbf {\bibinfo {volume} {713}},\ \bibinfo {pages} {1322}
  (\bibinfo {year} {2010})},\ \Eprint {http://arxiv.org/abs/0912.4490}
  {arXiv:0912.4490 [astro-ph.CO]} \BibitemShut {NoStop}%
\bibitem [{\citenamefont {Jeong}(2010)}]{jeongThesis}%
  \BibitemOpen
  \bibfield  {author} {\bibinfo {author} {\bibfnamefont {D.}~\bibnamefont
  {Jeong}},\ }\href
  {http://www.personal.psu.edu/duj13/dissertation/djeong_diss.pdf} {\bibfield
  {journal} {\bibinfo  {journal} {PhD dissertation,
  \url{http://www.personal.psu.edu/duj13/dissertation/djeong_diss.pdf}}\ }
  (\bibinfo {year} {2010})}\BibitemShut {NoStop}%
\bibitem [{\citenamefont {{Baldauf}}\ \emph {et~al.}(2013)\citenamefont
  {{Baldauf}}, \citenamefont {{Seljak}}, \citenamefont {{Smith}}, \citenamefont
  {{Hamaus}},\ and\ \citenamefont {{Desjacques}}}]{tobias1305}%
  \BibitemOpen
  \bibfield  {author} {\bibinfo {author} {\bibfnamefont {T.}~\bibnamefont
  {{Baldauf}}}, \bibinfo {author} {\bibfnamefont {U.}~\bibnamefont {{Seljak}}},
  \bibinfo {author} {\bibfnamefont {R.~E.}\ \bibnamefont {{Smith}}}, \bibinfo
  {author} {\bibfnamefont {N.}~\bibnamefont {{Hamaus}}}, \ and\ \bibinfo
  {author} {\bibfnamefont {V.}~\bibnamefont {{Desjacques}}},\ }\href {\doibase
  10.1103/PhysRevD.88.083507} {\bibfield  {journal} {\bibinfo  {journal}
  {\prd}\ }\textbf {\bibinfo {volume} {88}},\ \bibinfo {eid} {083507} (\bibinfo
  {year} {2013})},\ \Eprint {http://arxiv.org/abs/1305.2917} {arXiv:1305.2917
  [astro-ph.CO]} \BibitemShut {NoStop}%
\bibitem [{\citenamefont {{Reid}}\ \emph {et~al.}(2014)\citenamefont {{Reid}},
  \citenamefont {{Seo}}, \citenamefont {{Leauthaud}}, \citenamefont
  {{Tinker}},\ and\ \citenamefont {{White}}}]{beth1404}%
  \BibitemOpen
  \bibfield  {author} {\bibinfo {author} {\bibfnamefont {B.~A.}\ \bibnamefont
  {{Reid}}}, \bibinfo {author} {\bibfnamefont {H.-J.}\ \bibnamefont {{Seo}}},
  \bibinfo {author} {\bibfnamefont {A.}~\bibnamefont {{Leauthaud}}}, \bibinfo
  {author} {\bibfnamefont {J.~L.}\ \bibnamefont {{Tinker}}}, \ and\ \bibinfo
  {author} {\bibfnamefont {M.}~\bibnamefont {{White}}},\ }\href {\doibase
  10.1093/mnras/stu1391} {\bibfield  {journal} {\bibinfo  {journal} {\mnras}\
  }\textbf {\bibinfo {volume} {444}},\ \bibinfo {pages} {476} (\bibinfo {year}
  {2014})},\ \Eprint {http://arxiv.org/abs/1404.3742} {arXiv:1404.3742}
  \BibitemShut {NoStop}%
\bibitem [{\citenamefont {{White}}\ \emph {et~al.}(2014)\citenamefont
  {{White}}, \citenamefont {{Reid}}, \citenamefont {{Chuang}}, \citenamefont
  {{Tinker}}, \citenamefont {{McBride}}, \citenamefont {{Prada}},\ and\
  \citenamefont {{Samushia}}}]{mwhite1408}%
  \BibitemOpen
  \bibfield  {author} {\bibinfo {author} {\bibfnamefont {M.}~\bibnamefont
  {{White}}}, \bibinfo {author} {\bibfnamefont {B.}~\bibnamefont {{Reid}}},
  \bibinfo {author} {\bibfnamefont {C.-H.}\ \bibnamefont {{Chuang}}}, \bibinfo
  {author} {\bibfnamefont {J.~L.}\ \bibnamefont {{Tinker}}}, \bibinfo {author}
  {\bibfnamefont {C.~K.}\ \bibnamefont {{McBride}}}, \bibinfo {author}
  {\bibfnamefont {F.}~\bibnamefont {{Prada}}}, \ and\ \bibinfo {author}
  {\bibfnamefont {L.}~\bibnamefont {{Samushia}}},\ }\href@noop {} {\bibfield
  {journal} {\bibinfo  {journal} {ArXiv e-prints}\ } (\bibinfo {year}
  {2014})},\ \Eprint {http://arxiv.org/abs/1408.5435} {arXiv:1408.5435}
  \BibitemShut {NoStop}%
\bibitem [{\citenamefont {{White}}(2002)}]{mwhiteTreePM2002}%
  \BibitemOpen
  \bibfield  {author} {\bibinfo {author} {\bibfnamefont {M.}~\bibnamefont
  {{White}}},\ }\href {\doibase 10.1086/342752} {\bibfield  {journal} {\bibinfo
   {journal} {\apjs}\ }\textbf {\bibinfo {volume} {143}},\ \bibinfo {pages}
  {241} (\bibinfo {year} {2002})},\ \Eprint
  {http://arxiv.org/abs/astro-ph/0207185} {astro-ph/0207185} \BibitemShut
  {NoStop}%
\bibitem [{\citenamefont {Lewis}\ \emph {et~al.}(2000)\citenamefont {Lewis},
  \citenamefont {Challinor},\ and\ \citenamefont {Lasenby}}]{camb}%
  \BibitemOpen
  \bibfield  {author} {\bibinfo {author} {\bibfnamefont {A.}~\bibnamefont
  {Lewis}}, \bibinfo {author} {\bibfnamefont {A.}~\bibnamefont {Challinor}}, \
  and\ \bibinfo {author} {\bibfnamefont {A.}~\bibnamefont {Lasenby}},\
  }\href@noop {} {\bibfield  {journal} {\bibinfo  {journal} {Astrophys. J.}\
  }\textbf {\bibinfo {volume} {538}},\ \bibinfo {pages} {473} (\bibinfo {year}
  {2000})},\ \Eprint {http://arxiv.org/abs/astro-ph/9911177} {astro-ph/9911177}
  \BibitemShut {NoStop}%
\bibitem [{\citenamefont {{Foreman-Mackey}}\ \emph {et~al.}(2013)\citenamefont
  {{Foreman-Mackey}}, \citenamefont {{Hogg}}, \citenamefont {{Lang}},\ and\
  \citenamefont {{Goodman}}}]{emcee}%
  \BibitemOpen
  \bibfield  {author} {\bibinfo {author} {\bibfnamefont {D.}~\bibnamefont
  {{Foreman-Mackey}}}, \bibinfo {author} {\bibfnamefont {D.~W.}\ \bibnamefont
  {{Hogg}}}, \bibinfo {author} {\bibfnamefont {D.}~\bibnamefont {{Lang}}}, \
  and\ \bibinfo {author} {\bibfnamefont {J.}~\bibnamefont {{Goodman}}},\ }\href
  {\doibase 10.1086/670067} {\bibfield  {journal} {\bibinfo  {journal}
  {Publications of the Astronomical Society of the Pacific}\ }\textbf {\bibinfo
  {volume} {125}},\ \bibinfo {pages} {306} (\bibinfo {year} {2013})},\ \Eprint
  {http://arxiv.org/abs/1202.3665} {arXiv:1202.3665 [astro-ph.IM]} \BibitemShut
  {NoStop}%
\bibitem [{\citenamefont {{Valageas}}(2014)}]{valageas1311}%
  \BibitemOpen
  \bibfield  {author} {\bibinfo {author} {\bibfnamefont {P.}~\bibnamefont
  {{Valageas}}},\ }\href {\doibase 10.1103/PhysRevD.89.123522} {\bibfield
  {journal} {\bibinfo  {journal} {\prd}\ }\textbf {\bibinfo {volume} {89}},\
  \bibinfo {eid} {123522} (\bibinfo {year} {2014})},\ \Eprint
  {http://arxiv.org/abs/1311.4286} {arXiv:1311.4286} \BibitemShut {NoStop}%
\bibitem [{\citenamefont {{Kehagias}}\ \emph {et~al.}(2013)\citenamefont
  {{Kehagias}}, \citenamefont {{Perrier}},\ and\ \citenamefont
  {{Riotto}}}]{Kehagias1311}%
  \BibitemOpen
  \bibfield  {author} {\bibinfo {author} {\bibfnamefont {A.}~\bibnamefont
  {{Kehagias}}}, \bibinfo {author} {\bibfnamefont {H.}~\bibnamefont
  {{Perrier}}}, \ and\ \bibinfo {author} {\bibfnamefont {A.}~\bibnamefont
  {{Riotto}}},\ }\href@noop {} {\bibfield  {journal} {\bibinfo  {journal}
  {ArXiv e-prints}\ } (\bibinfo {year} {2013})},\ \Eprint
  {http://arxiv.org/abs/1311.5524} {arXiv:1311.5524 [astro-ph.CO]} \BibitemShut
  {NoStop}%
\bibitem [{\citenamefont {{Li}}\ \emph {et~al.}(2014)\citenamefont {{Li}},
  \citenamefont {{Hu}},\ and\ \citenamefont {{Takada}}}]{LiHu1401}%
  \BibitemOpen
  \bibfield  {author} {\bibinfo {author} {\bibfnamefont {Y.}~\bibnamefont
  {{Li}}}, \bibinfo {author} {\bibfnamefont {W.}~\bibnamefont {{Hu}}}, \ and\
  \bibinfo {author} {\bibfnamefont {M.}~\bibnamefont {{Takada}}},\ }\href
  {\doibase 10.1103/PhysRevD.89.083519} {\bibfield  {journal} {\bibinfo
  {journal} {\prd}\ }\textbf {\bibinfo {volume} {89}},\ \bibinfo {eid} {083519}
  (\bibinfo {year} {2014})},\ \Eprint {http://arxiv.org/abs/1401.0385}
  {arXiv:1401.0385} \BibitemShut {NoStop}%
\bibitem [{\citenamefont {{Figueroa}}\ \emph {et~al.}(2012)\citenamefont
  {{Figueroa}}, \citenamefont {{Sefusatti}}, \citenamefont {{Riotto}},\ and\
  \citenamefont {{Vernizzi}}}]{figueroa1205}%
  \BibitemOpen
  \bibfield  {author} {\bibinfo {author} {\bibfnamefont {D.~G.}\ \bibnamefont
  {{Figueroa}}}, \bibinfo {author} {\bibfnamefont {E.}~\bibnamefont
  {{Sefusatti}}}, \bibinfo {author} {\bibfnamefont {A.}~\bibnamefont
  {{Riotto}}}, \ and\ \bibinfo {author} {\bibfnamefont {F.}~\bibnamefont
  {{Vernizzi}}},\ }\href {\doibase 10.1088/1475-7516/2012/08/036} {\bibfield
  {journal} {\bibinfo  {journal} {\jcap}\ }\textbf {\bibinfo {volume} {8}},\
  \bibinfo {eid} {036} (\bibinfo {year} {2012})},\ \Eprint
  {http://arxiv.org/abs/1205.2015} {arXiv:1205.2015 [astro-ph.CO]} \BibitemShut
  {NoStop}%
\end{thebibliography}%

\end{document}